\documentclass[reprint,aps,prd,a4paper,showpacs]{revtex4-1}

\usepackage{atlasphysics}
\usepackage{graphicx}
\usepackage{subfigure}
\usepackage{mathrsfs}
\usepackage{multirow}
\usepackage{url}
\usepackage{verbatim}
\usepackage{slashed}
\usepackage[hyperindex,breaklinks,colorlinks]{hyperref} 
\usepackage{mathptmx} 
\usepackage{lineno}

\usepackage{preprintcover}  
\PreprintCoverPaperTitle{Search for pair-produced third-generation squarks decaying via charm quarks or in compressed supersymmetric scenarios
in $pp$ collisions at $\sqrt{s}=8~$TeV with the ATLAS detector}
\PreprintIdNumber{CERN-PH-EP-2014-141}  
\PreprintCoverAbstract{
Results of a search for supersymmetry via direct production of third-generation squarks 
are reported, using $20.3$~fb$^{-1}$ of proton-proton collision data
at $\sqrt{s} = 8$~TeV recorded by the ATLAS experiment at the LHC in 2012.
Two different analysis strategies based on monojet-like and
$c$-tagged event selections are
carried out to optimize the sensitivity for 
direct top squark pair production
in the decay channel to
a charm quark and the lightest  neutralino ($\tilde{t}_1 \to c + \tilde{\chi}_{1}^{0}$)   
across the top squark--neutralino mass parameter space. 
No excess above the Standard Model background expectation is
observed. 
The results are
interpreted in the context of direct pair production of top
squarks and presented in terms of exclusion limits
in the  ($m_{\tilde{t}_1}$, $m_{\tilde{\chi}_{1}^{0}}$) parameter space.
A top squark of mass up to  about 240~GeV is excluded at 95$\%$ confidence level
for arbitrary neutralino masses, within the kinematic boundaries.
Top squark masses up to 270~GeV are excluded for a neutralino mass of 200~GeV. 
In a scenario where the top squark  and the lightest neutralino are nearly 
degenerate in mass, top squark  
masses up to 260~GeV are excluded.  
The results from the monojet-like analysis are also 
interpreted in terms of compressed scenarios for 
top squark pair production in the decay channel $\tilde{t}_1 \to b + ff^{'} + \tilde{\chi}^{0}_{1}$ and 
sbottom pair production with $\tilde{b}_1 \to b + \tilde{\chi}^{0}_{1}$, leading to 
a similar exclusion for nearly mass-degenerate third-generation squarks and the lightest neutralino. The results in this paper 
significantly extend previous results at colliders. }
\PreprintJournalName{Physical Review D}  

\begin{document}
\title{Search for pair-produced third-generation squarks decaying via charm quarks or in compressed supersymmetric scenarios
in $pp$ collisions at $\sqrt{s}=8~$TeV with the ATLAS detector}

\author{The ATLAS Collaboration}

\date{\today}

\begin{abstract}

Results of a search for supersymmetry via direct production of third-generation squarks 
are reported, using $20.3$~fb$^{-1}$ of proton-proton collision data
at $\sqrt{s} = 8$~TeV recorded by the ATLAS experiment at the LHC in 2012.
Two different analysis strategies based on monojet-like and
$c$-tagged event selections are
carried out to optimize the sensitivity for
direct top squark pair production
in the decay channel to
a charm quark and the lightest  neutralino ($\tilde{t}_1 \to c + \tilde{\chi}_{1}^{0}$)
across the top squark--neutralino mass parameter space.
No excess above the Standard Model background expectation is
observed.
The results are
interpreted in the context of direct pair production of top
squarks and presented in terms of exclusion limits
in the  ($m_{\tilde{t}_1}$, $m_{\tilde{\chi}_{1}^{0}}$) parameter space.
A top squark of mass up to  about 240~GeV is excluded at 95$\%$ confidence level
for arbitrary neutralino masses, within the kinematic boundaries.
Top squark masses up to 270~GeV are excluded for a neutralino mass of 200~GeV.
In a scenario where the top squark  and the lightest neutralino are nearly
degenerate in mass, top squark
masses up to 260~GeV are excluded.
The results from the monojet-like analysis are also
interpreted in terms of compressed scenarios for
top squark pair production in the decay channel $\tilde{t}_1 \to b + ff^{'} + \tilde{\chi}^{0}_{1}$ and
sbottom pair production with $\tilde{b}_1 \to b + \tilde{\chi}^{0}_{1}$, leading to
a similar exclusion for nearly mass-degenerate third-generation squarks and the lightest neutralino. The results in this paper
significantly extend previous results at colliders.

\end{abstract}


\maketitle

\hyphenation{ATLAS}

\newcommand{\pho}{\phantom{0}}
\newcommand{\bslash}{\ensuremath{\backslash}}
\newcommand{\BibTeX}{{\sc Bib\TeX}}
\newcommand{\ptj}{p_{\rm T}^{\rm jet}}
\newcommand{\ptjet}{p_{\rm  T}}
\newcommand{\ptmi}{{\bf{p}}_{\rm T}^{\rm miss}}
\newcommand{\ptjetem}{p_{\rm T}^{\rm jet,em}}
\newcommand{\etajet}{\eta^{\rm jet}}
\newcommand{\phijet}{\phi}
\newcommand{\rapjet}{y}
\newcommand{\pthat}{\hat{p}_{\rm  T}}
\newcommand{\akt}{\hbox{anti-${k_t}$} }
\newcommand{\njet}{N_{\rm jet}}
\newcommand{\pttrk}{\pt^{\rm \ track}}
\newcommand{\etatrk}{\eta^{\rm \ track}}
\newcommand{\ttb}{t\bar{t}}
\newcommand{\zee}{Z/\gamma^*~(\rightarrow e^+e^-)}
\newcommand{\zmm}{Z/\gamma^*~(\rightarrow \mu^+\mu^-)}
\newcommand{\znn}{Z(\rightarrow \nu \bar{\nu})}
\newcommand{\ztt}{Z/\gamma^*(\rightarrow \tau^+ \tau^-)}
\newcommand{\zll}{Z/\gamma^*(\rightarrow \ell^+ \ell^-)}
\newcommand{\wen}{W(\rightarrow e \nu)}
\newcommand{\wmn}{W(\rightarrow \mu \nu)}
\newcommand{\wtn}{W(\rightarrow \tau \nu)}
\newcommand{\wln}{W(\rightarrow \ell \nu)}
\newcommand{\ete}{E_{\rm T}^{e}}
\newcommand{\etae}{\eta^{e}}
\newcommand{\ptm}{p_{\rm T}^{\mu}}
\newcommand{\etam}{\eta^{\mu}}
\newcommand{\mee}{M_{e^+ e^-}}
\newcommand{\mmm}{M_{\mu^+ \mu^-}}
\newcommand{\mchi}{m_{\chi}}
\newcommand{\gra}{\tilde{G}}
\newcommand{\mgra}{m_{\gra}}
\newcommand{\glu}{\tilde{g}}
\newcommand{\mglu}{m_{\glu}}
\newcommand{\squ}{\tilde{q}}
\newcommand{\msqu}{m_{\squ}}
\newcommand{\msqgl}{m_{\squ,\glu}}
\newcommand{\mtop}{m_{t}}
\newcommand{\mb}{m_{b}}

\newcommand{\light}{u}
\newcommand{\charm}{c}
\newcommand{\bottom}{b}
\newcommand{\pu}{P_{\light}}
\newcommand{\pc}{P_{\charm}}
\newcommand{\pb}{P_{\bottom}}
\newcommand{\ISR}{ISR}

\def\chaj{\ensuremath{\mathchoice%
      {\displaystyle\raise.4ex\hbox{$\displaystyle\tilde\chi^\pm_j$}}%
         {\textstyle\raise.4ex\hbox{$\textstyle\tilde\chi^\pm_j$}}%
       {\scriptstyle\raise.3ex\hbox{$\scriptstyle\tilde\chi^\pm_j$}}%
 {\scriptscriptstyle\raise.3ex\hbox{$\scriptscriptstyle\tilde\chi^\pm_j$}}}}

\def\neuj{\ensuremath{\mathchoice%
      {\displaystyle\raise.4ex\hbox{$\displaystyle\tilde\chi^0_j$}}%
         {\textstyle\raise.4ex\hbox{$\textstyle\tilde\chi^0_j$}}%
       {\scriptstyle\raise.3ex\hbox{$\scriptstyle\tilde\chi^0_j$}}%
 {\scriptscriptstyle\raise.3ex\hbox{$\scriptscriptstyle\tilde\chi^0_j$}}}}

\section{Introduction}
\label{sec:intro}

Supersymmetry
(SUSY)~\cite{Miyazawa:1966,Ramond:1971gb,Golfand:1971iw,Neveu:1971rx,Neveu:1971iv,Gervais:1971ji,Volkov:1973ix,Wess:1973kz,Wess:1974tw}
is a theoretically favored candidate
for physics beyond the Standard Model (SM).  It 
naturally solves the hierarchy
problem and provides a possible candidate for dark
matter in the universe. SUSY enlarges the SM spectrum
of particles by introducing a new supersymmetric partner
(sparticle) for each particle in the SM. In particular,
a new scalar field is associated with each left- and right-handed
quark state, and two squark mass eigenstates $\tilde{q}_1$
and $\tilde{q}_2$ result from the mixing of the scalar fields. 
In some
SUSY scenarios, 
a significant mass difference between the two eigenstates
in the bottom squark and top squark sectors 
can occur, leading to 
rather light sbottom $\tilde{b}_1$ and stop $\tilde{t}_1$ mass states, 
where sbottom and stop are the SUSY  partners of the SM bottom and top quarks, respectively. 
In addition, naturalness arguments suggest that the third generation squarks 
should be light with masses below 1~TeV~\cite{Barbieri:1987fn,deCarlos:1993yy}. 
In a generic 
supersymmetric extension of the SM   that assumes 
R-parity conservation~\cite{Fayet:1976et,Fayet:1977yc,Farrar:1978xj,Fayet:1979sa,Dimopoulos:1981zb}, sparticles are produced
in pairs and the lightest supersymmetric particle (LSP)
is stable. In this paper the LSP is assumed to be the lightest neutralino~\footnote{
Neutralinos ($\neuj$, j=1--4 in the order of increasing mass) and charginos ($\chaj$, j=1,2)
are SUSY mass eigenstates formed from the mixing of the SUSY partners of the charged and neutral Higgs
and electroweak gauge bosons.} ($\ninoone$). 

\begin{figure}[!h]
   \begin{center}
\mbox{
   \includegraphics[width=0.21\textwidth]{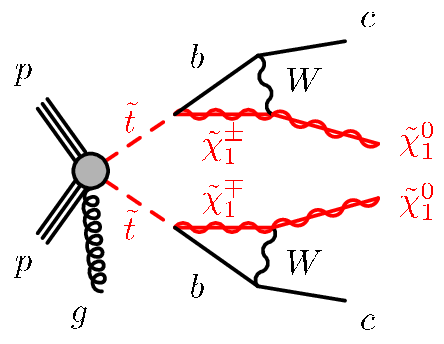}
   \includegraphics[width=0.21\textwidth]{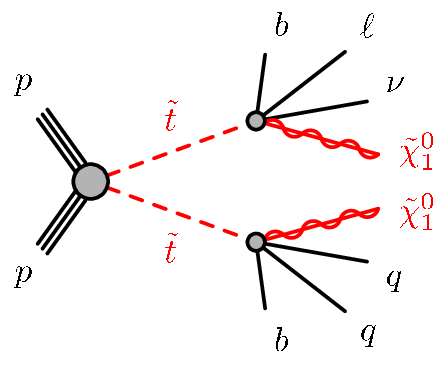}
}
   \includegraphics[width=0.21\textwidth]{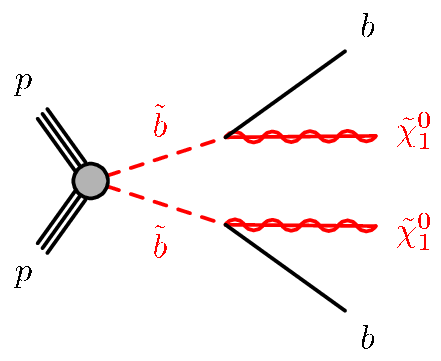}
   \end{center}
   \caption{Diagrams for the pair production of top squarks with the decay modes 
$\tilde{t}_1 \to c + \ninoone$ or $ \tilde{t}_1 \to  b + ff^{'} + \ninoone$,   and the pair production of sbottom squarks with the decay mode 
 $\tilde{b}_1 \to b + \ninoone$.
In one case, the presence of a jet from initial-state radiation is also indicated for illustration purposes.}
   \label{fig:stoptocharm}
\end{figure}

For a mass difference 
 $\Delta m \equiv m_{\tilde{t}_1} 
- m_{\ninoone}  >  m_{t}$ and depending on the SUSY
parameters and sparticle mass hierarchy,   
the dominant decay channels  are expected to be $\tilde{t}_1 \to t + \ninoone$ or  $\tilde{t}_1 \to b + \chinoonepm $,  
 where the latter decay mode involves charginos (\chinoonepm) that subsequently can  
decay into the lightest neutralino via  $W^{\rm (*)}$ emission, leading to a four-body decay $ \tilde{t}_1 \to  b + ff^{'} + \ninoone$, where
$ff^{'}$ denotes a pair of fermions (see Fig.~\ref{fig:stoptocharm}).  
If the chargino is heavier than the stop 
and $\mW\ + \mb < \Delta m  < m_{t} $, 
the dominant decay mode is expected to be the three-body $Wb\ninoone$ decay. 
Several searches on 7 TeV data have been carried out in
these decay channels in 0-, 1-, and 2-lepton final states~\cite{Aad:2012ywa, 
Aad:2012xqa, Aad:2012uu,Chatrchyan:2014goa} and have been extended using
8 TeV data~\cite{Aad:2014qaa, Aad:2013ija,Chatrchyan:2013xna,Aad:2014bva}.

In the scenario for which $\Delta m < \mW\ + \mb$,  the four-body decay mode above 
competes with the  stop decay to a charm quark and the LSP ($\tilde{t}_1 \to c + \ninoone$), 
which proceeds via a loop decay (see Fig.~\ref{fig:stoptocharm}).
 The corresponding final state 
 is characterized by the
 presence of two jets from the hadronization of
 the charm quarks and missing transverse momentum  ($\ptmi$, denoting its magnitude by $\met$) 
 from the two undetected LSPs.
However, given the relatively small mass
difference ($\Delta m$),  both the transverse momenta of the two charm
jets and the \met\ are low, making it very difficult to extract the signal from the large
multijet background.  
In this study, the event selection makes use of the presence of initial-state radiation (ISR) jets to identify signal events. In this case, 
the squark-pair system is boosted leading to larger $\met$. 
As an example, for a stop with a mass of 200 GeV and $\Delta m$ of 5 GeV, about 18$\%$ of the events have $\met> 150$~GeV 
and a jet with $\pt > 150$~GeV.
%
%
%
Two different approaches are used to maximize the sensitivity of the analysis across  the different $\Delta m$ regions.
A  ``monojet-like'' analysis is carried out, 
where events with low jet multiplicity and large $\met$ are selected, that is optimized for small $\Delta m$ 
($\Delta m \leq 20$ GeV).
For $\Delta m \geq 20$ GeV,  the charm jets receive a large enough boost to be
detected.  
In addition to the requirements on the presence of ISR jets, the identification of jets containing 
the decay products of charm hadrons ($c$-tagging) 
is used,  leading to a ``$c$-tagged'' analysis that 
further
enhances the sensitivity to 
the SUSY signal in the region $m_{\tilde{t}_1} > 200$~GeV and $\Delta m \geq 20$ GeV.   
 Results for
 searches in this channel  have been
 previously reported by collider  experiments~\cite{lepsusy,Aaltonen:2012tq,Abazov:2008rc}. 
In addition  to  the decay channel $\tilde{t}_1 \to c  + \ninoone$,
 the monojet-like results are  re-interpreted in terms of the search for stop 
pair production with $ \tilde{t}_1 \to  b + ff^{'} + \ninoone$ and small $\Delta m$.     
In such a scenario,  the
decay products of the top squark are too soft to be identified in the final state, and the signal selection relies
on the presence of an ISR jet.

In the case of sbottom pair production, assuming
a SUSY particle mass hierarchy such that the sbottom
decays exclusively as $\tilde{b}_1 \to b + \ninoone$ (see Fig.~\ref{fig:stoptocharm}),
the expected signal for
direct sbottom pair production is characterized by the
presence of two energetic jets from the hadronization of
the bottom quarks and large missing transverse momentum 
from the two LSPs in the final state.
Results on
searches in this channel at colliders have been reported~\cite{Aaltonen:2010dy,Abazov:2010wq,Aad:2011cw,Aad:2013ija,Chatrchyan:2014goa}.
In this study, 
the monojet-like results are also  re-interpreted in terms of the search for sbottom  
pair production with $\tilde{b}_1 \to b  + \ninoone$  
in a compressed scenario (small sbottom-neutralino mass difference) 
with two soft $b$-jets and an energetic ISR jet in the final state.

%
%

The paper is organized as follows. The ATLAS detector is described in the next section. 
Section~\ref{sec:mc} provides details
of the simulations used in the analysis for background and signal processes.
Section~\ref{sec:recons} discusses the  reconstruction of jets, leptons and 
the $\met$, while   
Sec.~\ref{sec:evt} 
describes the event selection.
The estimation of background contributions  and  
the study of systematic uncertainties are discussed in Secs.~\ref{sec:backg} and~\ref{sec:syst}. 
The results are presented in  Sec.~\ref{sec:results}, and are interpreted in terms 
of the search for stop and sbottom pair production. Finally, Sec.~\ref{sec:sum} is devoted to the conclusions.

\section{Experimental setup}
\label{sec:atlas}

The ATLAS detector~\cite{atlas-det} covers almost the whole solid angle around
the collision point with layers of tracking detectors, calorimeters and muon
chambers. The ATLAS inner detector (ID) has full coverage~\footnote{ATLAS uses a right-handed
  coordinate system with its origin at the nominal interaction point
  (IP) in the center of the detector and the $z$-axis along the beam
  pipe. The $x$-axis points from the IP to the center of the LHC ring,
  and the $y$-axis points upward.  The azimuthal angle $\phi$ is 
measured around the beam axis, and the polar angle $\theta$ is measured with respect to the $z$-axis.
We define transverse energy 
$E_{\rm T} = E \ {\rm sin}  \theta$,  
transverse momentum $\pt = p \ {\rm sin} \theta$, 
and pseudorapidity $\eta = - {\rm ln}({\rm tan}(\theta/2))$.}  in $\phi$
and covers the pseudorapidity range $|\eta|<2.5$. 
It consists of a silicon pixel detector, a silicon microstrip detector, and a 
straw tube tracker that also measures transition radiation for particle identification, all immersed in a 2 T axial  magnetic field produced by a solenoid.

High-granularity liquid-argon (LAr) electromagnetic sampling calorimeters, with excellent 
energy and position resolution, cover the pseudorapidity
range $|\eta|<$~3.2. The hadronic calorimetry in the range $|\eta|<$~1.7 is provided by a scintillator-tile calorimeter, consisting of a large barrel and 
 two smaller extended barrel cylinders, one on either side of
the central barrel. In the endcaps ($|\eta|>$~1.5), LAr hadronic
calorimeters match the outer $|\eta|$ limits of the endcap electromagnetic calorimeters. The LAr
forward calorimeters provide both the electromagnetic and hadronic energy measurements, and extend
the coverage to $|\eta| < 4.9$.

 The muon spectrometer measures the deflection of muon tracks in the large superconducting air-core toroid magnets
in the pseudorapidity range  $|\eta|<2.7$, using separate trigger and high-precision tracking chambers. 
Over most of the $\eta$-range, a precise  measurement of the track coordinates in the principal bending direction of the magnetic 
field is provided
 by monitored drift tubes. At large pseudorapidities, cathode strip chambers  with higher granularity are used in the innermost plane over $2.0 < |\eta| < 2.7$.
The muon trigger system covers the pseudorapidity range $|\eta| < 2.4$.


\section{Monte Carlo simulation}
\label{sec:mc}

Monte Carlo (MC) simulated event samples are used to assist in 
computing detector acceptance and reconstruction efficiencies,
determine signal and background contributions, and estimate systematic uncertainties on the final results. 

Samples of simulated $W$+jets and $Z$+jets events are generated using 
 {SHERPA}-1.4.1~\cite{sherpa}, including  LO matrix elements for up to 5 partons in the final state and  
using massive $b$/$c$-quarks,  with CT10~\cite{ct10} parton distribution functions (PDF) and its own
model for hadronization. 
Similar samples are generated using 
the  {ALPGEN}-v2.14~\cite{alpgen} generator and are employed to assess the 
corresponding modeling uncertainties.  
The MC predictions are initially 
normalized to next-to-next-to-leading-order (NNLO) predictions according to 
DYNNLO~\cite{wcrosssection1, wcrosssection}
using  MSTW2008 NNLO PDF sets~\cite{mstw}.

%
%

The production of top quark pairs ($\ttb$) is simulated using the 
 {POWHEG}-r2129~\cite{powheg} MC generator.  {ALPGEN}  and  {MC@NLO}-4.06~\cite{Frixione:2002ik}  
MC simulated samples are  
used to assess $\ttb$ modeling uncertainties. 
Single top production samples are generated with
 {POWHEG} for the $s$-- and $Wt$--channel and  {MC@NLO} is used to determine systematic uncertainties,  
while  {AcerMC}-v3.8~\cite{Kersevan:2002dd} is used for single top
production in the $t$--channel.  Finally,  samples of $\ttb$ production  associated with  additional vector
bosons ($\ttb + W$  and $\ttb + Z$ processes)  are 
generated with  {MADGRAPH}-5.1.4.8~\cite{madgraph}. 
In the case of  {POWHEG} and  {MADGRAPH}, parton showers are
implemented using  {PYTHIA}-6.426~\cite{pythia}, while  {HERWIG}-6.5.20~\cite{herwig} interfaced to
 {JIMMY}~\cite{Butterworth:1996zw}
is used for the  {ALPGEN} and  {MC@NLO} generators.
A top quark mass of 172.5~\GeV\ and the CTEQ6L1 PDFs are used. The Perugia 2011C~\cite{perugia} 
and AUET2B~\cite{ATLAStunes} tunes for the underlying event are used 
for the $\ttbar$, single top and  $\ttb + W/Z$ processes, respectively. 
The cross section prediction at NNLO+NNLL (next-to-next-to-leading-logarithm) accuracy, as determined by Top++2.0~\cite{Czakon:2011xx}, is used in the normalization 
of the  $\ttb$ ~\cite{Czakon:2013goa} sample. 
An approximate NLO+NNLL prediction is used for the $Wt$~\cite{Kidonakis:2010ux} process and NLO cross sections are considered for $\ttb + W$  and $\ttb + Z$  processes. 
%
%

Diboson samples ($WW$, $WZ$ and $ZZ$ production) are generated using
 {SHERPA}, using massive $b$/$c$-quarks, with CT10 PDFs, and are normalized 
to NLO predictions~\cite{Campbell:2011bn}. Additional samples are generated with  {HERWIG}  to assess
uncertainties.  
Finally,  Higgs boson production including $ZH$, $WH$  and $\ttbar H$ processes is generated using  {PYTHIA}-8.165~\cite{Sjostrand:2007gs} with CTEQ6L1 PDFs.

%
Stop pair production with $\tilde{t}_1 \to c + \tilde{\chi}_{1}^{0}$ is
modeled with  {MADGRAPH} with one
additional jet from the matrix element.  The showering is done with
 {PYTHIA}-6 and using the AUET2B tune for the underlying event,  which involves CTEQ6L1 PDFs. 
Samples are produced with stop masses between 100~\GeV\ and
400~\GeV\  and $\tilde{\chi}_{1}^{0}$ masses between
70~\GeV\ and 390~GeV. 
The $\Delta m$ step size increases with $\Delta m$ from 2~GeV to 30~GeV and
the maximum $\Delta m$ considered is 82~GeV. The region  $\Delta m < 2$~GeV 
is not considered since in this regime the stop can become long-lived
leading to the signature studied in Ref.~\cite{long-lived}.
Similarly, MC simulated samples are produced separately for 
$\tilde{t}_1 \to  b + ff^{'} + \ninoone$ and $\tilde{b}_1 \to  b + \ninoone$ processes
across the stop--neutralino and sbottom--neutralino mass planes. 
In the case of the $\tilde{t}_1 \to  b + ff^{'} + \ninoone$ process, samples are produced 
with stop masses in the range between 100~GeV and 300~GeV and 
$\Delta m$ that varies between 10~GeV and 80~GeV. For sbottom pair production 
with $\tilde{b}_1 \to  b + \ninoone$, samples are produced with 
sbottom masses in the range between 100~GeV and 350~GeV and $\tilde{\chi}_{1}^{0}$ masses
in the range between 1~GeV and 340~GeV, 
 with a sbottom--neutralino mass difference that 
varies between 10~GeV and 50~GeV.
Signal cross sections are calculated to NLO 
in the strong coupling constant, adding the resummation of soft gluon
emission at next-to-leading-logarithmic 
(NLO+NLL) accuracy~\cite{Beenakker:1997ut,Beenakker:2010nq,Beenakker:2011fu}. The
nominal cross section and the uncertainty are taken from an envelope
of cross section predictions using different PDF sets and
factorization and renormalization scales, as described in
Ref.~\cite{Kramer:2012bx}.

Differing pileup (multiple proton-proton interactions in the same or neighboring bunch-crossings) 
conditions as a function of the instantaneous luminosity are taken into account by overlaying simulated 
minimum-bias events generated with   {PYTHIA}-8 onto the hard-scattering process and re-weighting 
them according to the distribution of the mean number of interactions observed.
The MC generated samples are processed either with a full ATLAS detector simulation~\cite{:2010wqa} based on
GEANT4~\cite{Agostinelli:2002hh} or a fast simulation based on the
parameterization of the response of the electromagnetic and hadronic
showers in the ATLAS calorimeters~\cite{FastCaloSim} and a simulation of the trigger system.
The results based on fast simulation are validated against fully simulated samples.
The simulated events are
reconstructed and analyzed with the same analysis chain as for the data, using the same trigger and event 
selection criteria discussed in Section~\ref{sec:evt}.

\section{Reconstruction of physics objects}
\label{sec:recons}

%
%
Jets are reconstructed from energy deposits in the calorimeters using 
the $\akt$ jet algorithm~\cite{paper:antikt} with the distance parameter (in $\eta$--$\phi$ space) 
$\Delta R = \sqrt{(\Delta \eta)^2 + (\Delta \phi)^2}$ 
set to 0.4. 
The measured 
jet transverse momentum ($\ptjet$)  
is corrected for detector effects,  including the non-compensating character of the calorimeter, by weighting
 energy deposits arising from electromagnetic and hadronic showers differently.   
In addition, jets are corrected for 
contributions from pileup,  
as described in Ref.~\cite{Aad:2011he}.  Jets with  corrected $p_{\rm T} > 20$~GeV and $|\eta| <2.8$ are considered
 in the analysis.
In order to remove jets originating from pileup collisions, central jets ($|\eta| < 2.4$) with $p_{\rm T} < 50$~GeV and 
with charged-particle tracks
associated  to them must have a jet vertex fraction (JVF) above 0.5, where the JVF  
is defined as the ratio of 
the sum of transverse momentum of matched tracks that originate from the primary vertex 
to the sum of transverse momentum of all tracks associated with the jet.

%
%
The presence of leptons (muons or electrons) in the final state  is used in the analysis 
to define control samples and to reject background contributions in the signal regions (see Secs.~\ref{sec:evt} and~\ref{sec:backg}). 
Muon candidates are formed by combining information from the muon spectrometer and inner 
tracking detectors as described in
Ref.~\cite{muon} and are 
required to have  $p_{\rm T} > 10$~GeV, $|\eta| < 2.4$ and $\Delta R  > 0.4$ with respect to any jet with $p_{\rm T} > 20~\GeV$. 
The latter requirement is increased to 30~GeV in the case of the monojet-like analysis.
This 
increases the efficiency for the selection of real muons from W boson decays. It also avoids
biases in the muon selection  due to the presence of low $\pt$ jets with 
large pileup contributions,  affecting the  $\wmn$+jets events, as determined by simulations.
This is particularly relevant for the monojet-like analysis since, 
as described in Sec.~\ref{sec:backg}, $\wmn$+jets control samples in 
data are used to constrain the irreducible 
$\znn$+jets background contribution in the signal regions.
In addition, muons are required to be isolated: the sum of the transverse momenta of the
tracks not associated with the muon in a cone of radius $\Delta R =  0.2$ around the muon direction 
is required to be less than 1.8~GeV. 

Electron candidates are initially required to have $p_{\rm T} > 10~\GeV$ and $|\eta| <2.47$, and to pass the {\it medium}
 electron shower shape and track selection criteria described in Ref.~\cite{Aad:2014fxa} and   reoptimized for 2012
data. 
Overlaps between identified electrons and jets in the final state are
resolved. Jets are discarded if their separation $\Delta R$ from
an identified electron is less than 0.2. 
The electrons separated by 
$\Delta R$ between 0.2 and  0.4 from any remaining jet are removed.
In the monojet-like analysis, electrons are selected with $p_{\rm T} > 20~\GeV$ in both the control and signal regions. The 
use of the same $\pt$ threshold in control and signal regions minimizes the impact from lepton reconstruction and 
identification uncertainties on the final results.
The 20~GeV $\pt$ requirement together with the monojet-like selection also applied to define the control regions brings the 
background from jets misidentified as electrons to negligible levels 
without the need for electron isolation requirements.
As detailed in Sec.~\ref{sec:evt} and Sec.~\ref{sec:backg}, slightly 
different requirements on the lepton $p_{\rm T}$ are applied in the $c$-tagged analysis 
to define signal regions and background control samples.  
In this case,  the electrons are required to 
have $\pt >10$~GeV and $\pt >20$~GeV for signal and control samples, respectively,  and to be isolated: 
the total track momentum not associated with the electron in a cone of radius 0.2 around the electron 
candidate is required to be less than 10$\%$ of the electron $\pt$. In the $c$-tagged analysis, the 
use of a tighter electron veto in the signal regions, compared to that in the monojet-like analysis, 
contributes to the reduction of the sizable background from top-quark-related processes. 
%

$\met$ is reconstructed using all energy deposits in the calorimeter up 
to a pseudorapidity $|\eta| <  4.9$ and without including information from identified 
muons in the final state. 
Clusters associated with either electrons or photons with $\pt >10$~GeV and those
associated with jets with $\pt >20$~GeV make use of the corresponding calibrations for these objects. Softer jets 
and clusters not associated with these objects are calibrated using both calorimeter and tracking information~\cite{Aad:2012re}.   
%
%

\begin{figure*}[!htb]
\mbox{
\includegraphics[width=0.42\textwidth]{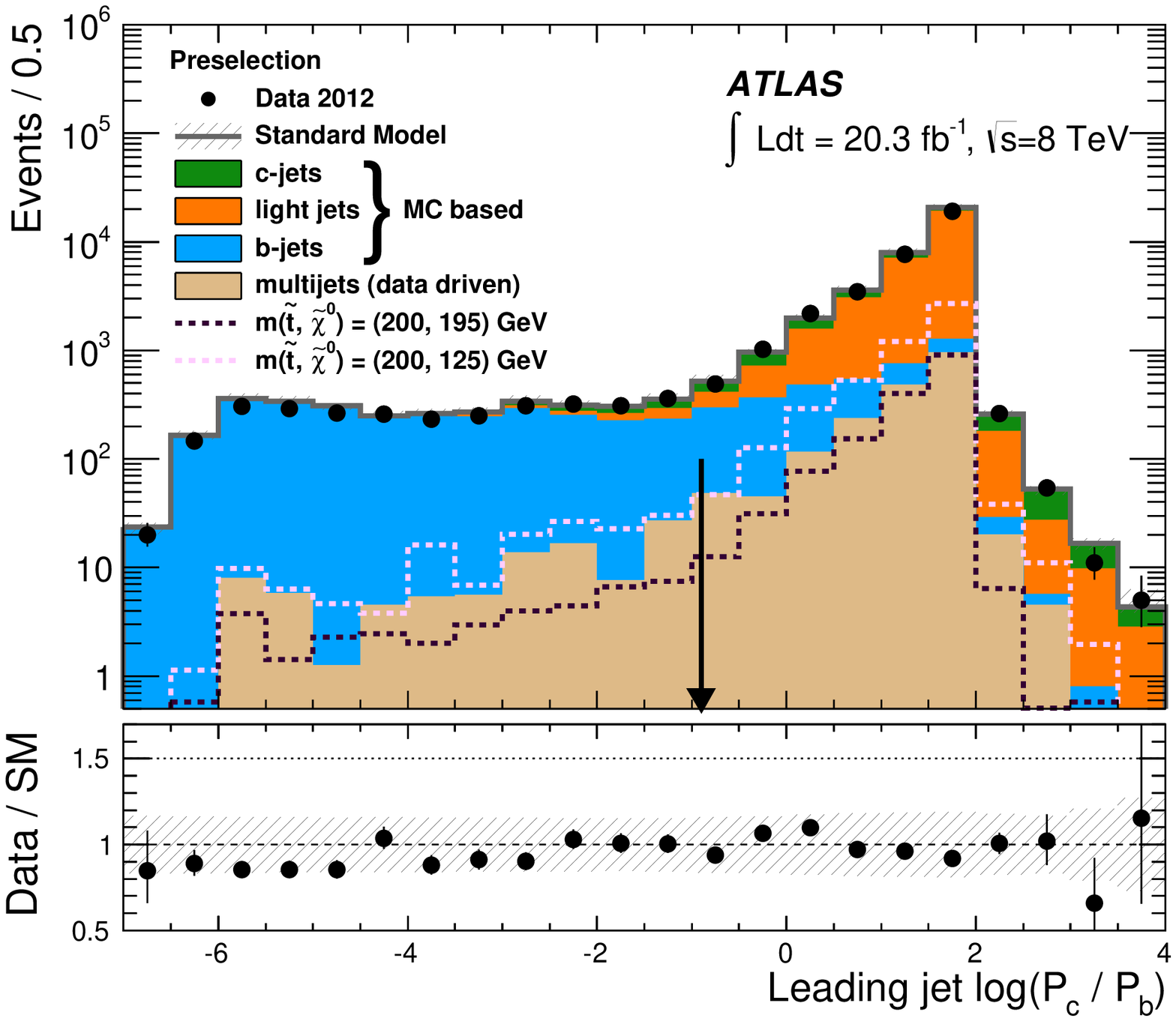}
\includegraphics[width=0.42\textwidth]{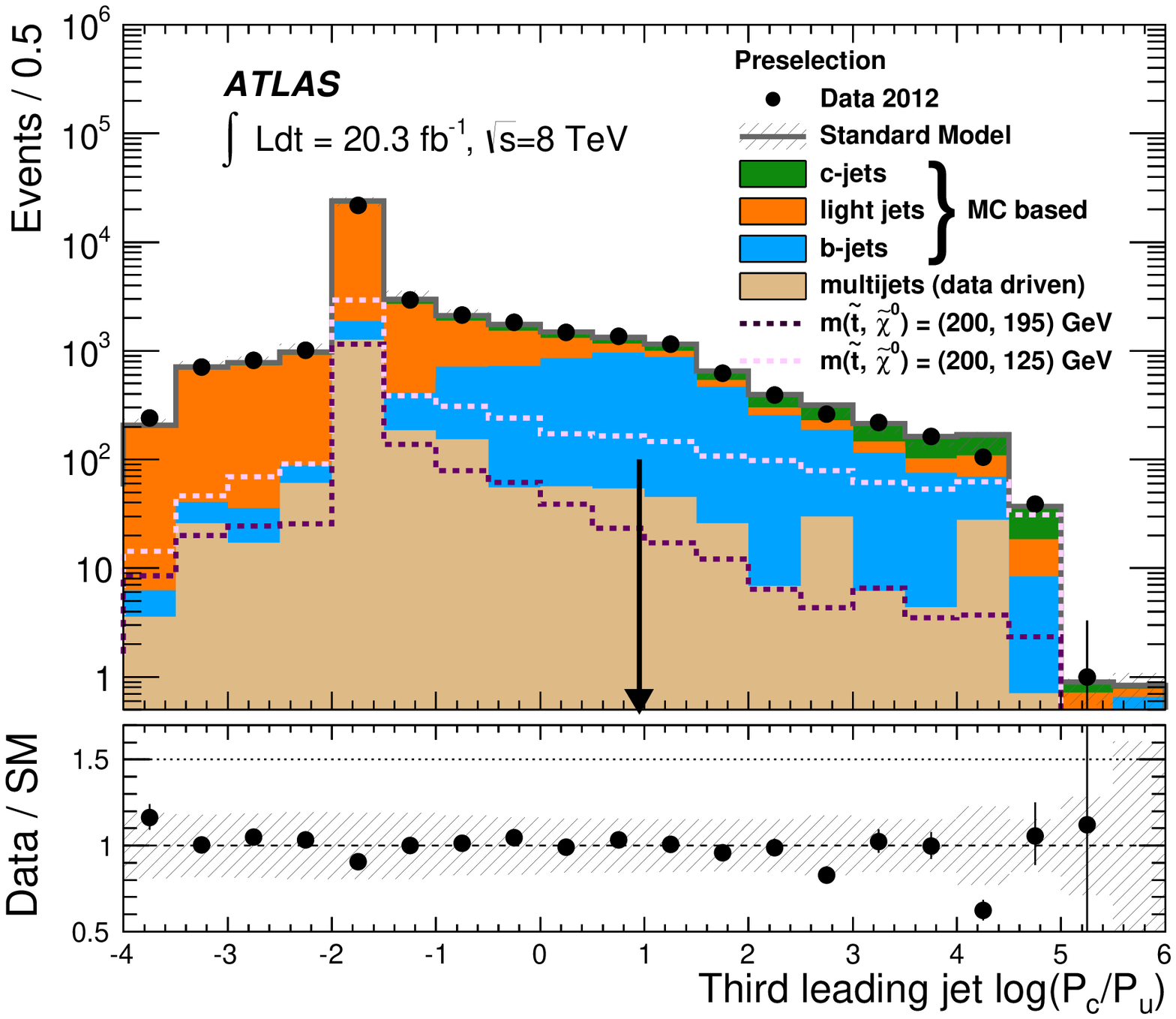}
}
\caption{Distribution of the discriminator against $b$-jets, 
  $\log(\pc/\pb)$,  for the first-leading  jet   and against light-jets, 
  $\log(\pc/\pu)$,  for the third-leading jet.
   The data are compared to MC simulations for the different SM processes, separated by jet flavor, 
and include the signal preselection defined in Sec.~\ref{sec:evt}
 without applying the tagging requirements, which are indicated by the arrows.
The bottom panels show the ratio between data and MC predictions. 
The error bands in the ratios include 
the statistical and experimental uncertainties in the predictions.
For illustration purposes, the distributions of two different SUSY scenarios for stop pair 
production with the decay mode $\tilde{t}_1 \to c + \ninoone$ are included.
In the SUSY signal, the first-leading jet mostly originates from ISR and the third-leading jet is expected to 
contain a large fraction of $c$-jets.  
}
\label{fig:cwei}
\end{figure*}

Jets are tagged as 
containing the decay products of charm hadrons
($c$-tagging) via a dedicated algorithm using multivariate
techniques. It combines information from the impact parameters of
displaced tracks and topological properties of secondary and
tertiary decay vertices reconstructed within the jet.  The algorithm
provides three probabilities: one targeted for light-flavor quarks and gluon jets ($\pu$), one for
charm jets ($\pc$) and one for $\bottom$-quark jets ($\pb$).  From these
probabilities,  anti-$\bottom$ and  anti-$\light$ discriminators are
calculated:
\begin{equation}
{\rm{anti-b}} \equiv \log\left(\frac{\pc}{\pb}\right)  \hspace*{10mm}{\rm
  and} \hspace*{10mm} 
{\rm{anti-u}} \equiv \log\left(\frac{\pc}{\pu}\right), 
\end{equation}
\noindent
 and used for the selected jets in the final state.  
Figure~\ref{fig:cwei} shows  the distributions of the anti-$b$ and anti-$u$ discriminators for the first- and the third-leading jets 
(sorted in decreasing jet $\pt$), respectively.   
   The data are compared to MC simulations for the different SM processes, separated by jet flavor~\footnote{The jet flavor in the MC samples is determined by associating jets to partons in $\eta$--$\phi$ space.}, and the 
data-driven multijet background prediction (see Sec.~\ref{sec:qcdback}), 
and include the signal preselection defined in Sec.~\ref{sec:evt}
 without applying the tagging requirements. Good agreement is observed between data and simulations. 
Two operating points specific to $c$-tagging are used.  The {\it medium} 
operating point ($\log\left({\pc}/{\pb}\right) > -0.9$, $\log\left({\pc}/{\pu}\right) > 0.95$) has a $\charm$-tagging 
efficiency of $\approx 20\%$,
and a rejection factor of $\approx 8$ for $\bottom$-jets,  $\approx
200$ for light-flavor jets, and $\approx 10$ for $\tau$ jets.  The {\it loose} operating point 
($\log\left({\pc}/{\pb}\right) > -0.9$) has a
$\charm$-tagging efficiency of $\approx 95\%$, with a factor of  2.5 
rejection of $\bottom$-jets but without any significant rejection for 
light-flavor or $\tau$ jets.  The efficiencies and rejections are quoted 
for jets with 30~GeV $< \pt < 200$~GeV and $|\eta|<2.5$ in simulated $\ttbar$ 
events, and reach a plateau at high jet $\pt$.

%
%

The $\charm$-tagging efficiency is calibrated using data with the method  described in Ref.~\cite{ctag} for 7 TeV collisions. This  method
makes use of a jet sample enriched in charm-quark-initiated jets  containing a  $D^{*+}$ meson identified in 
the $D^0(\to K^- \pi^+)\pi^+$ decay mode~\footnote{Charge conjugate states are included.}. 
The same calibration method applied to the 8~TeV data leads to reduced uncertainties.
The standard calibration techniques are used for the
$\bottom$-jet~\cite{btag,ttbtag} and light-jet~\cite{mistag} rejections:  
 a data-to-simulation multiplicative scale factor of about 0.9, with a very moderate jet $\pt$ dependence, 
is applied to the simulated heavy-flavor tagging efficiencies in the MC samples.  
The 
total uncertainty for the $c$-tagging efficiency varies between 
20$\%$ at low $\pt$ and 9$\%$ at high $\pt$ and includes: uncertainties on the heavy-flavor content of the 
charm-quark jet enriched sample and on the $b$-tagging scale factors; uncertainties on the $D^{*+}$ mass fit; 
uncertainties on the jet energy scale and resolution; and uncertainties on the extrapolation of the 
results to inclusive  charm-quark jets.   Similarly, data-to-simulation multiplicative 
scale factors of order 1.5 are applied to 
the simulated efficiency for tagging light-jets (mistags). They  are 
determined with a precision in the range between 20$\%$ and 40$\%$ depending on jet $\pt$ and $\eta$.  

\section{Event selection}
\label{sec:evt}

The data sample considered in this paper was collected with
tracking detectors, calorimeters, muon chambers, and magnets 
fully operational, and corresponds  
to a total integrated luminosity of $20.3 \ \rm fb^{-1}$. 
The uncertainty on the integrated luminosity is $2.8\%$, and it is estimated, following the same methodology 
detailed in Ref.~\cite{Aad:2013ucp}, from a preliminary calibration of the luminosity scale derived from beam-separation
scans performed in November 2012. 
The data were selected online using 
a trigger logic that selects events with  
$\met$ above 80~GeV, as computed at the final stage of the three-level
trigger system of ATLAS~\cite{Aad:2012xs}.
With respect to the final analysis requirements,  
the trigger selection is fully  efficient for $\met > 150$~GeV, 
as determined using a data sample with muons in the final state.
Table~\ref{tab:sr} summarizes the different event selection criteria applied in the signal
regions. The following preselection criteria are applied.

\begin{itemize}

\item Events are required to have a reconstructed primary vertex  
 consistent with the beamspot envelope and having at least five  associated  
tracks; when more than one such vertex is found, the vertex with the largest summed $\pt^2$
of the associated tracks is chosen.

\item Events are required to have $\met > 150$~GeV and at least one  
jet with $\ptjet > 150$~GeV and $|\eta| < 2.8$ ($|\eta| < 2.5$) in  the final state
for the monojet-like ($c$-tagged) selection.

\item Events are rejected if they contain any jet with $\ptjet > 20$~GeV and 
$|\eta| < 4.5$ that presents a charged fraction~\footnote{The charged fraction is defined as 
$f_{{\textrm{ch}}}=\sum p_{\rm T}^{{\textrm{track,jet}}} /p_{\rm T}^{{\textrm{jet}}}$, 
where $\sum p_{\rm T}^{{\textrm{track,jet}}}$ is the scalar
sum of the transverse momenta of tracks associated with the primary vertex
within a cone of radius $\Delta R=0.4$ around the jet axis, and 
$p_{\rm T}^{{ \textrm{jet}}}$ is the jet transverse momentum as determined 
from calorimetric measurements.},    
electromagnetic fraction 
in the calorimeter, 
or sampling fraction 
inconsistent with the requirement that they originate   
from a proton-proton collision~\cite{Aad:2013zwa}.
Additional requirements 
based on the timing and the pulse shape of the cells in the calorimeter
are applied to 
suppress coherent noise and electronic noise bursts in the calorimeter producing
anomalous energy deposits~\cite{ATLAS-CONF-2012-020}, which have a negligible effect on the signal efficiency. 

\item Events with isolated muons with $\pt > 10$~GeV are vetoed.  
Similarly, events with electrons with  $\pt > 20$~GeV ($\pt > 10$~GeV) are vetoed in the 
monojet-like ($c$-tagged) selection.

\end{itemize}

\subsection{Monojet-like selection}

The monojet-like analysis targets the region in which the stop and the lightest neutralino are nearly degenerate in mass so that 
the jets from the charm-quark fragmentation ($c$-jets) are too soft to be 
identified. Stop pair production events are then characterized by large $\met$ and 
a small number of jets, and can be identified via the 
presence of an energetic jet from initial-state radiation.
A maximum of three jets with $\pt > 30$~GeV and $|\eta| < 2.8$ in the event are allowed.
An additional requirement on
  the azimuthal separation of  $\Delta\phi(\textrm{jet},{\ptmi}) > 0.4$
  between the missing transverse momentum direction  and that of each of the selected jets is
  imposed. This requirement reduces the multijet background contribution where the
  large $\met$  originates  mainly from jet energy mismeasurement.
Three separate signal regions (here denoted by M1, M2 and M3) are defined with 
increasing lower thresholds on the leading jet $\pt$ and $\met$,  as 
the result of an optimization performed across the stop--neutralino mass plane with
increasing $\tilde{t}$
and $\tilde{\chi}_{1}^{0}$ masses.  For  the M1  selection, events are
required to have $\met > 220$~GeV  and  leading jet $\pt > 280$~GeV. 
For the M2 (M3) selection,  the thresholds are increased to $\met > 340$~GeV ($\met > 450$~GeV) and leading jet $\pt > 340$~GeV
($\pt > 450$~GeV).    

\subsection{$c$-tagged selection}

The kinematics of the charm jets from the stop decays depend mainly on $\Delta m$.
As $\Delta m$ decreases,  
the $\pt$  of the charm jets become softer
and it is more likely that other jets from initial state
radiation have a higher transverse momentum than the charm jets.
As a consequence, the stop signal is expected to have relatively large jet multiplicities and 
a $c$-tagged jet can be found among any of the subleading jets. 
An optimization of the $c$-tagged selection criteria is performed across 
the $\tilde{t}$ and $\tilde{\chi}_{1}^{0}$ mass plane
to maximize the sensitivity to a SUSY signal. 
In the $c$-tagged analysis,  the events are required to have at least four jets with 
$\pt > 30$~GeV, $|\eta| < 2.5$ and $\Delta\phi(\textrm{jet},{\ptmi}) > 0.4$. 
A veto against $b$-jets is applied to the selected jets 
in the event by using a loose $c$-tag requirement. 
In addition, at least one of the three 
subleading jets is required to be $c$-tagged using the medium  criteria.
The leading jet is required to have $\pt > 290$~GeV and two separate signal regions, here
denoted by C1 and C2, are defined with $\met > 250$~GeV and $\met > 350$~GeV, respectively.
The tighter requirement on $\met$ for the C2 signal region targets models with  
larger stop and neutralino masses. 
%
%

\begin{table*}[!ht]
\renewcommand{\baselinestretch}{1}
\caption{Event selection criteria applied for monojet-like (M1--M3) and $c$-tagged (C1,C2) analyses, as described in Sec.~\ref{sec:evt}.}
\begin{center}
\begin{small}
\begin{tabular*}{\textwidth}{@{\extracolsep{\fill}}p{0.52\linewidth}ccc}\hline
\multicolumn{4}{c}{\small{Selection criteria}} \\\hline\hline
\multicolumn{4}{c}{{\small{Preselection}} } \\\hline 
\multicolumn{4}{l}{Primary vertex}\\
\multicolumn{4}{l}{$\met > 150$~GeV }\\
\multicolumn{4}{l}{At least one jet with $\pt >150$~GeV and $|\eta|< 2.8$}\\
\multicolumn{4}{l}{Jet quality requirements}   \\
\multicolumn{4}{l}{Lepton vetoes}\\ \hline
\multicolumn{4}{c}{\small{Monojet-like selection}}\\\hline
\multicolumn{4}{l}{At most three jets with $\pt > 30$~GeV and $|\eta|<2.8$}\\
\multicolumn{4}{l}{$\Delta\phi(\textrm{jet},\ptmi) > 0.4$}\\\hline 
Signal region        & M1 & M2 & M3  \\ 
Minimum leading jet $\pt$ (GeV) & 280 & 340 & 450  \\ 
Minimum $\met$ (GeV) & 220 & 340  & 450 \\ \hline
\multicolumn{4}{c}{\small{$c$-tagged selection}}\\\hline
\multicolumn{4}{l}{At least four jets with $\pt > 30$~GeV and $|\eta|<2.5$}\\
\multicolumn{4}{l}{$\Delta\phi(\textrm{jet},\ptmi) > 0.4$}\\
\multicolumn{4}{l}{All four jets must pass loose tag requirements ($b$-jet vetoes)}\\
\multicolumn{4}{l}{At least one medium charm tag in the three subleading jets}\\\hline
Signal region        & C1 & C2 &   \\
Minimum leading jet $\pt$ (GeV) & 290 & 290 &   \\
Minimum $\met$ (GeV) & 250 & 350  &  \\ \hline
\end{tabular*}
\end{small}
\end{center}
\label{tab:sr}
\end{table*}

\section{Background estimation}
\label{sec:backg}

The expected SM background 
is dominated by $\znn$+jets, $\ttbar$ and $\wln$+jets ($\ell = e, \mu, \tau$)
production, and  includes small contributions from $\zll$+jets,   
single top, $\ttbar + V$, diboson ($WW,WZ,ZZ$) and multijet
processes.  
In the monojet-like analysis, the $\znn$+jets processes constitute more than 
50$\%$ -- 60$\%$ of the total background, followed by a 30$\%$ -- 40$\%$ contribution from $\wln$+jets processes. 
In the $c$-tagged selection, the background contributions from $\znn$+jets, $\wln$+jets, and top-quark 
related processes are similar, and each constitutes about $25\%$ to $30\%$ of the total background.   

The $W/Z$+jets backgrounds are estimated using 
MC event samples normalized using data in control regions.
The simulated $W/Z$+jets events are re-weighted to data as a
function of the generated $p_{\rm T}$ of the vector boson, following a procedure similar to that 
in Ref.~\cite{Aad:2012ms} based on the comparison of data and simulation in an event sample enriched in $Z$+jets events, 
which is found to
improve the agreement between data and  simulation. The weights applied to the simulation result 
from the comparison of the reconstructed boson $\pt$ distribution in 
data and SHERPA MC simulation in $W$+jets and $Z$+jets 
control samples 
where the jet and $\met$ preselection requirements (see Table~\ref{tab:sr}) have been applied.
The weights are defined in several bins in  boson $\pt$. 
Due to the limited number of data events
at large boson $\pt$, an
inclusive last bin with boson $\pt > 400$~GeV is used.  
The uncertainties of the re-weighting procedure are taken into account in the
final results.    

The top-quark background contribution to the monojet-like analysis is
very small and is determined using MC simulated samples.
In the case of the $c$-tagged analysis, 
the top-quark background is sizable, as it is enhanced by the jet multiplicity and 
$c$-tag requirements, and is estimated  using 
MC simulated samples normalized in  a top-quark-enriched control region.
The simulated $\ttbar$ events are re-weighted based on
the measurement in the data~\cite{Aad:2012hg},  indicating that the differential cross section as a
function of the $\pt$ of the $\ttbar$ system is 
softer than that predicted by the MC simulation.  

The normalization factors for $W/Z$+jets and $\ttbar$ background contributions  are
extracted simultaneously using a global fit to all control regions and include systematic uncertainties,
to properly take into account correlations.
The remaining SM backgrounds from $\ttbar +W/Z$, single top, diboson and Higgs processes   
are determined using Monte Carlo simulated samples, while the multijet background  contribution
is extracted from data. 
Finally, the potential
contributions from beam-related background and cosmic rays are
estimated in data using jet timing information and are found to be negligible.  

In the following subsections, details 
on the definition of $W/Z$+jets and $\ttbar$ control regions and on the data-driven determination of
the multijet background are given. This is followed by a description of the background fits and the validation 
of the resulting background estimations.

\subsection{$W/Z$+jets background}
\label{sec:wzback}

In the monojet-like analysis, control samples in data, orthogonal to the signal regions,  
with identified electrons or muons in the final state  and with the same requirements on 
the jet $\ptjet$, subleading jet vetoes, and $\met$, are used to determine 
the  $W/Z$+jets electroweak background contributions from data.
A $\wmn$+jets control sample is defined using events with a muon with $\pt >10$~GeV and  $W$
transverse mass~\footnote{The transverse mass $m_{\rm T}$  is defined by the lepton ($\ell$) and neutrino ($\nu$) $\pt$ and direction as
$m_{\rm T}=\sqrt{2\pt^{\ell}\pt^{\nu}(1-\cos(\phi^{\ell}-\phi^{\nu}))}$,
where the $(x,y)$ components of the neutrino momentum are taken to be
the same as the corresponding $\ptmi$ components.}
in the range  $30$~GeV~$<  m_{\rm T} <  100$~GeV. Similarly, a $\zmm$+jets control
sample is selected,  requiring the presence of two muons with invariant mass in the range
$66$~GeV~$< m_{\mu \mu} < 116$~GeV. 
The $\met$-based online trigger used in the analysis 
does not include muon information in the $\met$ calculation. This allows 
the $\wmn$+jets and $\zmm$+jets control samples to be collected 
with the same trigger as for the signal regions.
Finally, a $\wen$+jets dominated control sample is defined 
with an electron candidate with $\pt >20$~GeV.  
The $\met$ calculation
includes the contribution of the energy cluster from the identified electron in the calorimeter, 
since $\wen$+jets processes contribute to the  background in the signal regions 
when the electron is not identified. 
In the $\wmn$+jets and  $\zmm$+jets control regions,  the $\met$ 
does not include muon momentum contributions, motivated by the fact that these control regions
are used to estimate the irreducible $\znn$+jets background in the signal regions.         

%
%
The definition of the  control regions
in the $c$-tagged analysis follows closely that of the monojet-like approach with differences motivated by the 
background composition and the contribution from heavy-flavor jets. 
A  tighter cut of $81$~GeV $ < m_{\mu \mu} < 101$~GeV 
is used to define the $\zmm$+jets control
sample, as required to further reject $\ttbar$ contamination. 
This is complemented with a corresponding $\zee$+jets control sample, 
with the same mass requirements, for which 
the energy clusters associated with the identified electrons are then removed from the calorimeter.
The $\zee$+jets control sample  
is collected using a trigger that selects events with an electron in the final state.
As in the monojet-like case, in the $\wen$+jets control region the $\met$ calculation includes  
 the contribution from the identified electron. 
The electron also contributes to the number of jets in the final state, 
since the presence of a misidentified electron in the signal region 
can potentially affect the $c$-tagging results.
The $c$-tagging and the heavy-flavor composition 
are two of the major uncertainties (of the order of 10$\%$ -- 30$\%$) 
in the $c$-tagged selection and  
the same tagging criteria as used in the signal
selection are therefore applied to the $\wmn$+jets, $\wen$+jets, 
$\zmm$+jets and $\zee$+jets control regions. 
Since this reduces significantly the selection efficiency related to these control regions,
 the kinematic selections on the
leading jet $\pt$ and \met\  are both
reduced to 150~GeV, where the trigger selection still remains fully efficient.
This introduces  the need for a MC-based extrapolation of the normalization factors,
as determined using data at relatively low leading jet $\pt$ and $\met$, to the signal regions.
This extrapolation is tested in dedicated validation regions as described in Sec.~\ref{sec:vali}.
%
%

Monte Carlo--based transfer factors, determined from the
{SHERPA} simulation and including the boson $\pt$ re-weighting explained above, are defined for each of the signal
selections to estimate the different electroweak background contributions in the signal regions.
As an example, in the case of the dominant $\znn$+jets background
process in the monojet-like selection, its contribution to a given signal region 
$N^{\znn}_{\rm{signal}}$ 
is determined using the  $\wmn$+jets control sample in data 
according to

\begin{equation}
N^{\znn}_{\rm{signal}} = (N^{\rm{data}}_{\wmn,{\rm{control}}} - N^{{\rm{non}}-W}_{\wmn,{\rm{control}}})  \times
\frac{N^{{\rm{MC}} (\znn)}_{\rm{signal}}}{N^{{\rm{MC}}}_{\wmn,{\rm{control}}}}, 
\end{equation}
\noindent

\noindent
where $N^{{\rm{MC}} (\znn)}_{\rm{signal}}$ denotes the background predicted 
by the MC simulation in the signal region, and $N^{\rm{data}}_{\wmn,{\rm{control}}}$,  
 $N^{\rm{MC}}_{\wmn,{\rm{control}}}$, and $N^{\rm{non-W}}_{\wmn,{\rm{control}}}$ denote, in the control region, the 
number of $\wmn$+jets  candidates  in data and  MC simulation,  and 
the  non-$\wmn$ background contribution, respectively. 
The $N^{\rm{non-W}}_{\wmn,{\rm{control}}}$ term  refers mainly to top-quark and diboson processes,  but also includes  contributions from 
other $W/Z$+jets processes.  
 The transfer factors
for each process (e.g, the last term in Eq.~(2)) 
are defined as the ratio of simulated events for the process in the signal region over the total number of simulated 
events  in the control region. 

In the monojet-like analysis, the $\wmn$+jets control sample is used to define transfer factors 
for $\wmn$+jets and $\znn$+jets processes. 
As discussed in Secs.~\ref{sec:fitback} and \ref{sec:syst}, the use of the $\wmn$+jets control sample to constrain the normalization of the 
$\znn$+jets process  
translates into a reduced uncertainty on the estimation
of the main irreducible background contribution,  due to a partial cancellation of systematic uncertainties and the 
statistical power of the 
$\wmn$+jets control sample in data,  about seven times larger than the $\zmm$+jets control sample.  
The $\wen$+jets control sample is used to constrain  
$\wen$+jets, $\wtn$+jets, $\ztt$+jets, and $\zee$+jets contributions. 
Finally,  the $\zmm$+jets control sample is used to constrain the $\zmm$+jets background contribution.

The $c$-tagged analysis follows a similar approach to determine the normalization factors for each of the
$W/Z$+jets background contributions.
However, in this case the $\znn$+jets, $\zee$+jets and $\zmm$+jets normalization factors are extracted from the combined $\zll$+jets ($\ell = e, \mu$)
control sample, motivated by the fact that these processes involve identical
heavy-flavor production mechanisms. Simulation studies indicate a very similar 
heavy-flavor composition in control and signal regions.

\begin{figure*}[!ht]
\begin{center}
\mbox{
  \includegraphics[width=0.42\textwidth]{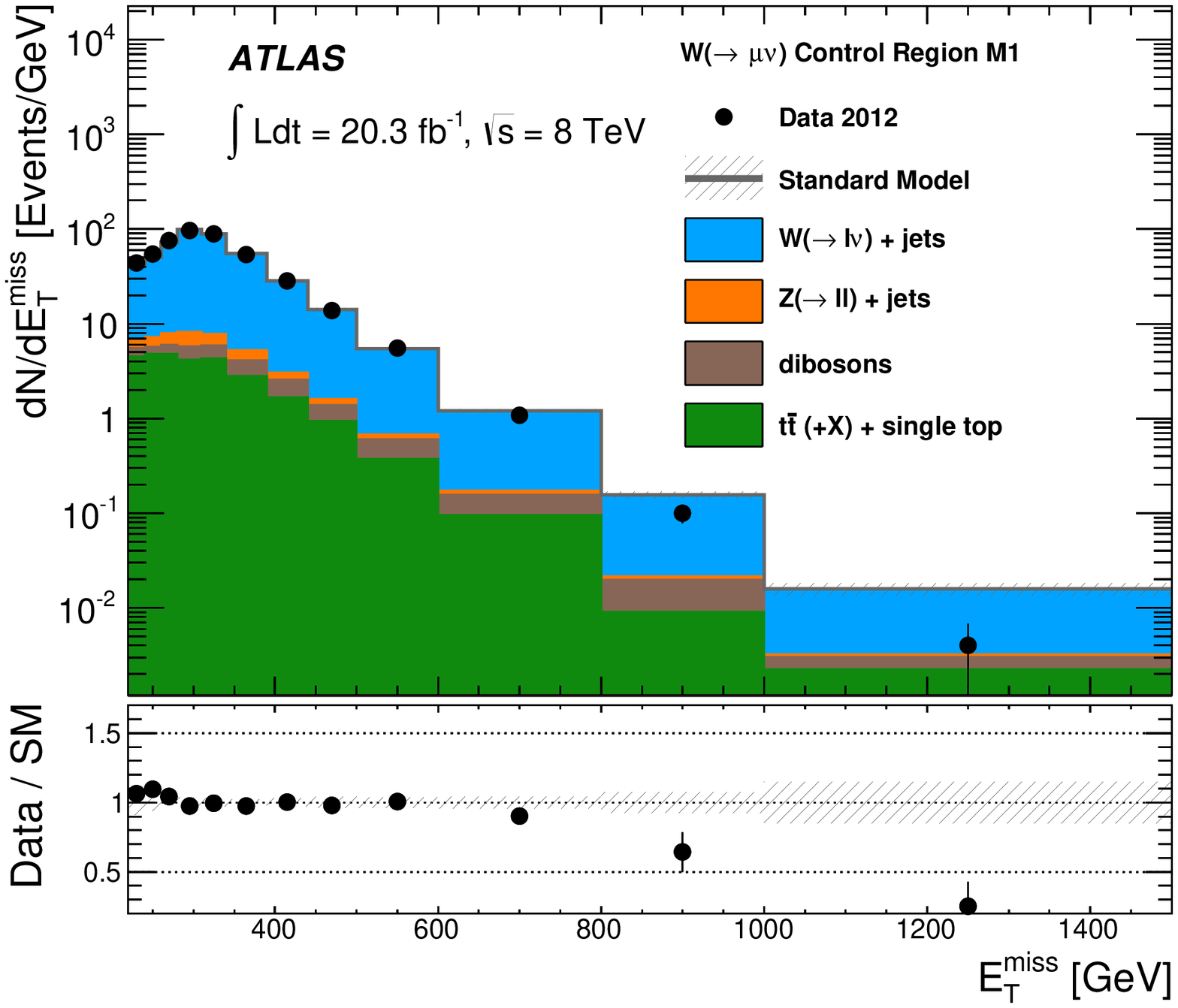}
  \includegraphics[width=0.42\textwidth]{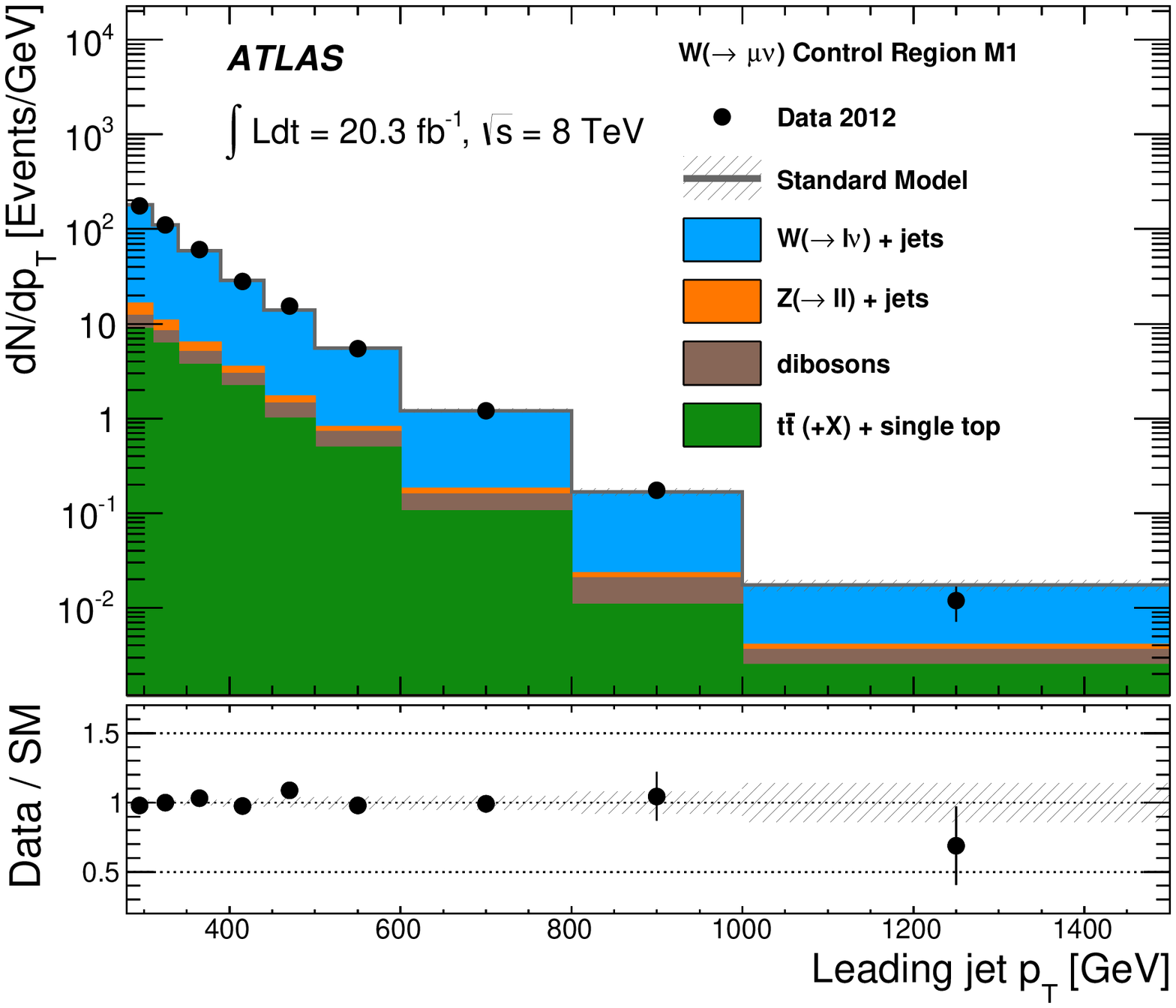}
}
\mbox{
  \includegraphics[width=0.42\textwidth]{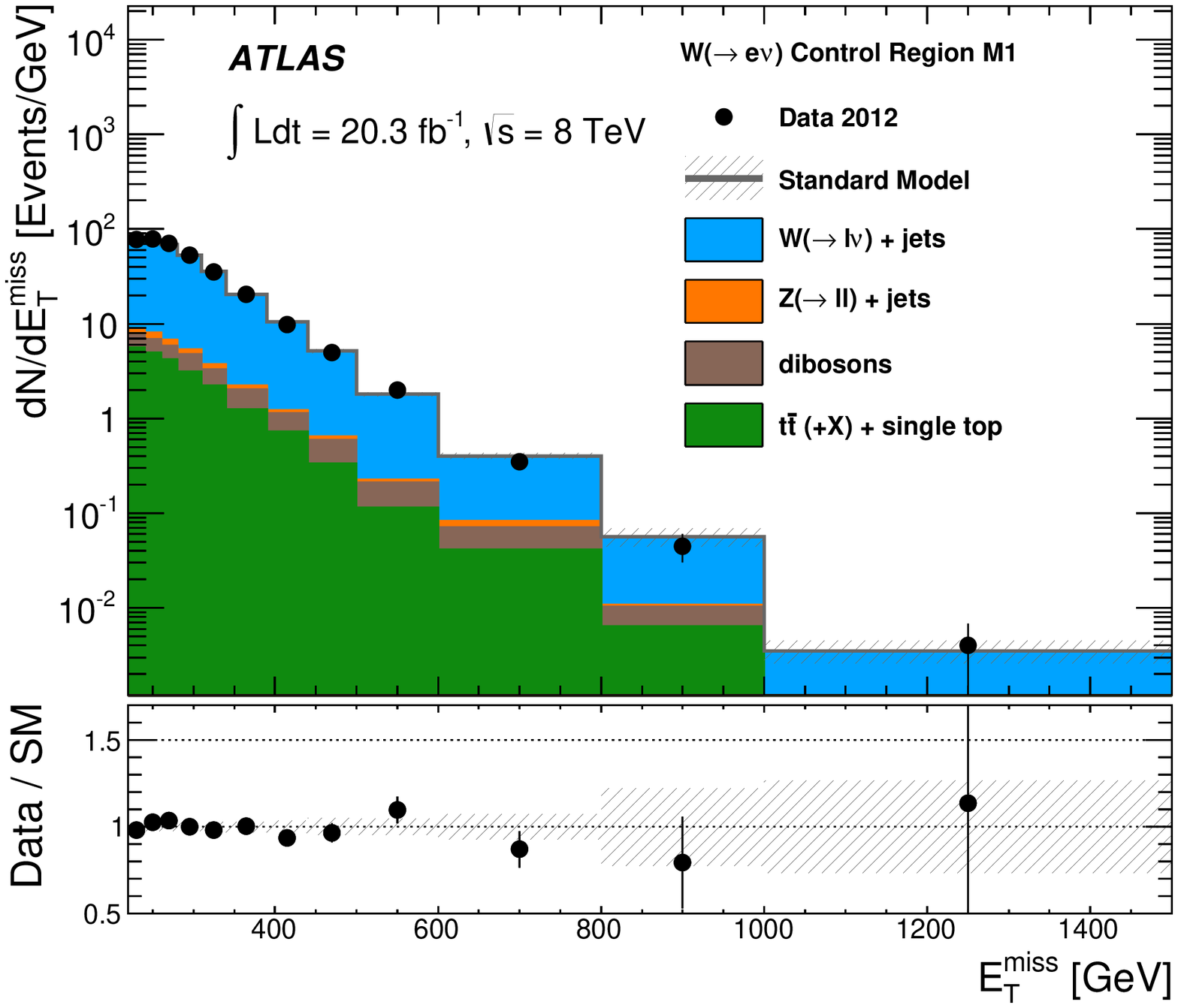}
  \includegraphics[width=0.42\textwidth]{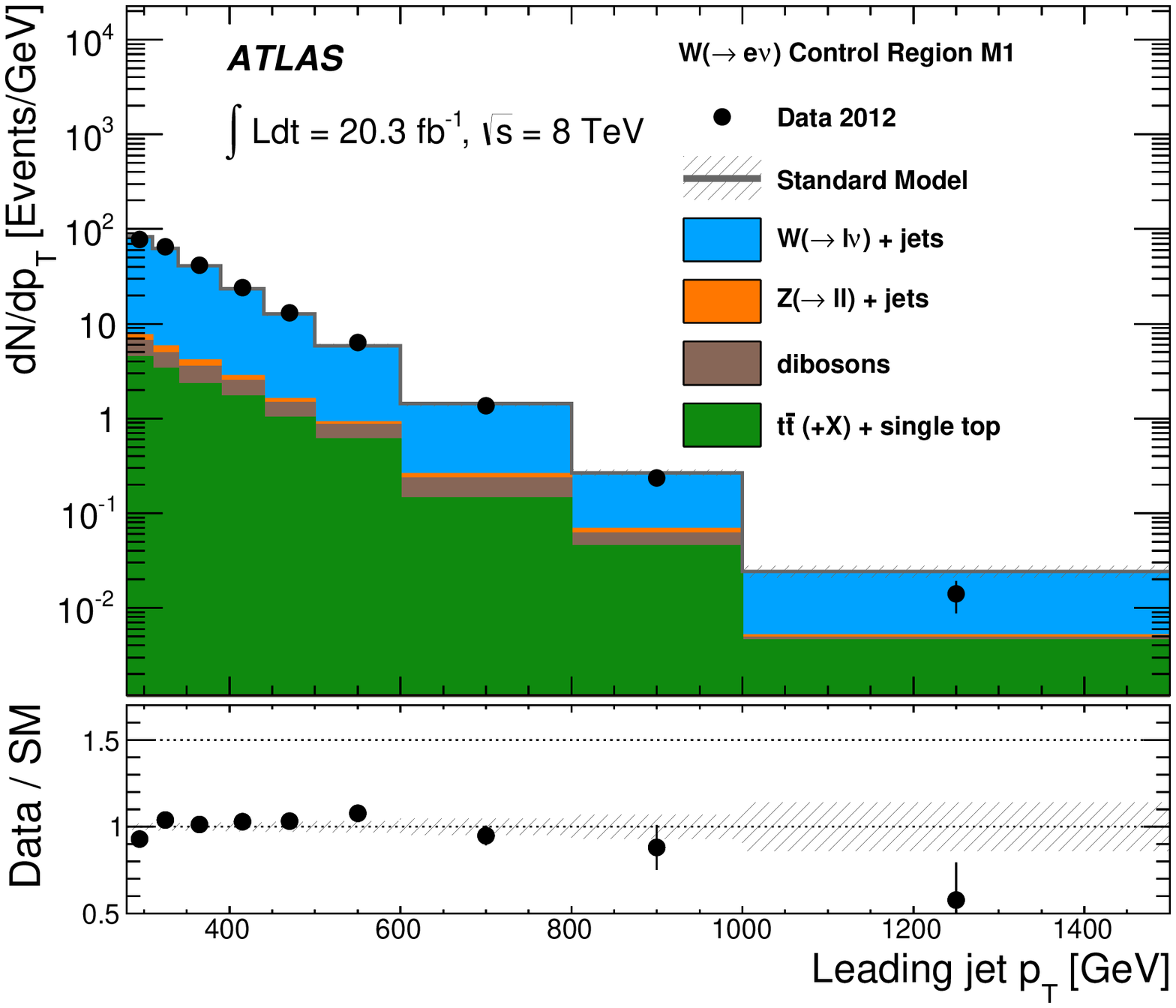}
}
\mbox{
  \includegraphics[width=0.42\textwidth]{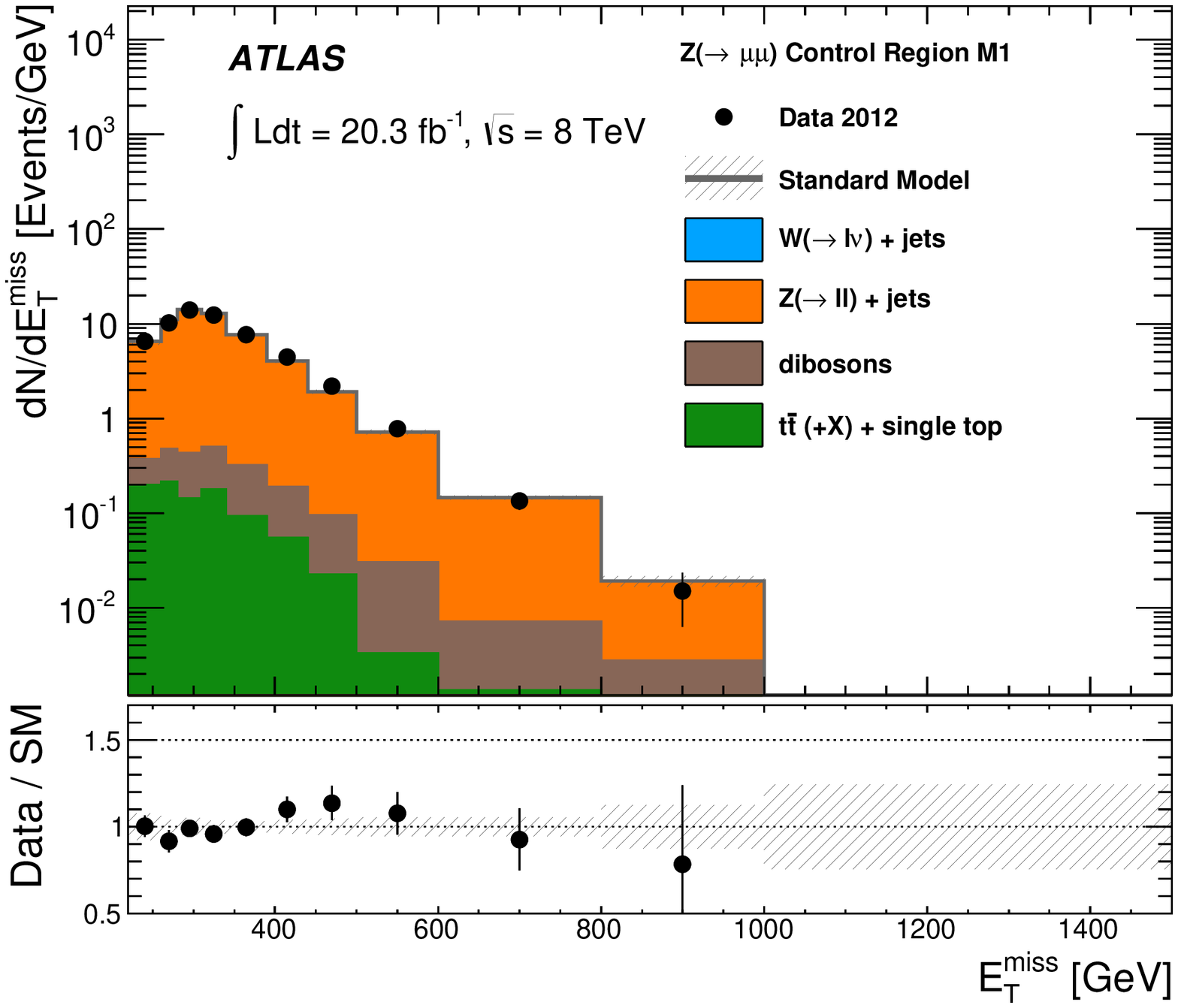}
  \includegraphics[width=0.42\textwidth]{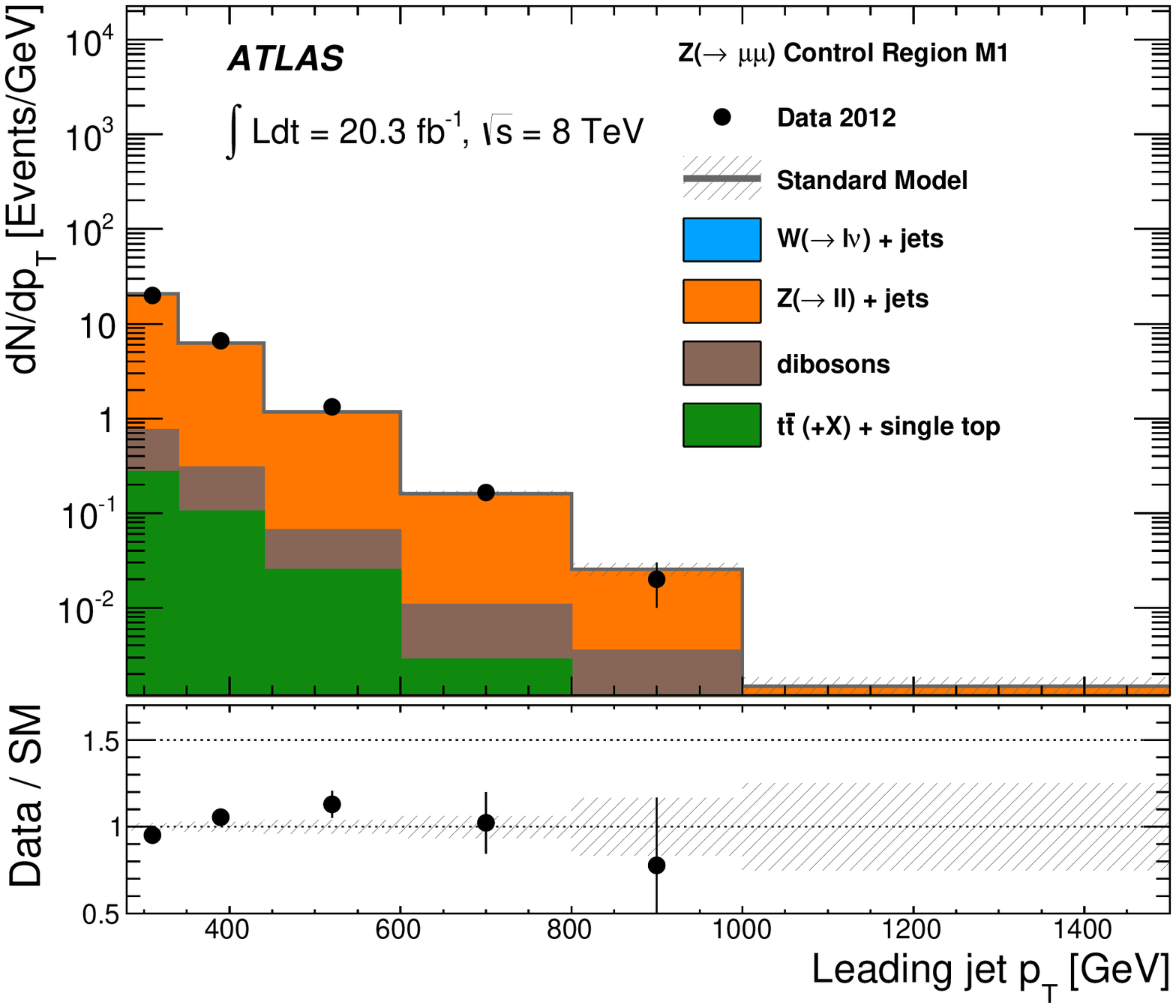}
}
\end{center}
\caption{
The measured $\met$ and leading jet $\pt$ distributions
in the $\wmn$+jets (top),  $\wen$+jets (middle), and $\zmm$+jets (bottom) control regions, for the M1 selection, compared to
the background predictions.
The latter include the global normalization factors extracted from the fit.
The error bands in the ratios include the statistical and experimental  uncertainties on the background predictions.
}
\label{fig:cr1}
\end{figure*}

\begin{figure*}[!ht]
\begin{center}
\mbox{
  \includegraphics[width=0.42\textwidth]{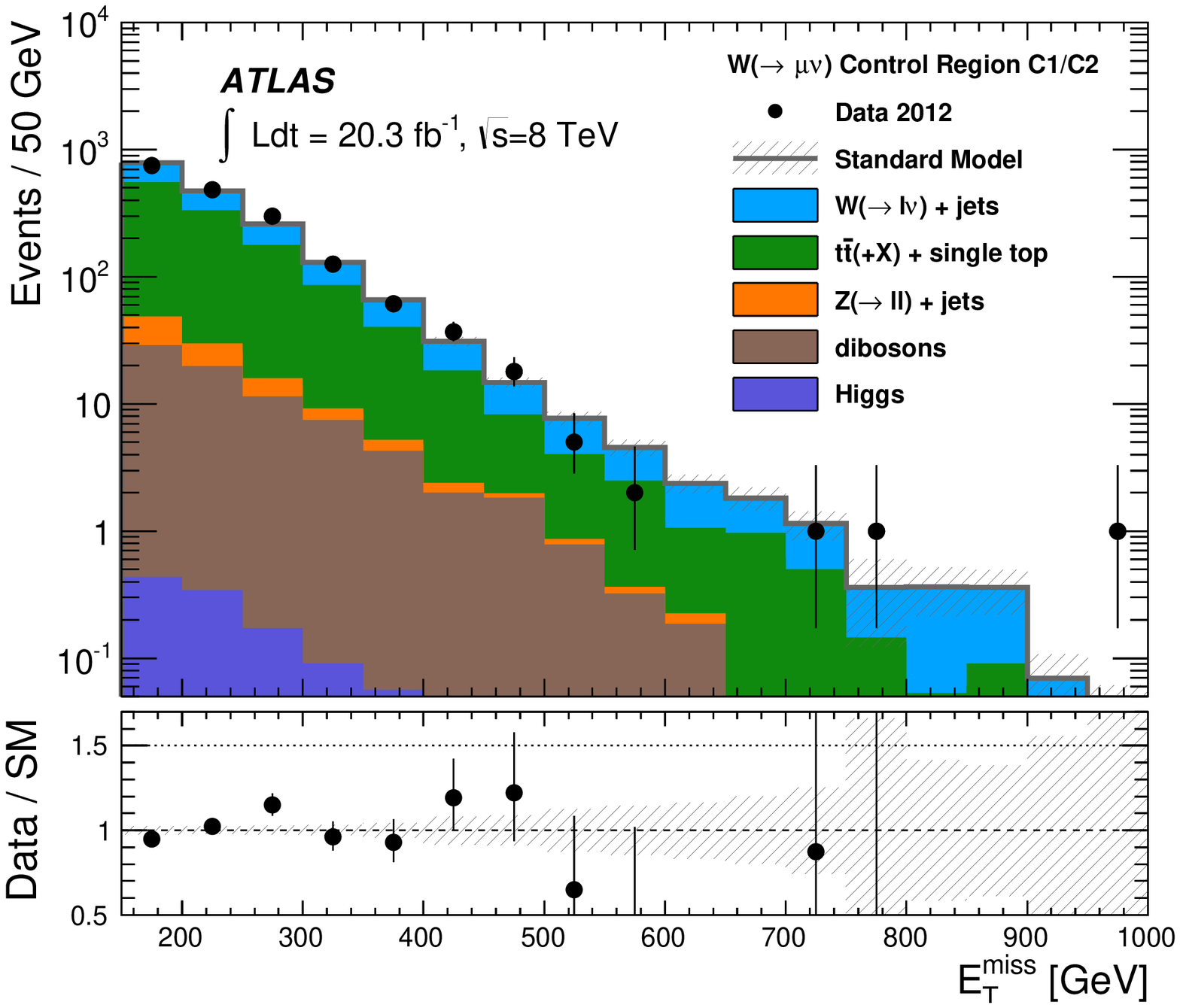}
  \includegraphics[width=0.42\textwidth]{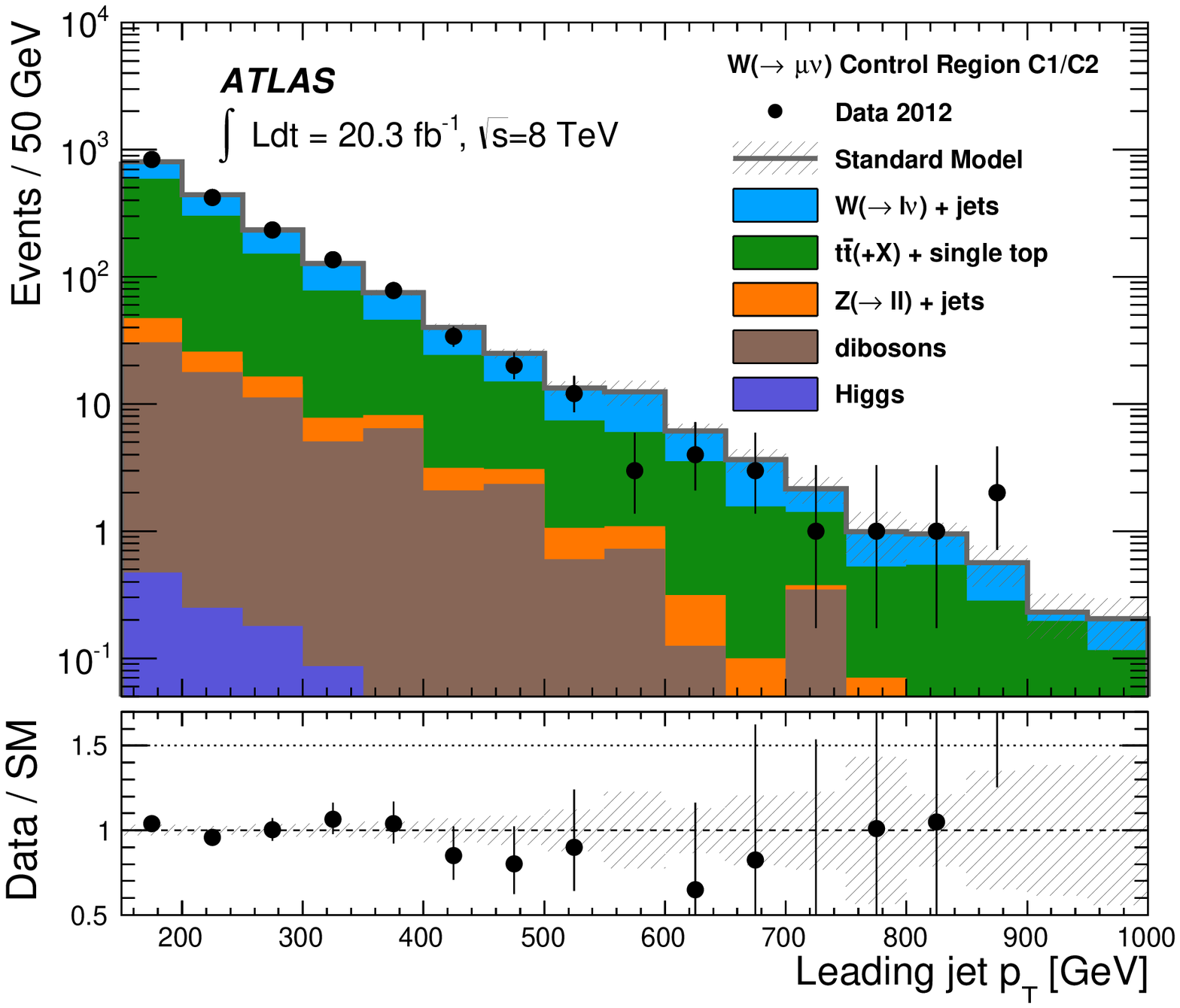}
}
\mbox{
  \includegraphics[width=0.42\textwidth]{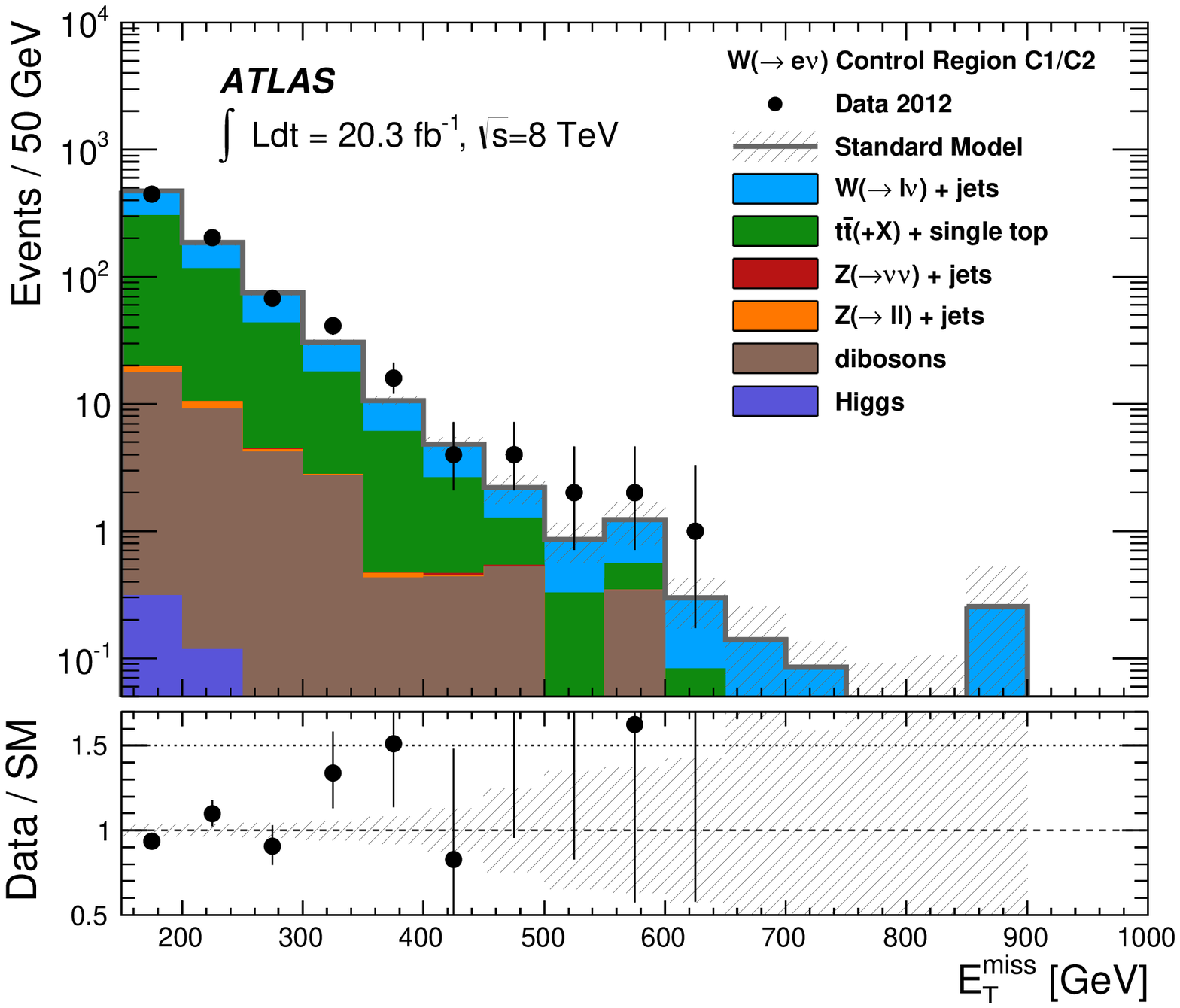}
  \includegraphics[width=0.42\textwidth]{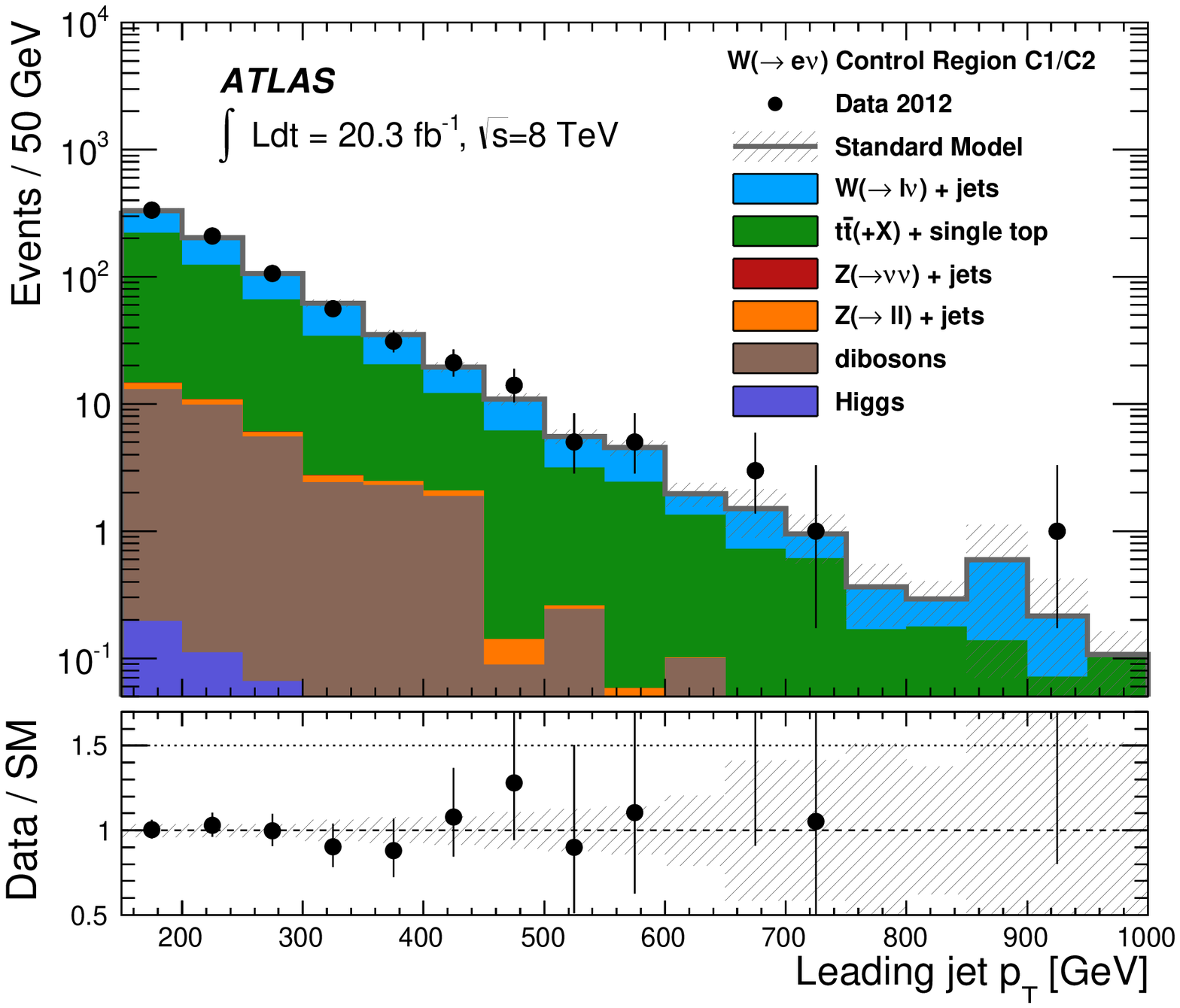}
}
\mbox{
  \includegraphics[width=0.42\textwidth]{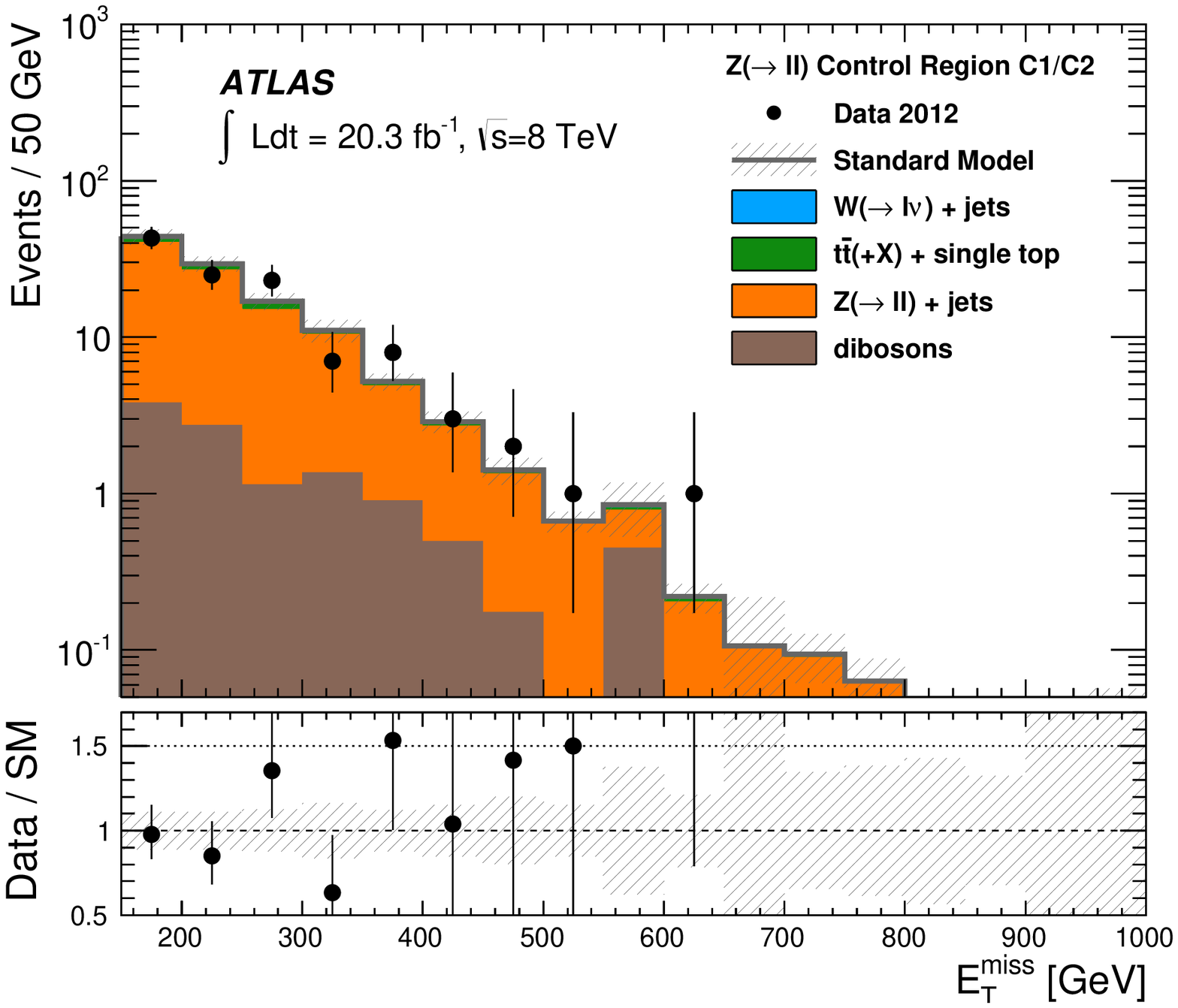}
  \includegraphics[width=0.42\textwidth]{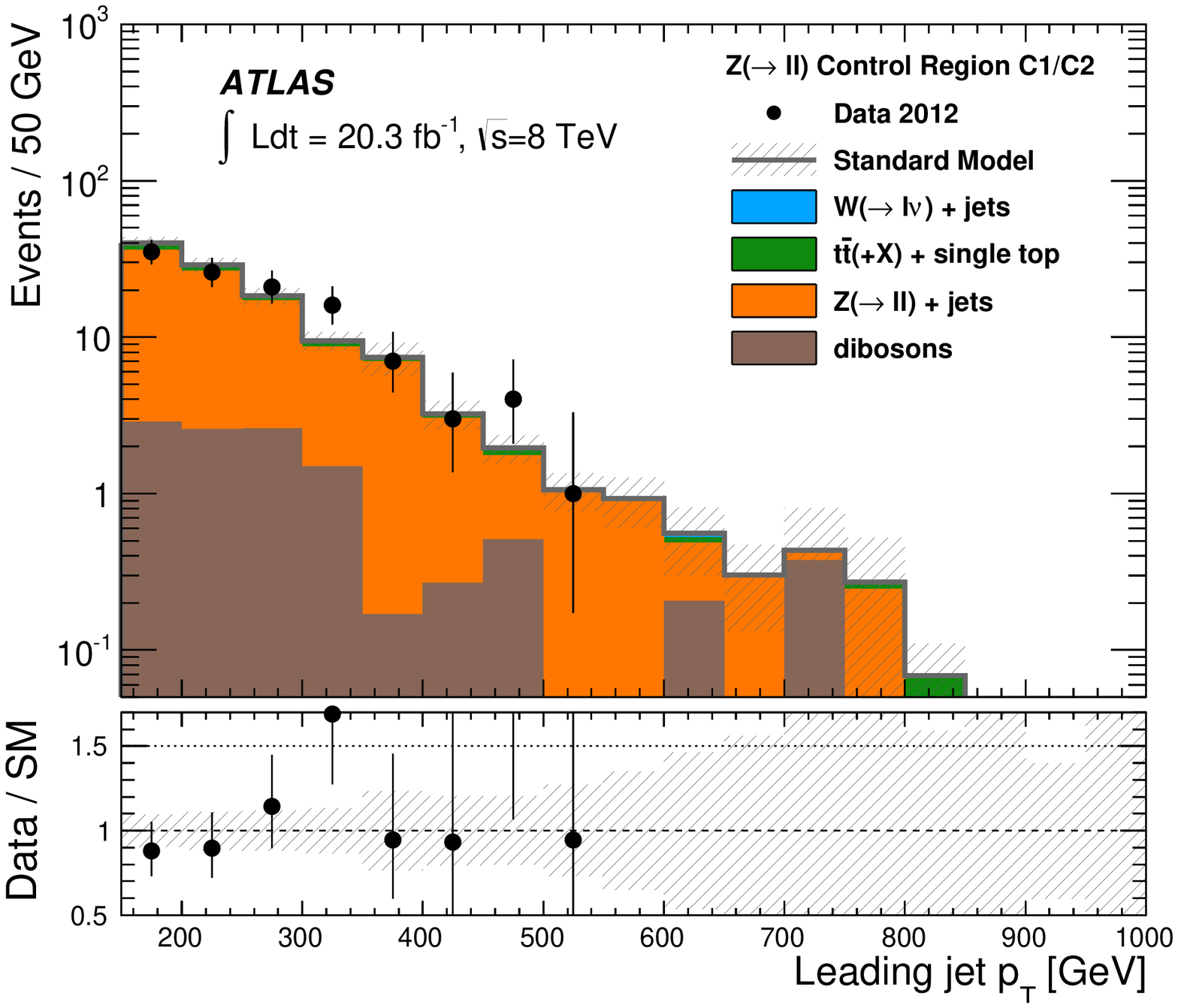}
}
\end{center}
\caption{The measured $\met$ and leading jet $\pt$ distributions
in the $\wmn$+jets (top), $\wen$+jets (middle), and $\zll$+jets (bottom) control regions, for the $c$-tagged selection, compared to
the background predictions.
The latter include the global normalization factors extracted from the fit.
The error bands in the ratios include the statistical and experimental  uncertainties on the background predictions.
}
\label{fig:cr4}
\end{figure*}

Figure~\ref{fig:cr1}
shows, for the M1 monojet-like kinematic selection and in the different control regions, 
the distributions of the $\met$ and the 
leading-jet $\pt$ in data and MC simulations.  The MC predictions
include data-driven normalization factors 
as a result of the use of transfer factors from control to signal regions discussed above.
Similarly, the distributions 
for events  in the $W/Z$+jets control regions of the $c$-tagged selection
are shown in Fig.~\ref{fig:cr4}. Altogether, the MC  simulation provides a good 
description of the shape of the measured distributions for both the monojet-like and $c$-tagged
selections in the different control regions.

\subsection{Top quark background}
\label{sec:topback}

\begin{figure*}[!ht]
\begin{center}
\mbox{
  \includegraphics[width=0.42\textwidth]{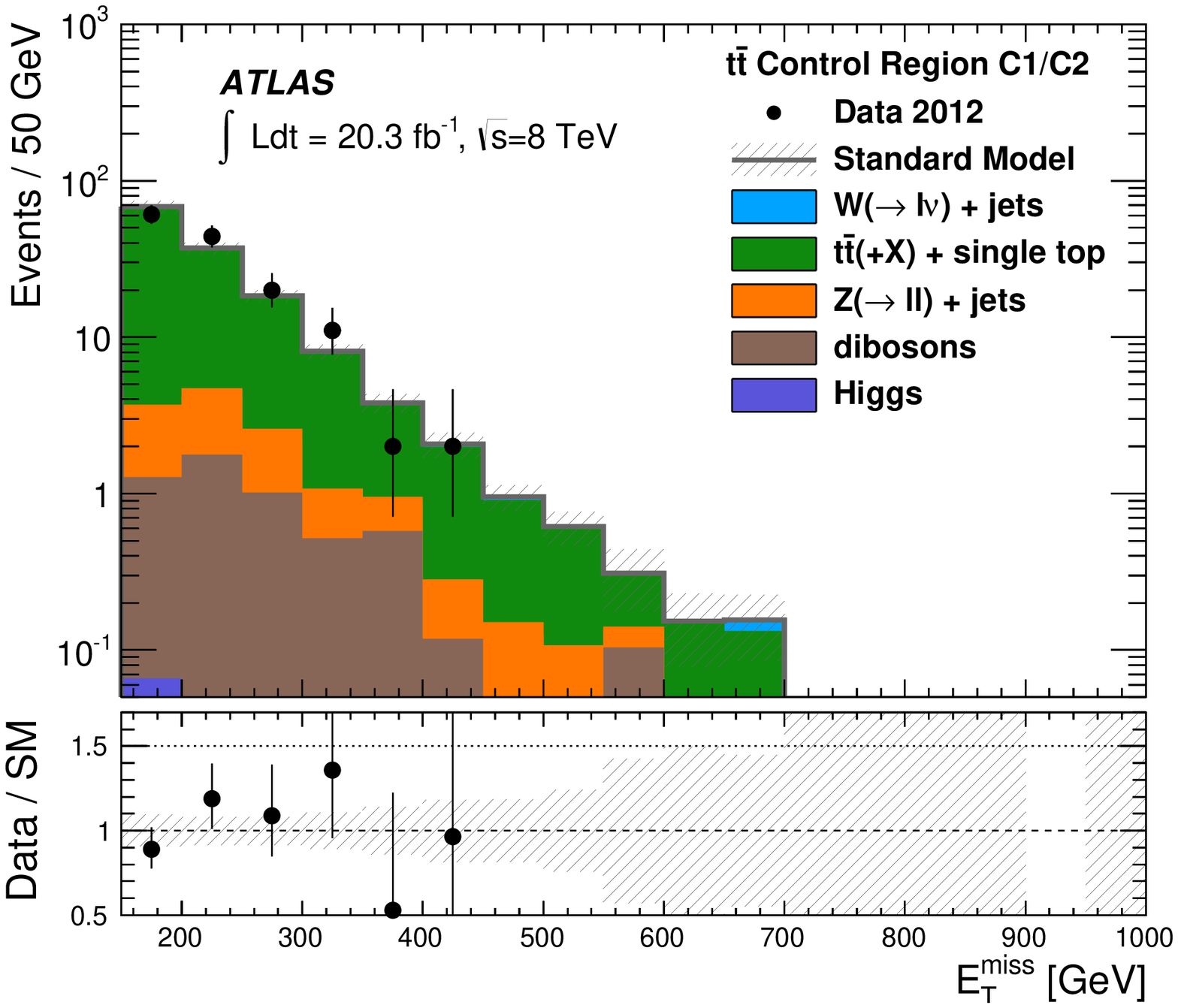}
  \includegraphics[width=0.42\textwidth]{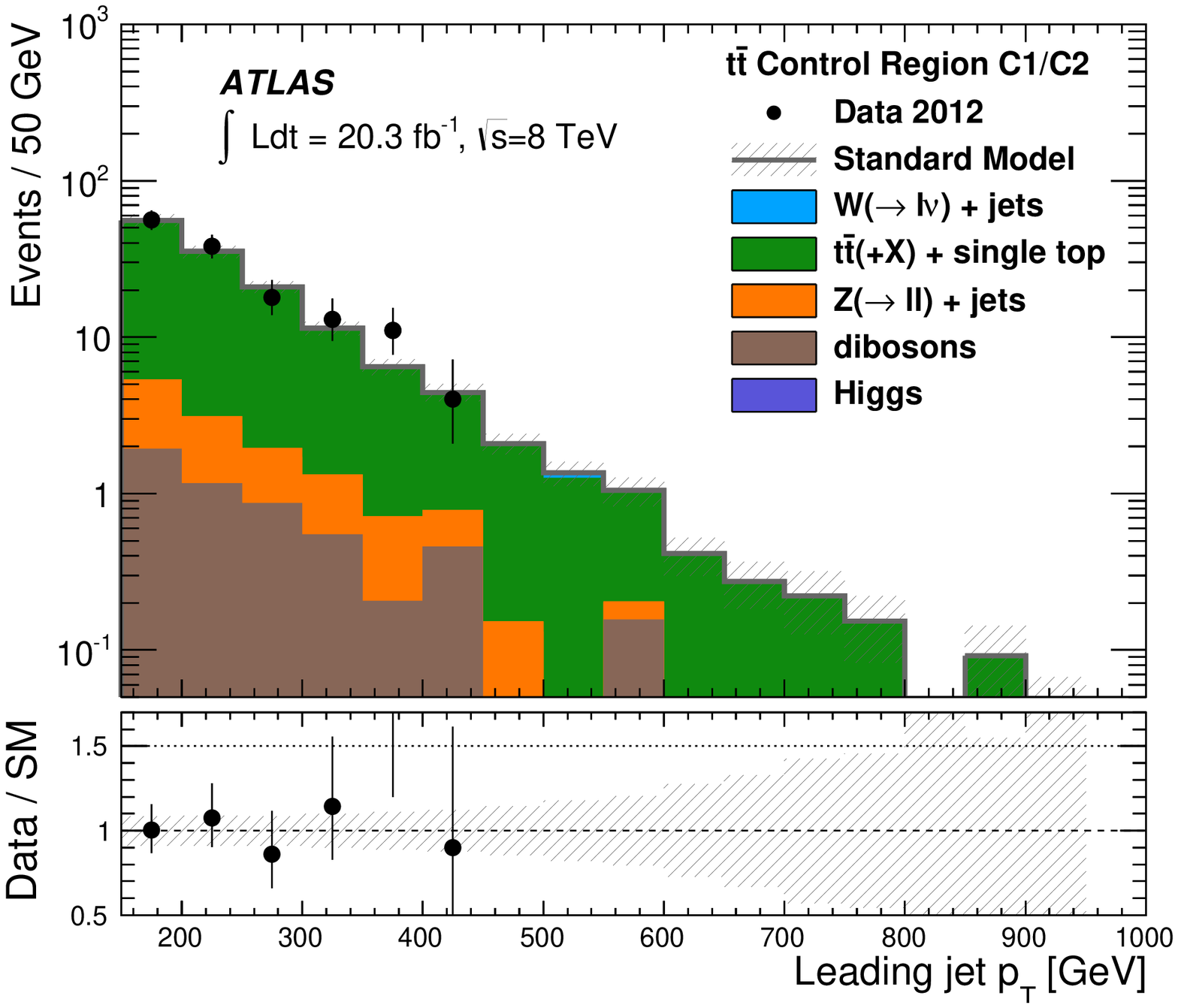}
}
\end{center}
\caption{
The measured $\met$ and leading jet $\pt$ distributions
in the $\ttb$ control region, for the $c$-tagged selection, compared to
the background predictions.
The latter include the global normalization factors extracted from the fit.
The error bands in the ratios include the statistical and experimental  uncertainties on the background predictions.
}
\label{fig:cr7}
\end{figure*}

The background contribution from top-quark-related production processes to the 
monojet-like selection is small and is entirely determined from MC simulations.  
In the case of the $c$-tagged analysis, single top and $\ttbar +W/Z$ processes are
directly taken from MC simulations and  
the $\ttbar$ MC predictions are 
normalized to the data in a separate control region.
The $\ttb$ background contribution is dominated by events with hadronic $\tau$-lepton decays and ISR jets in the final state.
A $\ttb$ control sample is selected with two opposite-charge leptons ($ee$, $\mu\mu$, or $e\mu$ configurations) in
the final state, the  same selection criteria  for jet multiplicity and $c$-tagging as in the signal region, and  
relaxed $\met > 150$~GeV and leading jet $\pt > 150$~GeV requirements.
In order to reduce the potential $\zee$+jets and $\zmm$+jets contamination in the $\ttb$ control sample, 
$ee$ and $\mu\mu$ events with a dilepton invariant mass within 15~GeV of the nominal $Z$ boson mass are rejected. 
Figure~\ref{fig:cr7} compares the distributions for
data and simulation in the  $\ttb$ control region. The MC simulation 
provides a good description of the shape of the measured distributions.
%

\subsection{Multijets background}
\label{sec:qcdback}

The multijet background with large $\met$ mainly originates 
from the misreconstruction of the energy of a 
jet in the calorimeter and to a lesser extent is due to the presence 
of neutrinos in the final state from heavy-flavor decays.  
In this analysis, the multijet background is determined from data, using a  {\it jet smearing} method as
described in Ref.~\cite{Aad:2012fqa}, which relies  on the assumption that the $\met$ of multijet events is dominated by  
fluctuations in the jet response in the detector that  can be measured in the data. 
Different response functions are used for untagged and heavy-flavor tagged jets.  
For the M1 monojet-like and C1 $c$-tagged analyses, the multijet background constitutes 
about $1\%$ of the total background, and  
 is negligible for the other signal regions.

\subsection{Background fits}
\label{sec:fitback}

The use of control regions to constrain the normalization of the dominant background contributions from 
$\znn$+jets, $W$+jets, (and $\ttbar$ in the case of the $c$-tagged analysis) reduces  significantly
the relatively large  theoretical and experimental 
systematic uncertainties, of the order of  $20\%$--$30\%$,  associated with purely MC-based 
background predictions in the signal regions.
A complete study of systematic uncertainties is carried out in the monojet-like and $c$-tagged analyses, 
as detailed in Sec.~\ref{sec:syst}.
To determine the final uncertainty on the total background,  
all systematic uncertainties are treated as nuisance
parameters with Gaussian shapes in a fit based on the profile
likelihood method~\cite{statforumlimits},  that takes into account correlations among systematic variations.
The fit takes also into account cross contamination between different background sources in the control regions. 

A simultaneous likelihood fit to the $\wmn$+jets, $\wen$+jets,
$\zll$+jets and \ttbar\ control regions (the latter only in the case of the $c$-tagged analysis)
is performed separately for each analysis to normalize and
constrain the corresponding background estimates in the signal regions.
The results of the background-only fits in the control regions 
are presented in Tables~\ref{tab:fitm1}--\ref{tab:fitm3} for the monojet-like 
selections, 
and in Table~\ref{tab:fitc1} for the $c$-tagged analysis.  
As the tables indicate, the $W/Z$+jets background predictions 
receive multiplicative  normalization factors that vary in the range 
between 1.1 and 0.9 for the monojet-like analysis, 
depending on the process and the kinematic selection, 
 and between 0.8 and 0.9 for the $c$-tagged analyses.
In the $c$-tagged analysis, the $\ttbar$ background predictions 
are normalized with a scale factor 1.1 for both the C1 and C2 selections.  
%
%

\begin{table*}[!ht]
\caption{Data and background predictions in the control regions before and after the fit is performed for the M1 selection.
The background predictions include both the statistical and systematic uncertainties.
The individual uncertainties are correlated, and do not necessarily add 
in quadrature to the total background uncertainty.
}
\begin{center}
\setlength{\tabcolsep}{0.0pc}
{\footnotesize
\begin{tabular*}{\textwidth}{@{\extracolsep{\fill}}lrrr}
\noalign{\smallskip}\hline\noalign{\smallskip}
{\bf  M1 control regions}                  & $\wen$
& $\wmn$           & $\zmm$      \\
\noalign{\smallskip}\hline\noalign{\smallskip}
Observed  events (20.3~fb${}^{-1}$)    & $9271$              & $14786$              & $2100$ \\
\noalign{\smallskip}\hline\noalign{\smallskip}
SM prediction (post-fit)    & $9270 \pm 110$          & $14780 \pm 150$          & $2100 \pm 50$              \\             \noalign{\smallskip}\hline\noalign{\smallskip}
        Fitted $\wen$          & $6580 \pm 130$          & $0.4 \pm 0.2$          & $-$              \\
        Fitted $\wmn$          & $39 \pm 5$          & $12110 \pm 200$          & $2.4 \pm 0.2$              \\
        Fitted $\wtn$          & $1640 \pm 40$          & $1130 \pm 30$          & $0.6 \pm 0.1$              \\
        Fitted $\zee$          & $0.04_{-0.04}^{+0.07}$          & $-$          & $-$              \\
        Fitted $\zmm$          & $3.6 \pm 0.5$          & $290 \pm 20$          & $2010 \pm 50$              \\
        Fitted $\ztt$          & $116 \pm 3$          & $43 \pm 3$          & $2.9 \pm 0.3$              \\
        Fitted $\znn$          & $17 \pm 3$          & $4.2 \pm 0.4$          & $-$              \\
        Expected $\ttbar$, single top, $\ttbar$+V          & $600 \pm 80$          & $880 \pm 90$          & $32 \pm 9$              \\
        Expected  dibosons         & $280 \pm 90$          & $330 \pm 110$          & $58 \pm 21$              \\
\noalign{\smallskip}\hline\noalign{\smallskip}
MC exp. SM events              & $9354$          & $15531$          & $2140$              \\
\noalign{\smallskip}\hline\noalign{\smallskip}
        Fit input $\wen$          & $6644$          & $0.4$          & $-$              \\
        Fit input $\wmn$          & $41$          & $12839$          & $2.5$              \\
        Fit input $\wtn$          & $1650$          & $1142$          & $0.6$              \\
        Fit input $\zee$          & $0.04$          & $-$          & $-$              \\
        Fit input $\zmm$          & $3.7$          & $291$          & $2044$              \\
        Fit input $\ztt$          & $117$          & $44$          & $3.0$              \\
        Fit input $\znn$          & $18$          & $4.5$          & $-$              \\
        Fit input $\ttbar$, single top, $\ttbar$+V        & $600$          & $880$          & $32$              \\
        Fit input dibosons         & $280$          & $330$          & $58$              \\
\noalign{\smallskip}\hline\noalign{\smallskip}
\end{tabular*}
}
\end{center}
\label{tab:fitm1}
\end{table*}

\begin{table*}[!ht]
\caption{Data and background predictions in the control regions before and after the fit is performed for the M2 selection.
The background predictions include both the statistical and systematic uncertainties.
The individual uncertainties are correlated, and do not necessarily add in
quadrature to the total background uncertainty.}
\begin{center}
\setlength{\tabcolsep}{0.0pc}
{\footnotesize
\begin{tabular*}{\textwidth}{@{\extracolsep{\fill}}lrrr}
\noalign{\smallskip}\hline\noalign{\smallskip}
{\bf  M2 control regions}           & $\wen$            & $\wmn$            & $\zmm$              \\[-0.05cm]
\noalign{\smallskip}\hline\noalign{\smallskip}
Observed events  (20.3 fb${}^{-1}$)        & $1835$              & $4285$              & $650$                    \\
\noalign{\smallskip}\hline\noalign{\smallskip}
SM prediction (post-fit)         & $1840 \pm 45$          & $4280 \pm 70$          & $650 \pm 26$              \\
\noalign{\smallskip}\hline\noalign{\smallskip}
        Fitted $\wen$          & $1260 \pm 43$          & $-$          & $-$              \\
        Fitted $\wmn$          & $10 \pm 2$          & $3500 \pm 90$          & $0.8 \pm 0.2$              \\
        Fitted $\wtn$          & $350 \pm 13$          & $330 \pm 15$          & $0.28 \pm 0.03$              \\
        Fitted $\zee$          & $0.03_{-0.03}^{+0.05}$          & $-$          & $-$              \\
        Fitted $\zmm$         & $1.2 \pm 0.2$          & $71 \pm 4$          & $620 \pm 27$              \\
        Fitted $\ztt$         & $17 \pm 1$          & $8.5 \pm 0.6$          & $1.0 \pm 0.1$              \\
        Fitted $\znn$         & $4.6 \pm 0.7$          & $0.8 \pm 0.1$          & $-$              \\
        Expected $\ttbar$, single top, $\ttbar$+V       & $120 \pm 20$          & $240 \pm 35$          & $8 \pm 2$              \\
        Expected dibosons         & $80 \pm 30$          & $130 \pm 53$          & $21 \pm 7$              \\
 \noalign{\smallskip}\hline\noalign{\smallskip}
SM prediction (pre-fit)    & $1873$          & $4513$          & $621$              \\
\noalign{\smallskip}\hline\noalign{\smallskip}
        Fit input $\wen$         & $1287$          & $-$          & $-$              \\
        Fit input $\wmn$         & $11$          & $3725$          & $0.8$              \\
        Fit input $\wtn$         & $352$          & $342$          & $0.3$              \\
        Fit input $\zee$         & $0.04$          & $-$          & $-$              \\
        Fit input $\zmm$         & $1.2$          & $67$          & $590$              \\
        Fit input $\ztt$         & $17$          & $8.7$          & $1.0$              \\
        Fit input $\znn$         & $4.9$          & $0.8$          & $-$              \\
        Fit input $\ttbar$, single top, $\ttbar$+V     & $120$          & $240$          & $8$              \\
        Fit input dibosons         & $80$          & $130$          & $21$              \\
\noalign{\smallskip}\hline\noalign{\smallskip}
\end{tabular*}
}
\end{center}
\label{tab:fitm2}
\end{table*}

%
%

\begin{table*}[!ht]
\caption{Data and background predictions in the control regions before and after the fit is performed for the M3 selection.
The background predictions include both the statistical and systematic uncertainties.
The individual uncertainties are correlated, and do not necessarily add in
quadrature to the total background uncertainty.}
\begin{center}
\setlength{\tabcolsep}{0.0pc}
{\footnotesize
\begin{tabular*}{\textwidth}{@{\extracolsep{\fill}}lrrr}
\noalign{\smallskip}\hline\noalign{\smallskip}
{\bf  M3 control regions}           & $\wen$            & $\wmn$            & $\zmm$              \\[-0.05cm]
\noalign{\smallskip}\hline\noalign{\smallskip}
Observed  (20.3 fb${}^{-1}$)            & $417$              & $946$              & $131$                    \\
\noalign{\smallskip}\hline\noalign{\smallskip}
SM prediction (post-fit)          & $420 \pm 20$          & $950 \pm 30$          & $130 \pm 12$              \\
\noalign{\smallskip}\hline\noalign{\smallskip}
        Fitted $\wen$          & $270 \pm 17$          & $-$          & $-$              \\
        Fitted $\wmn$          & $2.2 \pm 0.4$          & $750 \pm 37$          & $0.3 \pm 0.1$              \\
        Fitted $\wtn$          & $84 \pm 6$          & $79 \pm 6$          & $0.02 \pm 0.01$              \\
        Fitted $\zee$          & $-$          & $-$          & $-$              \\
        Fitted $\zmm$          & $0.7 \pm 0.1$          & $13 \pm 1$          & $120 \pm 12$              \\
        Fitted $\ztt$          & $4.7 \pm 0.4$          & $1.8 \pm 0.3$          & $0.28 \pm 0.03$              \\
        Fitted $\znn$          & $1.2 \pm 0.2$          & $0.08 \pm 0.02$          & $-$              \\
        Expected $\ttbar$, single top, $\ttbar$+V        & $31 \pm 5$          & $65 \pm 10$          & $1 \pm 1$              \\
        Expected dibosons          & $22 \pm 8$          & $40 \pm 17$          & $5 \pm 3$              \\
 \noalign{\smallskip}\hline\noalign{\smallskip}
    SM prediction (pre-fit)             & $416$          & $1023$          & $132$              \\
\noalign{\smallskip}\hline\noalign{\smallskip}
        Fit input $\wen$          & $271$          & $-$          & $-$              \\
        Fit input $\wmn$          & $2.4$          & $824$          & $0.3$              \\
        Fit input $\wtn$          & $83$          & $79$          & $0.02$              \\
        Fit input $\zee$          & $-$          & $-$          & $-$              \\
        Fit input $\zmm$          & $0.7$          & $13$          & $125$              \\
        Fit input $\ztt$          & $4.6$          & $1.8$          & $0.3$              \\
        Fit input $\znn$          & $1.3$          & $0.10$          & $-$              \\
        Fit input $\ttbar$, single top, $\ttbar$+V         & $31$          & $65$          & $1$              \\
        Fit input dibosons          & $22$          & $40$          & $5$              \\
\noalign{\smallskip}\hline\noalign{\smallskip}
\end{tabular*}
}
\end{center}
\label{tab:fitm3}
\end{table*}

%
%

%
%
  
\begin{table*}[!ht]
\caption{
 Data and background predictions in the $W/Z$+jets and $\ttbar$ control regions before and after the fit is performed for the $c$-tagged selection.
The background predictions include both the statistical and systematic uncertainties.
The individual uncertainties are correlated, and do not necessarily add in
quadrature to the total background uncertainty.
}
\begin{center}
\setlength{\tabcolsep}{0.0pc}
{\footnotesize
\begin{tabular*}{\textwidth}{@{\extracolsep{\fill}}lrrrr}
\noalign{\smallskip}\hline\noalign{\smallskip}
{\bf  $c$-tagged control regions}           & $\wmn$ & $\wen$  & $\Zll$  & $\ttbar$ \\[-0.05cm]
\noalign{\smallskip}\hline\noalign{\smallskip}
Observed events (20.3~fb${}^{-1}$)  & $1783$              & $785$              & $113$              & $140$                    \\
\noalign{\smallskip}\hline\noalign{\smallskip}
SM prediction  (post-fit) & $1780 \pm 42$          & $790 \pm 28$          & $110 \pm 11$          & $140 \pm 12$ \\
\noalign{\smallskip}\hline\noalign{\smallskip}
Fitted $\wen$  & $-$          & $260 \pm 49$          & $0.08 \pm 0.02$          & $0.19 \pm 0.05$ \\
Fitted $\wmn$  & $480 \pm 110$          & $0.1 \pm 0.1$          & $0.01 \pm 0.01$          & $0.6 \pm 0.1$  \\
Fitted $\wtn$  & $70 \pm 14$          & $29 \pm 6$          & $-$          & $0.06 \pm 0.02$  \\
Fitted $\znn$  & $-$          & $0.35 \pm 0.05$          & $-$          & $-$ \\
Fitted $\zee$  & $-$          & $-$          & $49 \pm 6$          & $-$ \\
Fitted $\zmm$  & $22 \pm 3$          & $-$          & $45 \pm 5$          & $6.4 \pm 0.8$ \\
Fitted $\ztt$  & $16 \pm 3$          & $3.7 \pm 0.7$          & $-$          & $1.9 \pm 0.4$ \\
Fitted $\ttbar$  & $1000 \pm 110$          & $400 \pm 43$          & $7.1 \pm 0.8$          & $120 \pm 12$ \\
Expected  $\ttbar + V$ & $9 \pm 1$          & $4.5 \pm 0.5$          & $1.0 \pm 0.1$          & $1.8 \pm 0.2$  \\
Expected single top         & $95 \pm 18$          & $49 \pm 9$          & $0.35 \pm 0.08$          & $7 \pm 1$ \\ 
Expected dibosons & $76 \pm 15$          & $35 \pm 8$          & $11 \pm 2$          & $5 \pm 1$ \\ 
Expected  Higgs   & $1.1 \pm 0.2$          & $0.5 \pm 0.1$          & $0.06 \pm 0.01$          & $0.14 \pm 0.02$  \\
 \noalign{\smallskip}\hline\noalign{\smallskip}
SM prediction (pre-fit)              & $1830$          & $790$          & $127$          & $132$ \\ 
\noalign{\smallskip}\hline\noalign{\smallskip}
        Fit input $\wen$         & $-$          & $290$          & $0.08$          & $0.20$ \\
        Fit input $\wmn$         & $588$          & $0.1$          & $0.02$          & $0.7$ \\
        Fit input $\wtn$         & $79$          & $32$          & $-$          & $0.10$ \\
        Fit input $\znn$         & $-$          & $0.40$          & $-$          & $-$ \\
        Fit input $\zee$         & $-$          & $-$          & $56$          & $-$ \\
        Fit input $\zmm$         & $25$          & $-$          & $52$          & $7.4$ \\
        Fit input $\ztt$         & $17$          & $4.1$          & $-$          & $2.2$ \\
        Fit input  $\ttbar$          & $940$          & $374$          & $6.7$          & $108$ \\
        Fit input $\ttbar + V$          & $9$          & $4.5$          & $1.0$          & $1.8$ \\
        Fit input single top          & $95$          & $49$          & $0.35$          & $7$ \\
        Fit input dibosons         & $76$          & $35$          & $11$          & $5$ \\
        Fit input Higgs         & $1.1$          & $0.5$          & $0.06$          & $0.14$  \\
\noalign{\smallskip}\hline\noalign{\smallskip}
\end{tabular*}
}
\end{center}
\label{tab:fitc1}
\end{table*}

\subsection{Validation of the background determination}
\label{sec:vali}

In the monojet-like analysis, the control regions are defined using the same
requirements for $\met$, leading jet $\pt$, event topologies, and jet vetoes as in the
signal regions,  such that no extrapolation in $\met$ and jet $\pt$ is needed from control to signal regions. The 
agreement between data and background predictions is confirmed in a low-$\pt$ validation region defined using 
the same monojet-like  selection criteria with $\met$ and  leading jet $\pt$ limited to the range $150$--$220$~GeV. 

\begin{figure*}[!ht]
\begin{center}
\mbox{
  \includegraphics[width=0.42\textwidth]{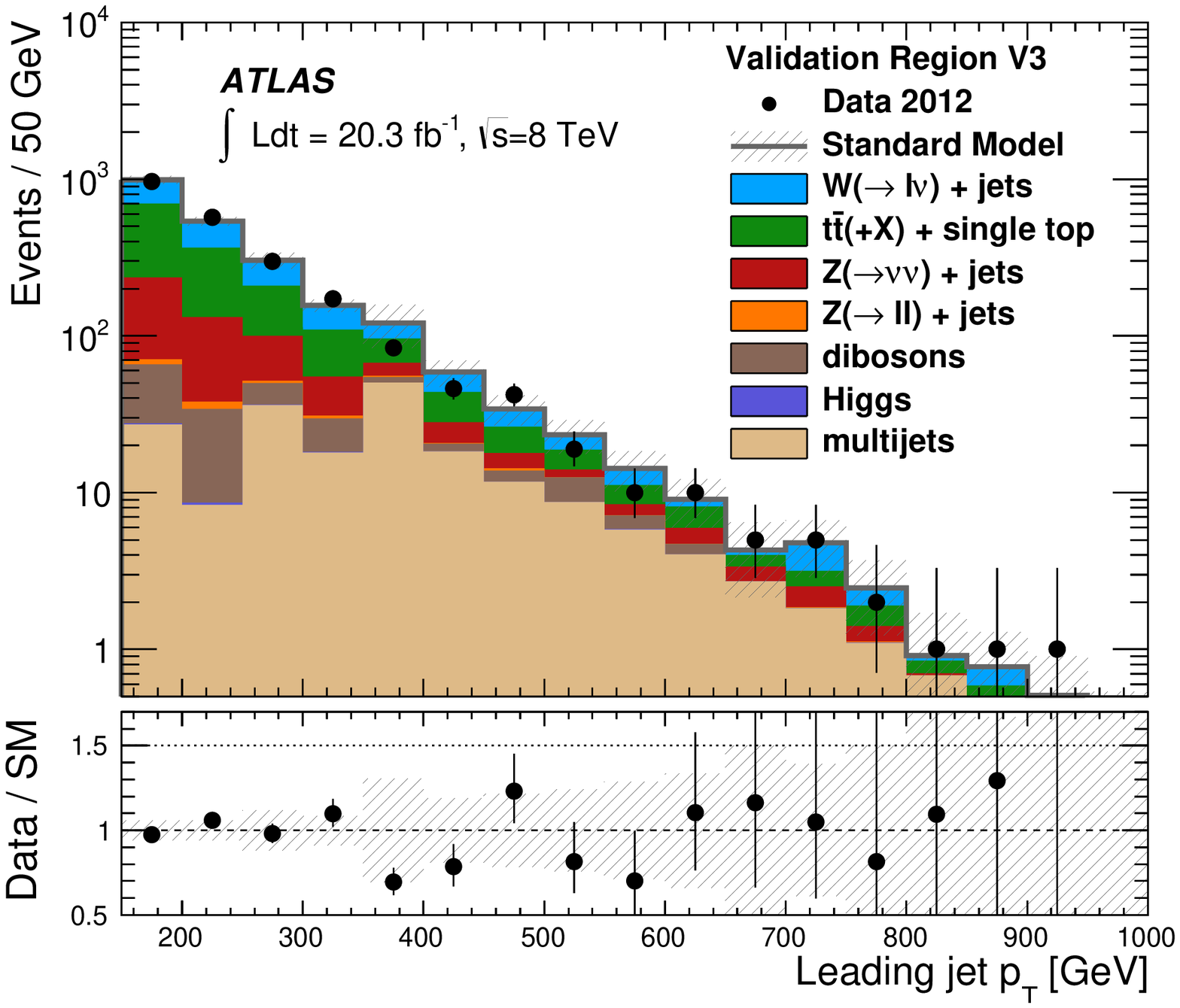}
  \includegraphics[width=0.42\textwidth]{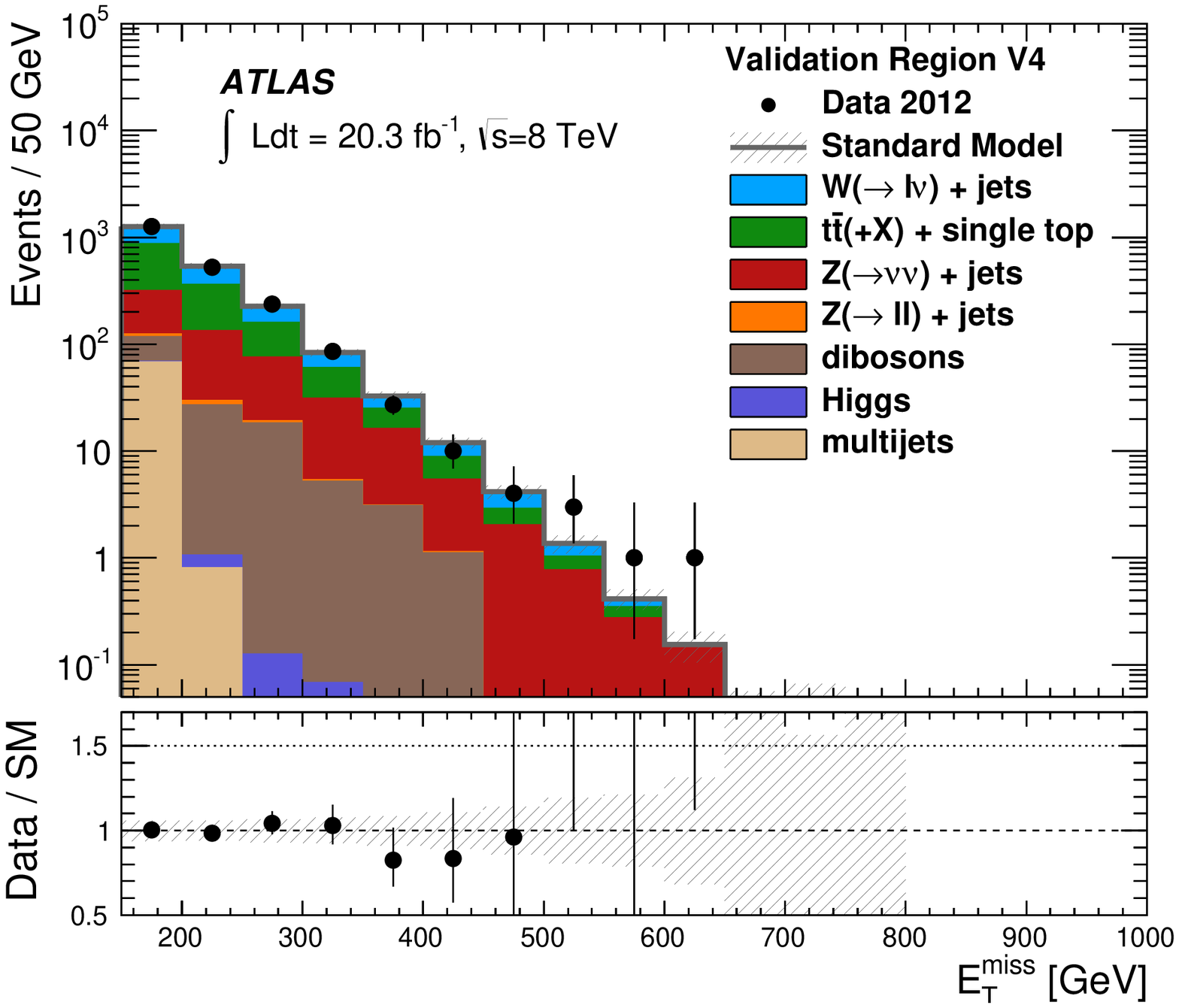}
}
\mbox{
  \includegraphics[width=0.42\textwidth]{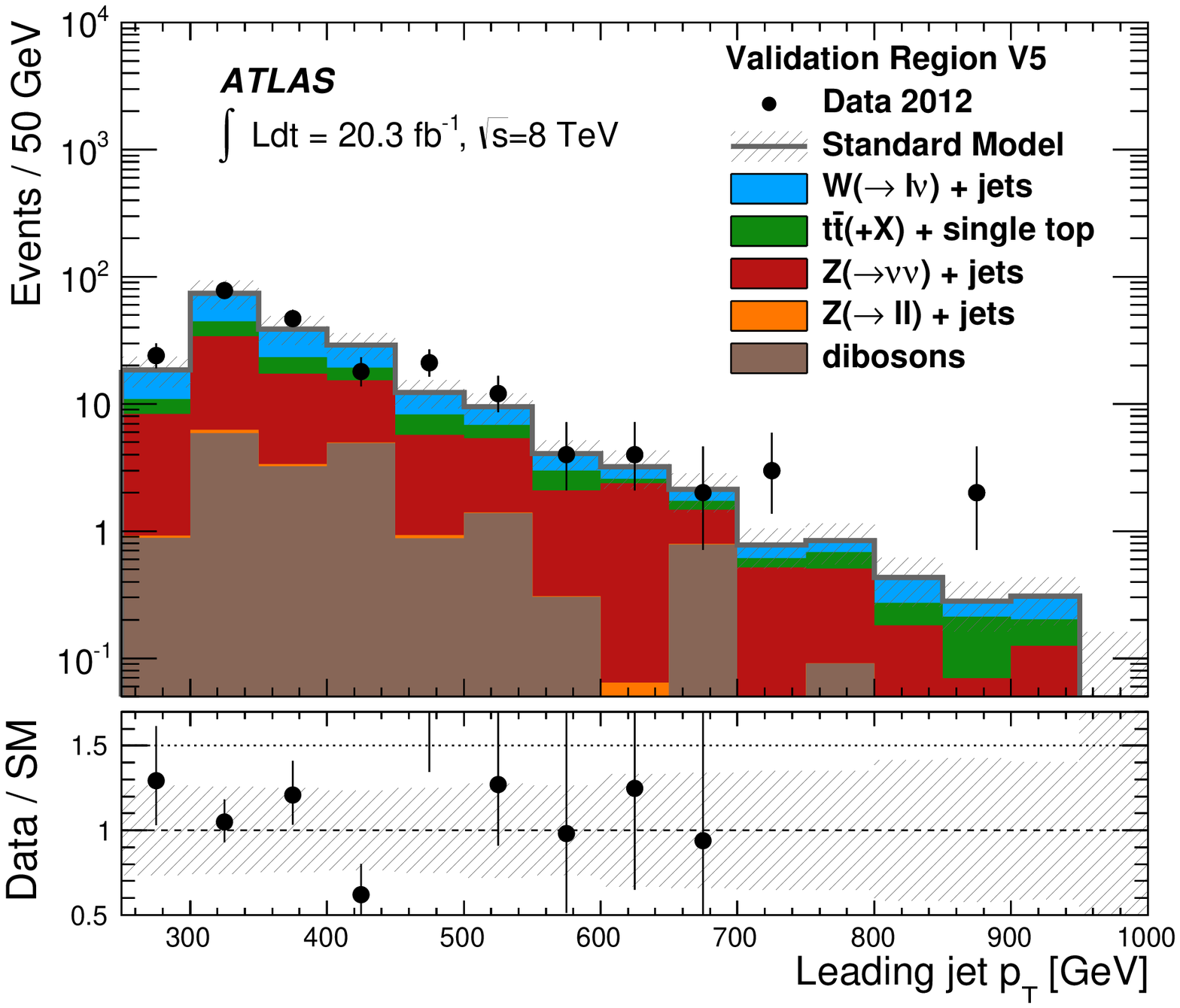}
  \includegraphics[width=0.42\textwidth]{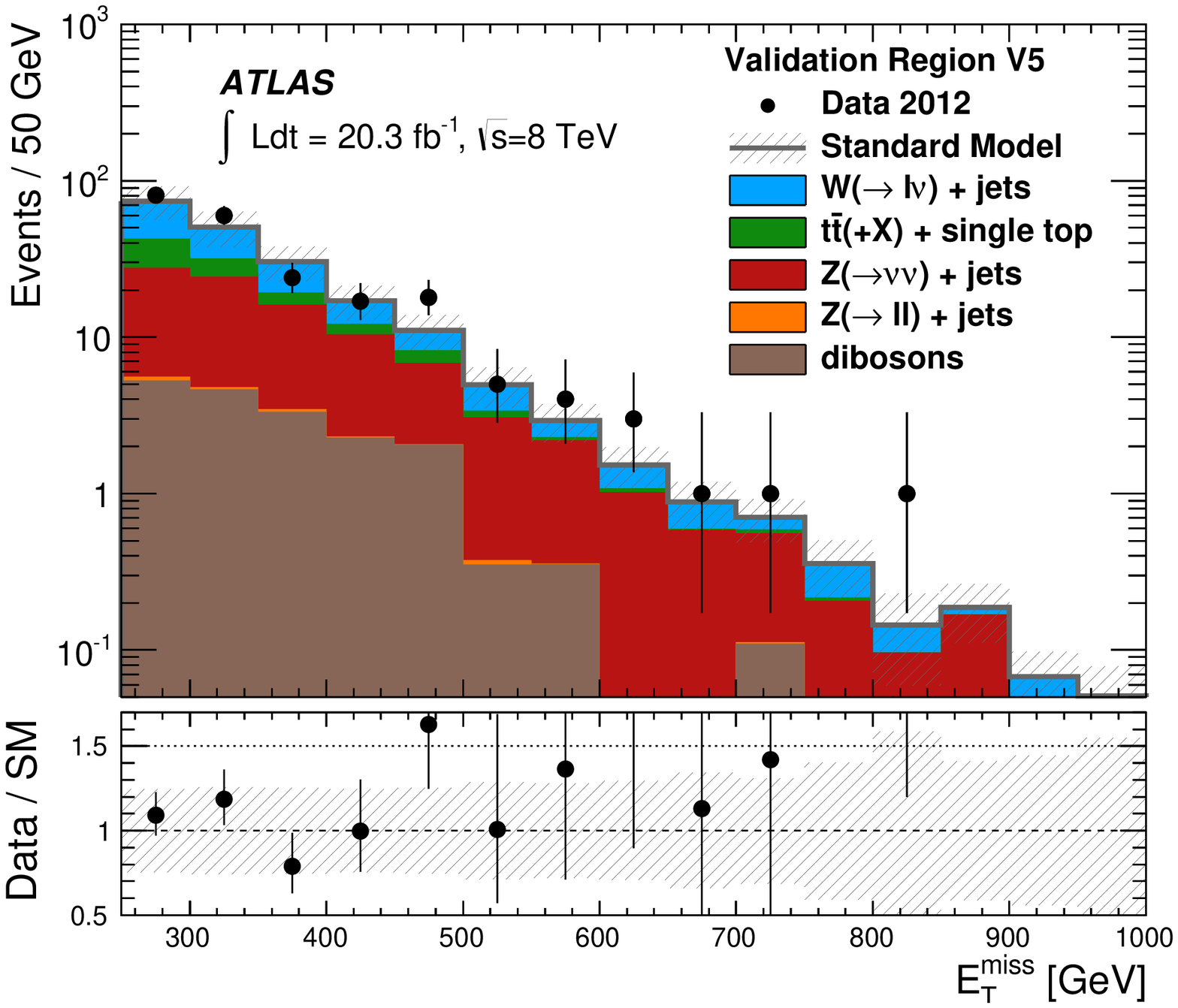}
}
\end{center}
\caption{
Measured leading jet $\pt$ and $\met$ distributions for the V3--V4 (top) and V5 (bottom)
selections compared to the SM predictions.
The error bands in the ratios include both the statistical and systematic uncertainties 
on the background predictions.
}
\label{fig:vr1}
\end{figure*}

In the case of the $c$-tagged analysis,  for which the control regions are defined with lower thresholds on the
leading jet $\pt$ and $\met$ compared to those of the signal regions,   the
$\wmn$+jets, $\wen$+jets, $\zll$+jet and $\ttbar$ yields fitted
in the control regions are then validated in dedicated validation regions (here denoted by V1--V5).
The definition of the validation regions is presented in 
Table~\ref{tab:vrdef} and is such that  there is  no 
overlap of events with the control and signal regions.  The validation regions V1--V4 differ 
from the signal regions only on the thresholds imposed on the $\met$ and leading jet
$\pt$. In the case of V5, the same requirements as one of the signal regions  on $\met$ and leading jet $\pt$ are imposed but 
the number of jets is limited to be exactly three. 
Similar to the transfer factors from control to signal regions, 
transfer factors from the control to the validation regions are also 
defined based on MC simulation. The same experimental
systematic uncertainties are evaluated and taken into account in the
extrapolation. These transfer factors are  subject to the 
modeling uncertainties of the simulation, which are also applied in
the validation regions. Hence, the extrapolation to the validation
regions is identical to that of the signal regions.
Table~\ref{tab:vr} presents the comparison between data and the scaled
MC predictions in the validation regions and Fig.~\ref{fig:vr1}  presents  
the $\met$ and leading jet $\pt$ distributions for V3 to V5 regions. 
 Good agreement, within uncertainties, is observed between data and predictions demonstrating a good understanding of the background yields.

\begin{table*}[!ht]
\caption{Definition of the validation regions for the $c$-tagged selection.}
\begin{center}
\setlength{\tabcolsep}{0.0pc}
{\small
\begin{tabular*}{\textwidth}{@{\extracolsep{\fill}}lccccc}
\noalign{\smallskip}\hline\noalign{\smallskip}
 & V1 & V2 & V3& V4 & V5\\\hline
& \multicolumn{5}{c}{Preselection} \\
\noalign{\smallskip}\hline\noalign{\smallskip}
Tagging & \multicolumn{5}{c}{One medium $c$-tag among jets 2--4(2--3) for V1--V4(V5)}\\
 & \multicolumn{5}{c}{Three (two) loose $c$-tags, acting as $b$-veto, for other 3(2) jets  for V1--V4(V5)}\\
\noalign{\smallskip}\hline\noalign{\smallskip}
$N_e$ & 0 & 0 & 0 & 0 & 0\\
$N_\mu$ & 0&0&0&0&0\\
$N_{\rm jet}$ & {$\geq$ 4} & {$\geq$ 4}& {$\geq$ 4}& {$\geq$ 4}& = 3\\
\met\ (\GeV) & $\in [150,250]$ & $\in [200,250]$ &  $\in [150,250]$ &
$>$ 150 & $>$ 250\\
Leading jet $\pt$ (\GeV) & $\in [150,250]$ & $\in [200,290]$ &  $>$ 150 & $\in
      [150,290]$ & $>$ 290\\
\noalign{\smallskip}\hline\noalign{\smallskip}
\end{tabular*}
}
\end{center}
\label{tab:vrdef}
\end{table*}

\begin{table*}[!ht]
\caption{Observed events and SM background predictions from the control regions for the 
V1 to V5 validation regions. The errors shown are the statistical plus systematic uncertainties.
The individual uncertainties are correlated, and do not necessarily add in
quadrature to the total background uncertainty.
}
\begin{center}
\setlength{\tabcolsep}{0.0pc}
{\footnotesize
\begin{tabular*}{\textwidth}{@{\extracolsep{\fill}}lrrrrr}
\noalign{\smallskip}\hline\noalign{\smallskip}
{\bf  $c$-tagged validation regions}           & V1           & V2            & V3
& V4  & V5             \\[-0.05cm]
\noalign{\smallskip}\hline\noalign{\smallskip}
Observed events  (20.3 fb${}^{-1}$)        & $1534$              & $257$              & $2233$              & $2157$       & $215$               \\
\noalign{\smallskip}\hline\noalign{\smallskip}
Fit prediction          & $1530 \pm 90$          & $260 \pm 20$          & $2300 \pm 190$          & $2200 \pm 190$          & $200 \pm 50$              \\
\noalign{\smallskip}\hline\noalign{\smallskip}
          $\wen$           & $70 \pm 13$          & $12 \pm 2$          & $100 \pm 20$          & $100 \pm 18$          & $9 \pm 3$              \\
          $\wmn$           & $60 \pm 14$          & $10 \pm 2$          & $90 \pm 20$          & $90 \pm 19$          & $10 \pm 3$              \\
          $\wtn$           & $330 \pm 60$          & $64 \pm 12$          & $470 \pm 86$          & $460 \pm 82$          & $50 \pm 19$              \\
          $\znn$           & $260 \pm 44$          & $52 \pm 12$          & $360 \pm 56$          & $410 \pm 95$          & $80 \pm 20$              \\
          $\zee$           & $-$          & $-$          & $-$          & $-$          & $-$              \\
          $\zmm$           & $1.1 \pm 0.1$          & $0.14 \pm 0.02$          & $1.6 \pm 0.2$          & $1.5 \pm 0.2$          & $0.11 \pm 0.03$              \\
          $\ztt$           & $8 \pm 1$          & $0.9 \pm 0.2$          & $12 \pm 2$          & $10 \pm 2$          & $0.5 \pm 0.2$              \\
          $\ttbar$            & $630 \pm 90$          & $92 \pm 14$          & $830 \pm 160$          & $830 \pm 170$          & $20 \pm 5$              \\
          $\ttbar +V$           & $6.3 \pm 0.7$          & $1.3 \pm 0.1$          & $10 \pm 1$          & $10 \pm 1$          & $0.16 \pm 0.05$              \\
          Single top           & $60 \pm 12$          & $9 \pm 2$          & $80 \pm 17$          & $80 \pm 16$          & $8 \pm 1$              \\
          Dibosons           & $60 \pm 14$          & $14 \pm 3$          & $100 \pm 22$          & $100 \pm 23$          & $18 \pm 3$              \\
          Higgs           & $0.7 \pm 0.1$          & $0.15 \pm 0.03$          & $1.1 \pm 0.2$          & $1.1 \pm 0.2$          & $0.09 \pm 0.02$              \\
          Multijets           & $40 \pm 19$          & $0.8 \pm 0.8$          & $200 \pm 99$          & $70 \pm 36$          & $-$              \\
 \noalign{\smallskip}\hline\noalign{\smallskip}
\end{tabular*}
}
\end{center}
\label{tab:vr}
\end{table*}

%
%

 \section{Systematic uncertainties  and background fits}
 \label{sec:syst}

In this section  the impact of each  source of systematic uncertainty on the total 
background prediction in the signal regions, as determined via the global fits 
explained in Sec.~\ref{sec:fitback}, is discussed separately for monojet-like and $c$-tagged selections. 
Finally, the experimental and theoretical uncertainties on the SUSY signal yields are discussed.  

%
%

\subsection{Monojet-like analysis}


%
%
Uncertainties on the absolute jet and $\met$ energy scale and resolution~\cite{Aad:2011he}
translate  into an uncertainty on
the total background that varies between $1.1\%$ for M1 and $1.3\%$ for M3.
Uncertainties related to jet quality requirements and pileup description and 
corrections to the jet $\pt$ and $\met$ introduce 
a $0.2\%$  to $0.3\%$ uncertainty on the background predictions.
%
%
Uncertainties on
the simulated lepton identification and reconstruction efficiencies, energy/momentum scale and resolution translate 
into a  $1.2\%$ and $0.9\%$ uncertainty in the total background for M1 and M3 selections, respectively.
%
%

Variations of the renormalization/factorization and parton-shower matching scales and PDFs 
in the  {SHERPA}  $W/Z$+jets background samples translate into a $1\%$ to $0.4\%$ uncertainty in the total background.
Variations within uncertainties in the re-weighting procedure for the simulated $W$ and $Z$ $\pt$ distributions 
introduce less than a  
0.2$\%$ uncertainty on the total background estimates. 

Model uncertainties, related to potential differences between $W$+jets and $Z$+jets final states, affecting the normalization of the 
dominant $\znn$+jets and the small $\ztt$+jets and  $\zee$+jets background contributions, as determined in $\wmn$+jets
and $\wen$+jets control regions,  are studied in detail. This includes uncertainties related to PDFs and  renormalization/factorization scale settings, the  
parton shower parameters and the hadronization model used in the MC simulations, and the dependence on the lepton reconstruction and 
acceptance. As a result, an additional  $3\%$ uncertainty on the $\znn$+jets, $\ztt$+jets and  $\zee$+jets contributions is included for all the selections.  
Separate studies using parton-level predictions for $W/Z$+jet production, as implemented in MCFM-6.8~\cite{mcfm}, indicate that NLO strong
corrections affect the $\wmn$+jets to $\znn$+jets ratio by less than 1$\%$ in the $\met$ and leading jet $\pt$ kinematic range considered.  
In addition, the effect from NLO electroweak corrections on the $W$+jets to $Z$+jets ratio is taken into account~\cite{Denner:2012ts,Denner:2011vu,Denner:2009gj}.
Dedicated parton-level calculations are performed with the same $\met$ and leading jet $\pt$ requirements as  in the M1 to M3
signal regions. The studies suggest an effect on the $W$+jets to $Z$+jets ratio that varies between about 2$\%$ for M1 and $3\%$ for M2 and M3, although the 
calculations suffer from large uncertainties, mainly due to the limited knowledge of the photon PDFs inside the proton. In this analysis, 
these results are conservatively adopted as an additional uncertainty on the  $\znn$+jets, $\ztt$+jets and  $\zee$+jets contributions.
Altogether, this translates into an uncertainty on the total background that varies from  $1.9\%$ and $2.1\%$  for the M1 and M2 selections, 
respectively, to about $2.6\%$ for the M3 selection.

%
%

Theoretical uncertainties on the predicted background yields for top-quark-related processes include: uncertainties on 
the absolute $\ttbar$, single top and $\ttbar + Z/W$ cross sections;  uncertainties on the MC generators and the modeling of parton showers employed (see Sec.~\ref{sec:mc}); variations in the set of parameters that govern the parton showers and the amount of initial- and final-state soft gluon radiation; and uncertainties 
due to the choice of renormalization and factorization scales and PDFs.  This introduces an uncertainty on the total background
prediction that varies between 1.6$\%$ and 1.0$\%$ for the M1 and M3 selections, respectively. 
Uncertainties on the diboson 
contribution are estimated in a similar way and translate into an uncertainty on the total background in the 
range between 0.7$\%$ and 1.3$\%$.  
A conservative 100$\%$ uncertainty on the multijet background 
 estimation  is adopted,  leading to a $1\%$ uncertainty on the total background for the M1 selection.
Finally, statistical uncertainties related to the data control regions and simulation samples lead to 
an additional uncertainty on the final background estimates in the signal regions 
that vary between 1.2$\%$ for M1 and 1.4$\%$ for M3 selections. 
Other uncertainties related to the trigger efficiency and 
the determination of the total integrated luminosity~\cite{Aad:2013ucp} are also included, which 
cancel out in the case of the dominant background contributions that are 
determined using data-driven methods, leading to a less than 0.3$\%$ uncertainty 
on the total background.

%
%

\subsection{$c$-tagged analysis}

In the $c$-tagged analysis, the jet energy scale uncertainty translates into a 0.3$\%$ to 2.2$\%$ uncertainty in the final background estimate.  
Uncertainties related to the {\it loose} and {\it medium} $c$-tag introduce a 2.8$\%$ and 2.5$\%$ 
uncertainty on the background yield for the C1 and C2 selections, respectively. 
Uncertainties related
to the jet energy resolution, soft contributions to $\met$, modeling of multiple $pp$ interactions,
trigger and lepton reconstruction and identification (momentum and
energy scales, resolutions and efficiencies) translate into   
about a 1.2$\%$ (1.4$\%$) uncertainty for the C1 (C2) selection. 
Variations of the renormalization/factorization and parton-shower matching scales and PDFs
in the  {SHERPA}  $W/Z$+jets background samples translate into a 3.0$\%$ and 
$3.3\%$ uncertainty in the total background for the C1 and C2 selections, respectively.
Uncertainties in the re-weighting of the simulated $W$ and $Z$ $\pt$ distributions, affecting 
the extrapolation of the MC normalization factors from the control to the signal regions,  
introduce a less than 0.6$\%$ uncertainty in the final background estimates.
In the $c$-tagged analysis,  the $Z$+jets and
$W$+jets background is enriched in heavy-flavor jets produced in
association with the vector boson and the same heavy-flavor processes 
are present in the signal region and the $V$+jets control regions.
Theoretical uncertainties on the background predictions for top-related processes and 
diboson contributions are computed following the same prescriptions 
as in the monojet-like analysis and constitute the dominant sources of systematic uncertainty.  
In the case of top-related processes,  this 
translates into an uncertainty on the total
background prediction of 5.2$\%$ and 5.0$\%$ for C1 and C2 selections, respectively. 
Similarly, the uncertainties on the diboson contributions lead to 
an uncertainty on the total background of 5.5$\%$ (11.5$\%$) for the 
C1 (C2) selection. 
The limited number of SM MC events and data events
in the control
regions lead to an additional uncertainty of 3.0$\%$ (4.4$\%$) for the C1 (C2) signal region.  
Finally, a conservative 100$\%$ uncertainty on the multijet background contribution in the control and signal 
regions is also adopted,  which translates into a 0.4$\%$ and 0.9$\%$ uncertainty on the total background for
the C1 and C2 selections, respectively.


\subsection{Signal systematic uncertainties}

Different sources of systematic uncertainty on the predicted SUSY signals are considered. 
Experimental uncertainties related to the jet and $\met$ reconstruction, energy
scales and resolutions introduce uncertainties in the signal yields in the range  
3$\%$ to 7$\%$ and 10$\%$ to 27$\%$   for the monojet-like and $c$-tagged analyses, respectively, depending 
on the stop and neutralino masses considered. 
In the $c$-tagged analysis, uncertainties on the simulated  $c$-tagging efficiencies 
for {\it loose}  and {\it medium} tags introduce  9$\%$ to 16$\%$
uncertainties in the signal yields.
In addition,  a 2.8$\%$
uncertainty on the integrated luminosity is included.
Uncertainties affecting the signal
acceptance times efficiency ($A \times \epsilon$)
related to the generation of the SUSY samples 
are determined using additional samples with modified
parameters. This includes uncertainties on the  
modeling of the initial- and final-state
gluon radiation,       
the choice of renormalization/factorization scales,  and the parton-shower matching scale settings. 
Altogether this translates into an uncertainty on the signal yields that 
tends to increase with 
decreasing $\Delta m$ and    
varies between 8$\%$ and 12$\%$ in the monojet-like analyses, and between 17$\%$ and 38$\%$ in the 
$c$-tagged selections, depending on the stop and neutralino masses. 
Finally, uncertainties on the predicted SUSY signal cross sections include PDF uncertainties, variations 
on the $\alpha_{s}(M_Z)$ value employed, as well as variations of the renormalization and factorization 
scales by factors of two and one-half. Altogether, this results in a total theoretical uncertainty 
on the cross section 
that
varies between 14$\%$ and 16$\%$ for stop masses in the range between 100~GeV and  400~GeV.       

\section{Results and interpretation}
\label{sec:results}

\begin{figure*}[!ht]
\begin{center}
\mbox{
  \includegraphics[width=0.42\textwidth]{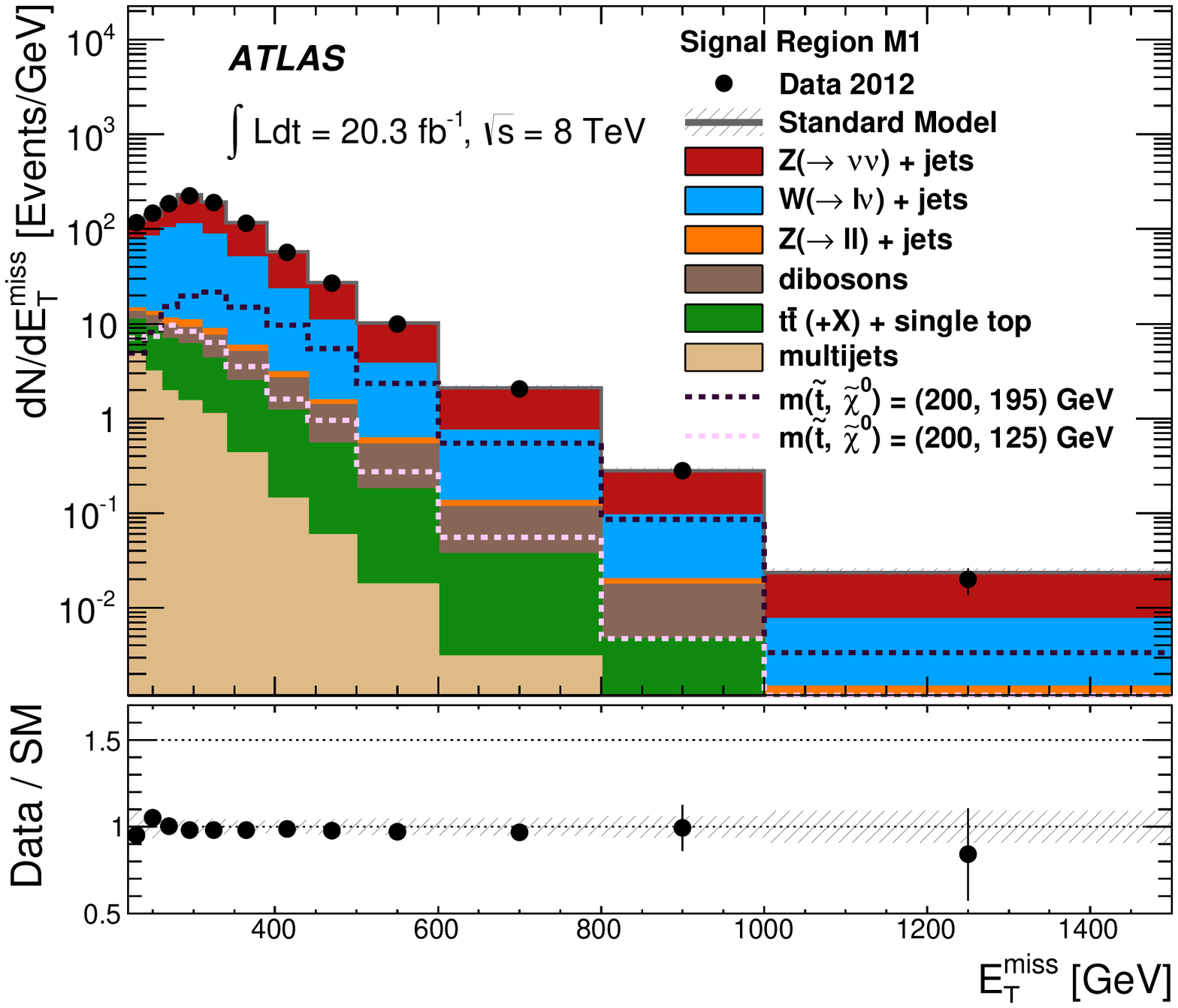}
  \includegraphics[width=0.42\textwidth]{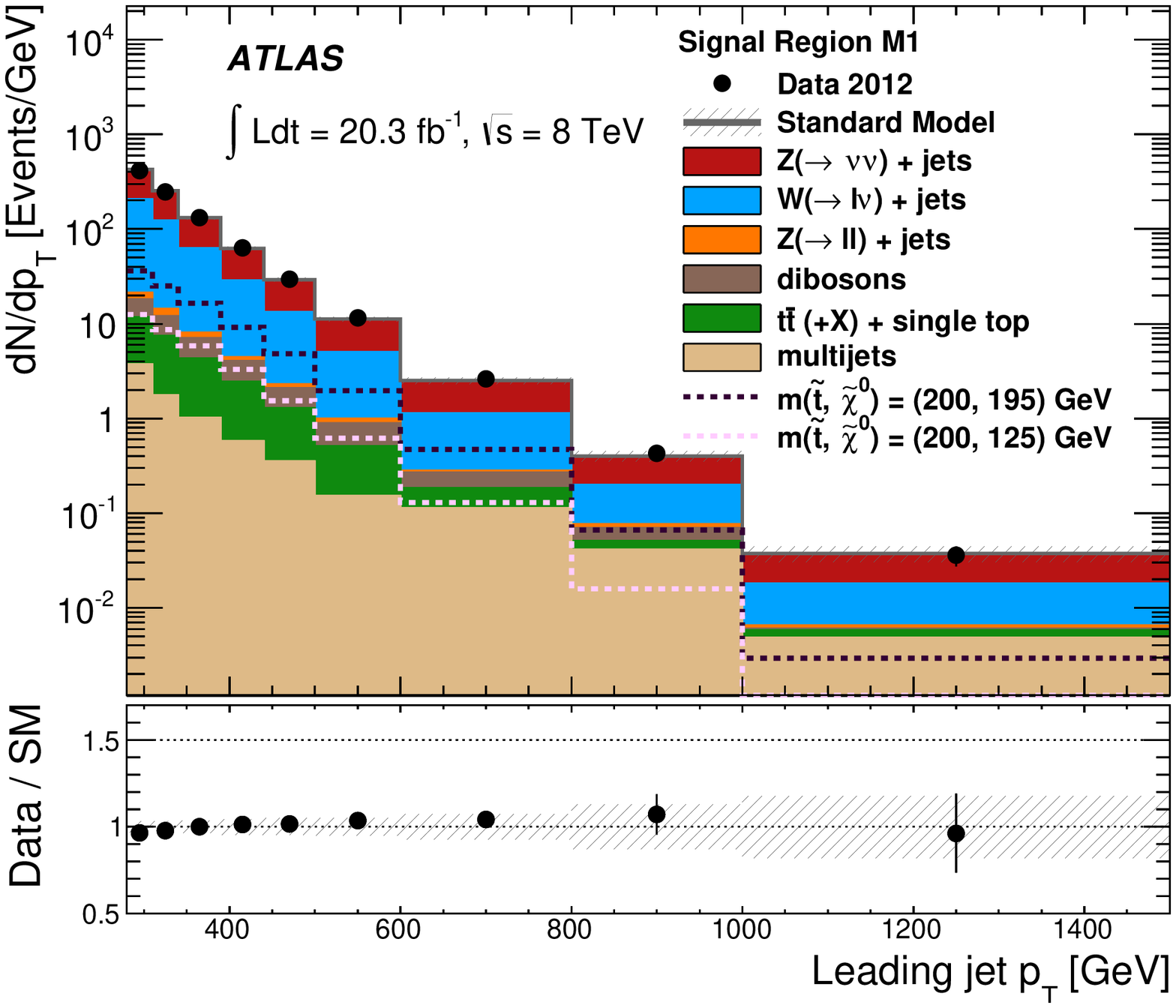}
}
\mbox{
  \includegraphics[width=0.42\textwidth]{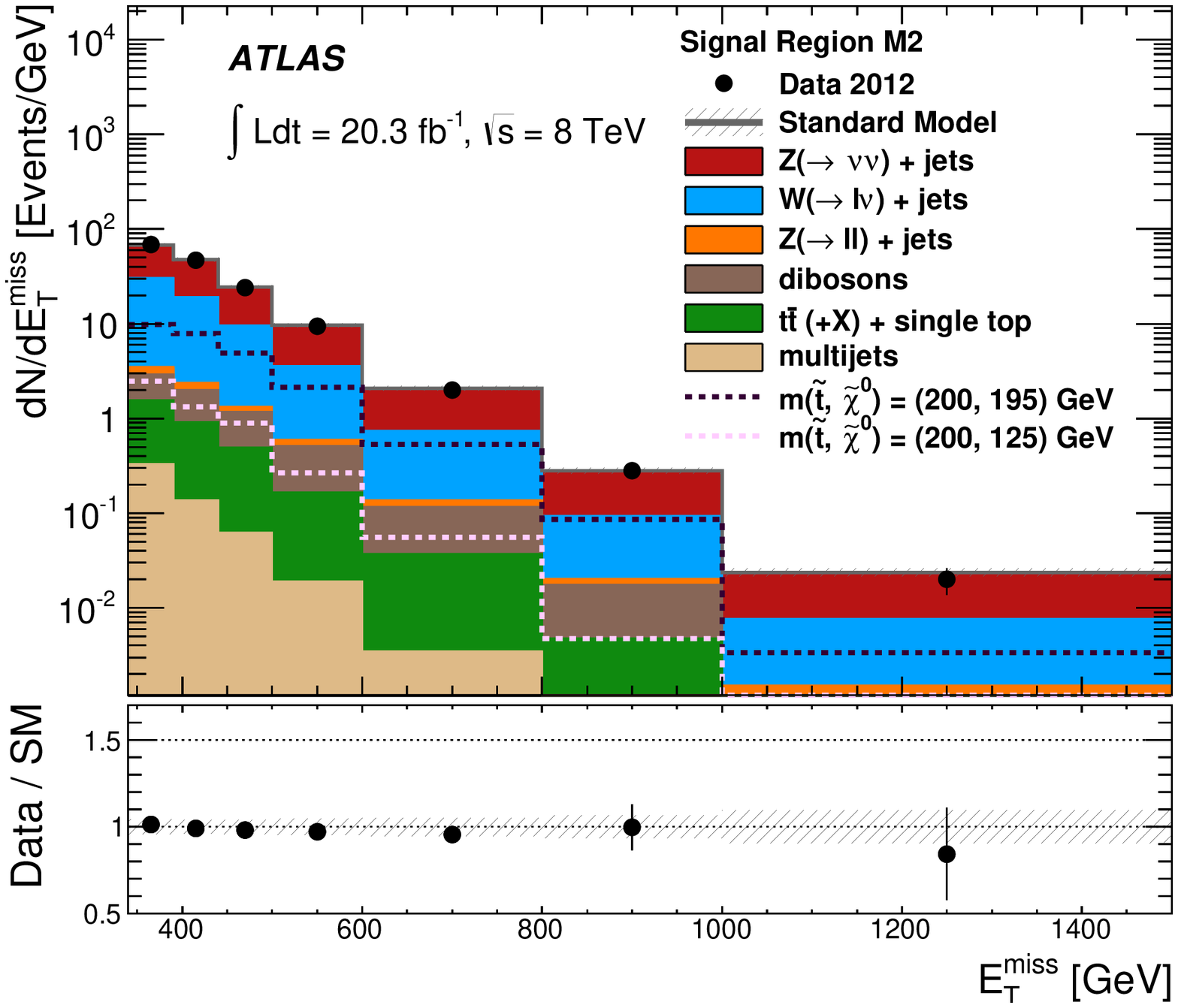}
  \includegraphics[width=0.42\textwidth]{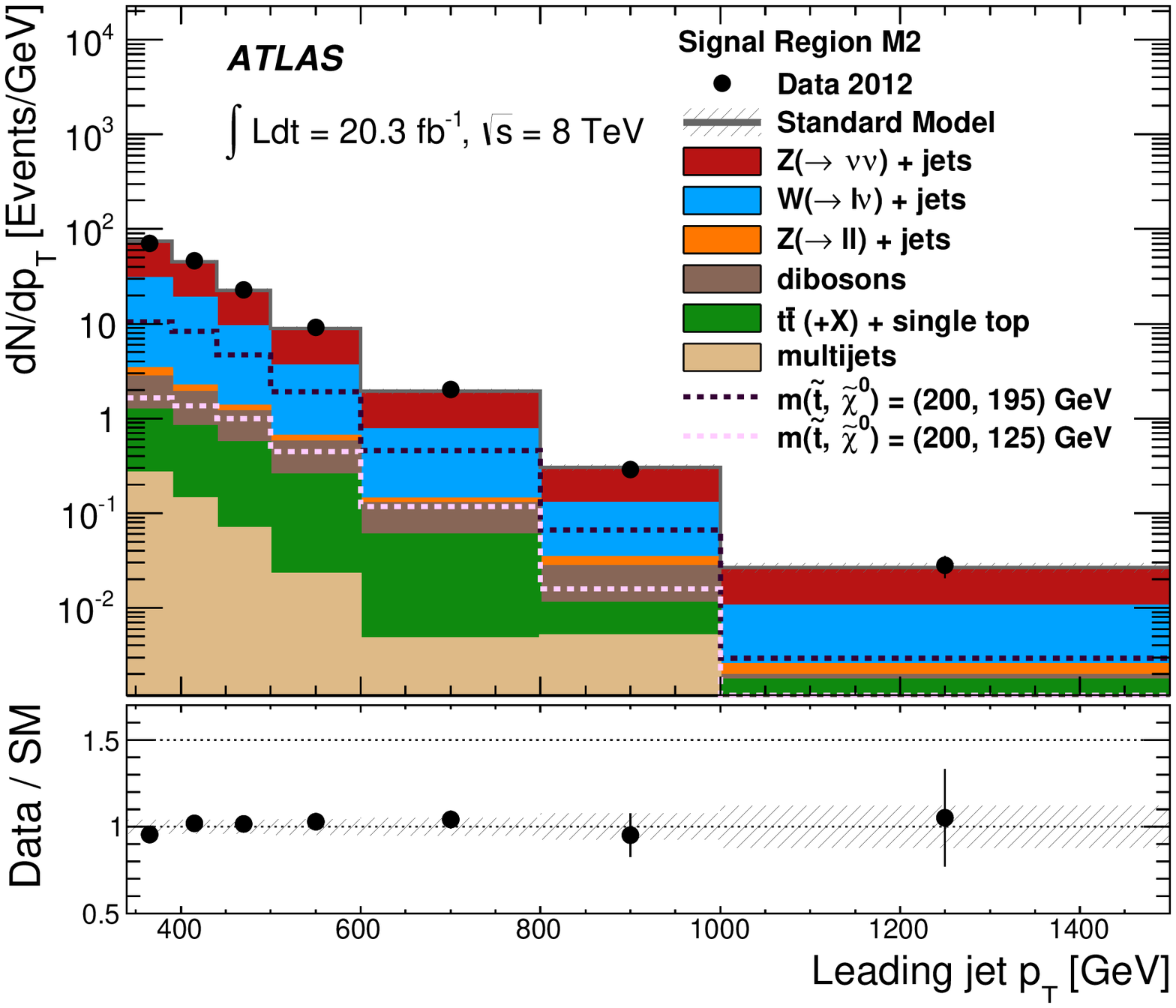}
}
\mbox{
  \includegraphics[width=0.42\textwidth]{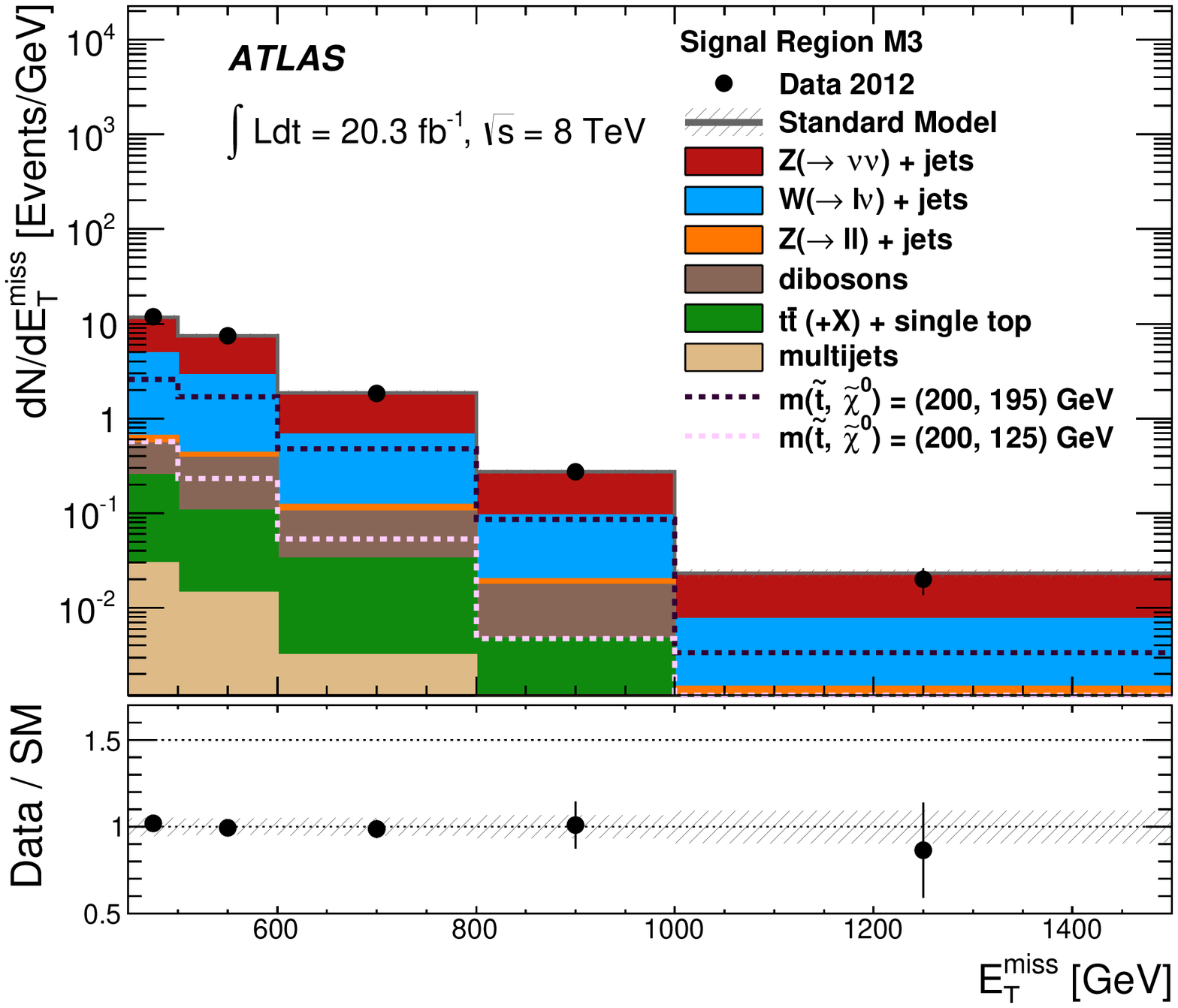}
  \includegraphics[width=0.42\textwidth]{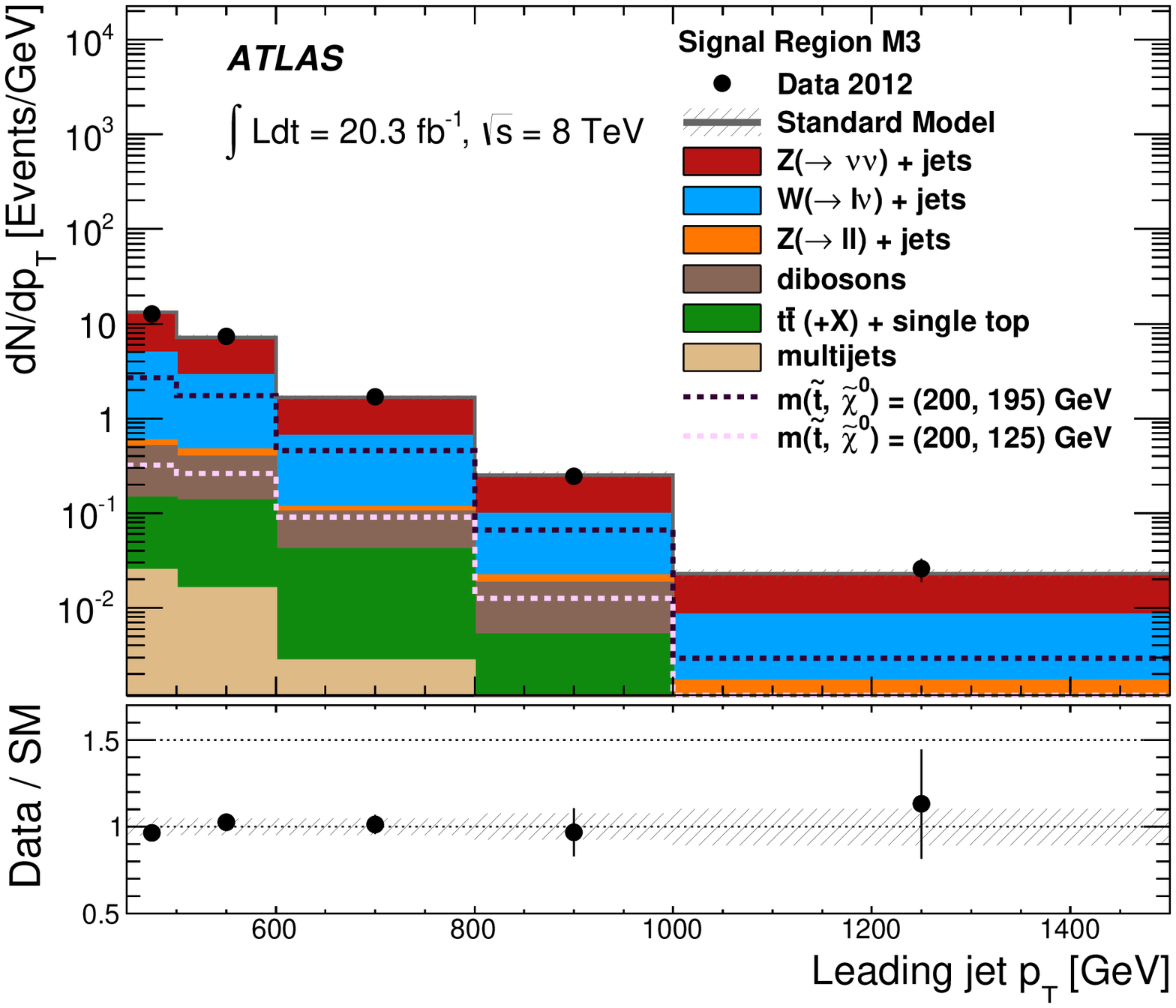}
}
\end{center}
\caption{
Measured $\met$ and leading jet $\pt$ distributions for the M1 (top), M2 (middle), and M3 (bottom) 
selections compared to the SM predictions. 
For illustration purposes, the distribution of two different SUSY scenarios are included.
The error bands in the ratios include both the statistical and systematic uncertainties 
on the background predictions.
}
\label{fig:sr1}
\end{figure*}

\begin{figure*}[!ht]
\begin{center}
\mbox{
  \includegraphics[width=0.42\textwidth]{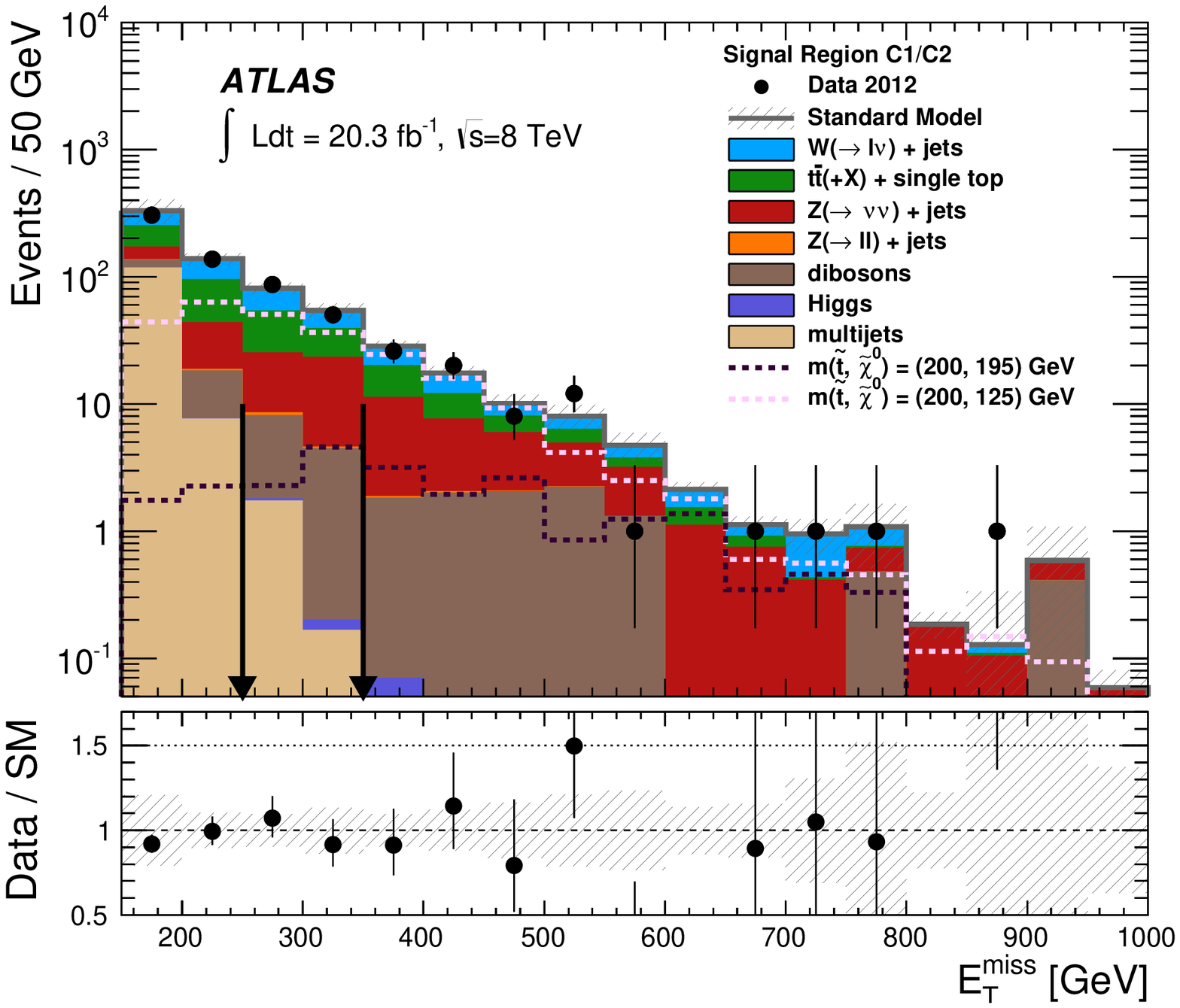}
  \includegraphics[width=0.42\textwidth]{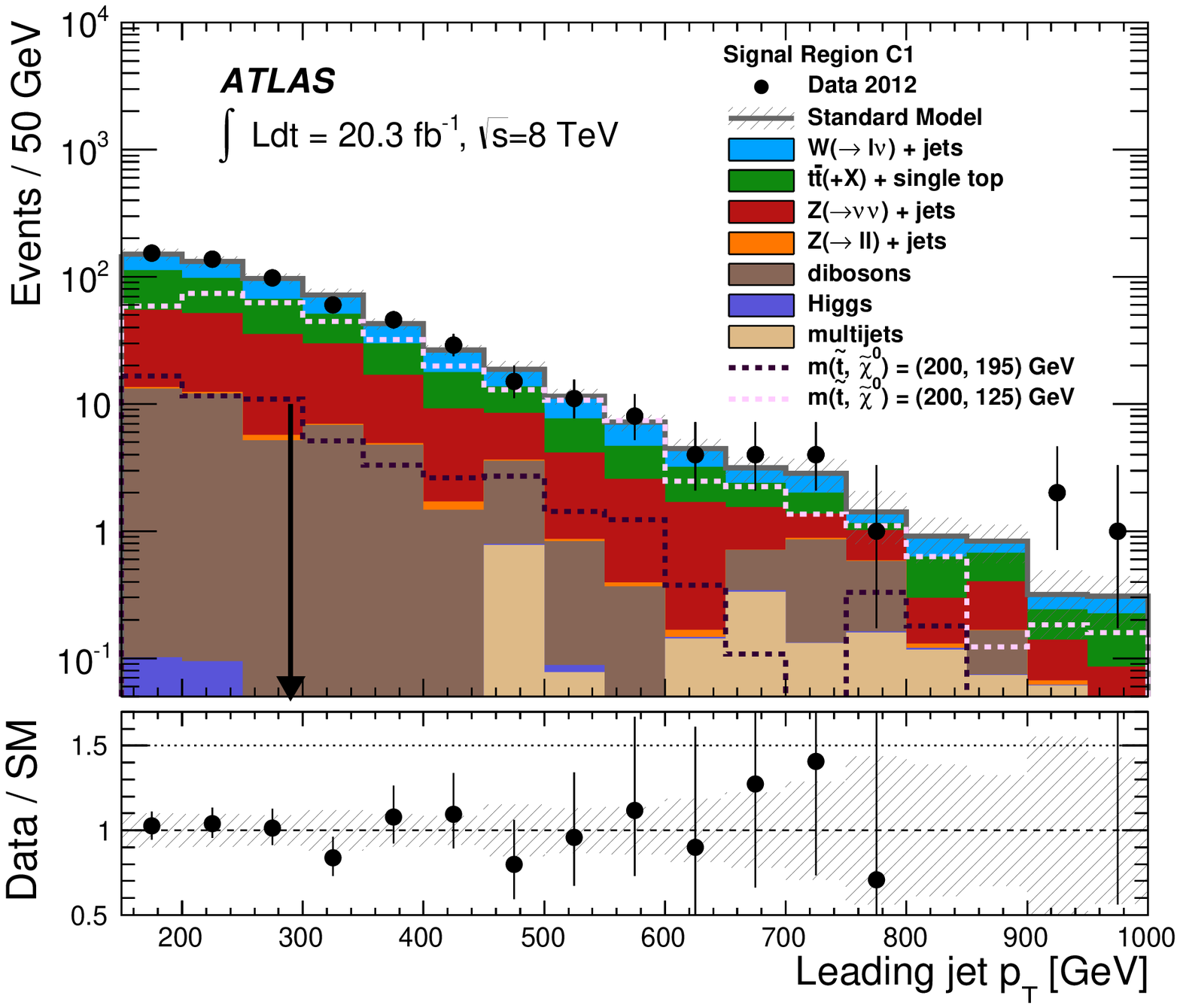}
}
\mbox{
  \includegraphics[width=0.42\textwidth]{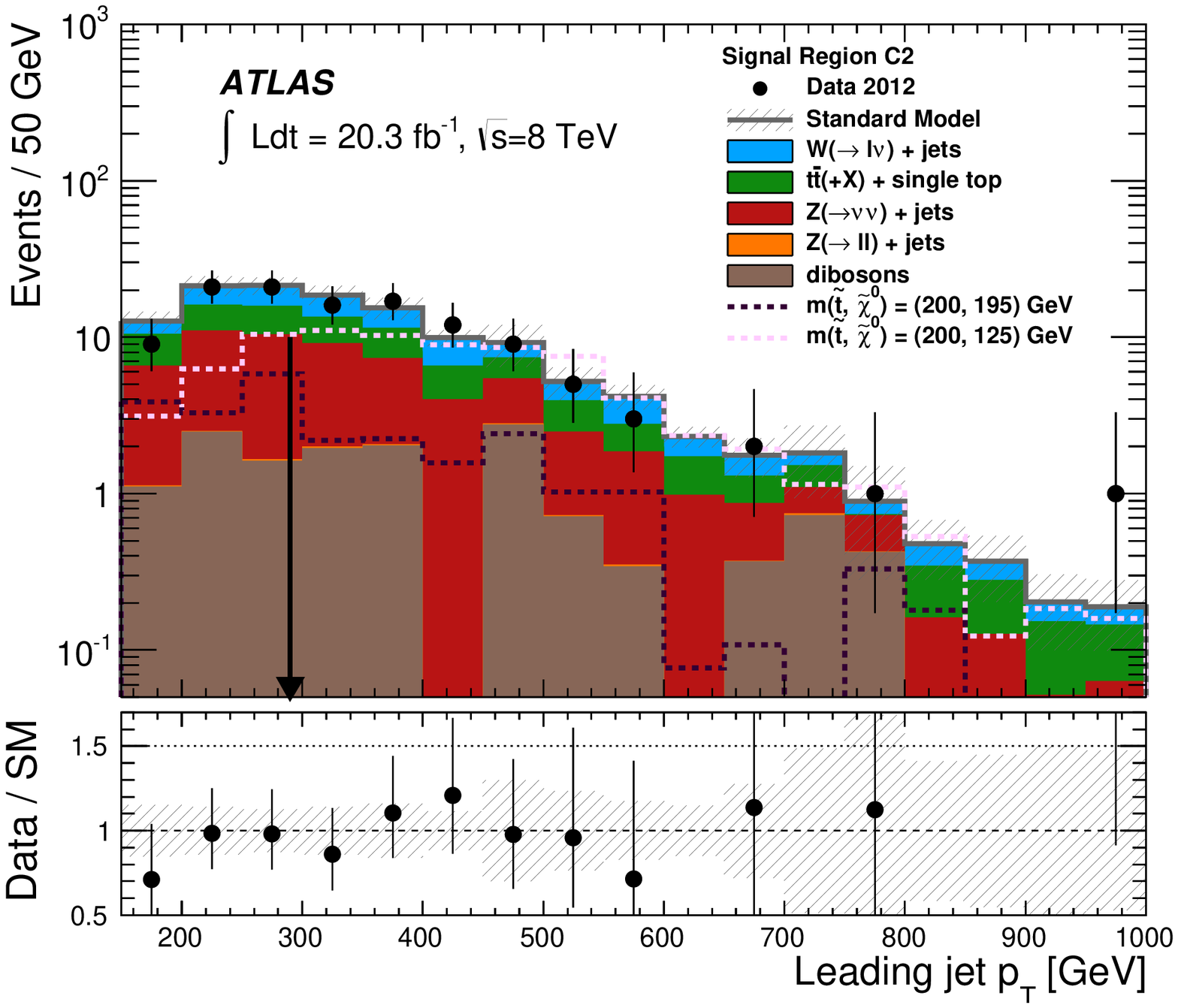}
  \includegraphics[width=0.42\textwidth]{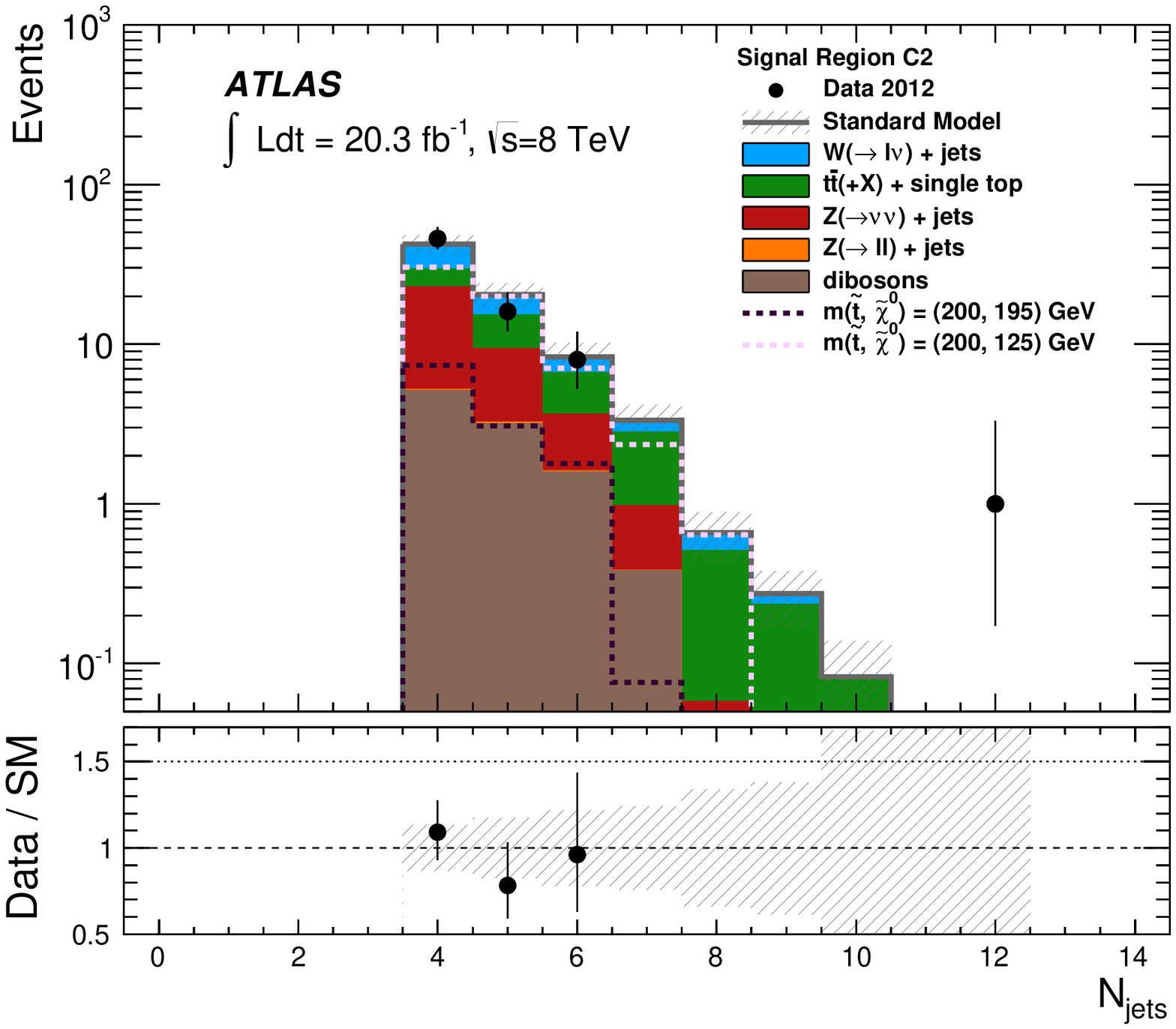}
}
\end{center}
\caption{
(top) Measured $\met$ and leading jet $\pt$ distributions for the C1 selection before the cut in the 
variable shown (as indicated by the vertical arrows) is applied.  In the case of the $\met$ distribution,  
the cuts corresponding to C1 and C2 selections are both indicated. (bottom) 
Measured leading jet $\pt$ and jet multiplicity for the C2 selection. The data are compared to the SM predictions. 
For illustration purposes, the distribution of two different SUSY scenarios are included.
The error bands in the ratios include both the statistical and systematic uncertainties 
on the background predictions.
}
\label{fig:sr2}
\end{figure*}

The data and the expected background predictions for the monojet-like and $c$-tagged  analyses are 
summarized in Table~\ref{tab:sr123}. Good agreement is observed between the data and the SM predictions in each case.
The  SM predictions for the monojet-like selections are 
determined with a total uncertainty of 2.9$\%$, 3.2$\%$,  and 4.6$\%$ for the M1, M2, M3 
signal regions, respectively, which include correlations between uncertainties on the individual 
background contributions. 
Similarly, the SM predictions for the $c$-tagged analyses are 
determined with a total uncertainty of 10$\%$  for C1 and 14$\%$ for C2 selections.
  Figure~\ref{fig:sr1} shows  the measured
leading jet $\ptjet$ and $\met$ distributions for the monojet-like  selections
compared to the background predictions. Similarly, Fig.~\ref{fig:sr2} presents
the leading jet $\ptjet$, $\met$ and jet multiplicity distributions for the
$c$-tagged  selections.
For illustration purposes, the distribution of two different SUSY scenarios for stop pair production 
in the $\tilde{t}_1 \to  c +  \tilde{\chi}^{0}_{1}$ decay channel with 
stop masses of 200~GeV and neutralino masses of 125~GeV and 195~GeV are included.   

%
%


\begin{table*}[!ht]
\caption{Data and SM background prediction in the signal region for the monojet-like and $c$-tagged selections.
For the SM predictions both the statistical and systematic uncertainties are included.
In each case the individual
uncertainties can be correlated, and do not necessarily add in quadrature to the total background uncertainty.
}
\begin{center}
\begin{small}
\begin{tabular*}{\textwidth}{@{\extracolsep{\fill}}lrrrrr}\hline
{\bf  Signal Region}  & M1 & M2 & M3 & C1 & C2\\
Observed events  (20.3 fb${}^{-1}$)&  $33054$  & $8606$  & $1776$  &  208  & 71   \\ \hline
SM prediction &  $33450 \pm 960$  & $8620 \pm 270$ & $1770 \pm 81$ & $210 \pm 21$   &  $75 \pm 11$  \\ \hline
$\wen$        &  $3300 \pm 140$ &  $700 \pm 43$   & $130 \pm 12$ & $11 \pm 2$ &  $3.0 \pm 0.7$   \\
$\wmn$        &  $3000 \pm 100$ &  $700 \pm 29$  & $133 \pm 8$ & $8 \pm 2$&  $3.0 \pm 0.7$   \\
$\wtn$        &  $7800 \pm 290$  & $1690 \pm 74$   & $320 \pm 24$ & $42 \pm 9$& $14 \pm 3$   \\
$\zee$        &  $-$   &  $-$ & $-$ & $-$  & $-$  \\
$\zmm$        &  $170 \pm 27$  & $53 \pm 9$  & $13 \pm 3$ & $0.07 \pm 0.01$& $0.04 \pm 0.01$   \\
$\ztt$        &  $95 \pm 6$    & $17 \pm 1$ & $1.8 \pm 0.3$ &  $0.7 \pm 0.1$& $0.15 \pm 0.03$ \\
$\znn$        &  $17400 \pm 720$ & $5100 \pm 240$  &$1090 \pm 72$  & $62 \pm 9$& $27 \pm 3$  \\ 
$\ttbar$, single top, $\ttbar$+V           &     $780 \pm 73$   &  $150 \pm 19$ & $27 \pm 4$  & $63 \pm 13$& $18 \pm 4$  \\
Dibosons      &  $650 \pm 99$  & $220 \pm 40$  &  $60 \pm 14$  & $21 \pm 13$  & $10 \pm 9$   \\ 
Higgs         & $-$   &  $-$ & $-$ & $0.16 \pm 0.03$ & $0.07 \pm 0.01$  \\
Multijets     &  $300 \pm 300$ &  $30 \pm 30$ &  $4 \pm 4$ & $2 \pm 2$ & $0.1 \pm 0.1$  \\ \hline\hline
\end{tabular*}
\end{small}
\end{center}
\label{tab:sr123}
\end{table*}

%
%

%
%

The agreement between the data and the SM predictions for the total number of events  
in the different signal regions is translated into
 95$\%$ confidence level (CL) upper limits 
on the visible cross section, $\sigma \times A \times \epsilon$,   
using the $CL_s$ modified frequentist approach~\cite{Read:2002hq},  
considering  the systematic 
uncertainties on the SM backgrounds 
and assuming there is no signal contamination in the control regions.
The upper limits are derived from pseudo-experiments and from an asymptotic approximation~\cite{statforumlimits}, 
which gives similar results.
For the monojet-like analysis, values of $\sigma \times A \times \epsilon$ 
in the range between 96~fb and 9.6~fb are excluded at 95$\%$ CL.  In the 
case of the $c$-tagged analysis, visible cross sections 
above 1.76~fb and 0.95~fb, for the C1 and the C2 selections, respectively,  are excluded at 95$\%$ CL,   
as shown in Table~\ref{tab:milimits}.

%
%

\begin{table*}[!ht]
\caption[Breakdown of upper limits.]{
Left to right: 95\% CL upper limits on the visible cross section
($\langle\sigma\rangle_{\rm obs}^{95}$) and on the number of
signal events ($S_{\rm obs}^{95}$).  The third column
($S_{\rm exp}^{95}$) shows the 95\% CL upper limit on the number of
signal events, given the expected number (and $\pm 1\sigma$
on the expectation) of background events.
The $CL_B$ value, i.e. the confidence level observed for
the background-only hypothesis, and the $p_0$ values, which represents
the probability of the background alone to fluctuate to the observed
numbers of events or higher, are also reported. The $p_0$-values
are truncated at 0.5 if the number of observed events is below the
number of expected events. The limits derived using an asymptotic
approximation instead of pseudo-experiments are given in parentheses.
}
\begin{center}
\begin{small}
\setlength{\tabcolsep}{0.0pc}
\begin{tabular*}{\textwidth}{@{\extracolsep{\fill}}cccccc}
\noalign{\smallskip}\hline\noalign{\smallskip}
{\bf Signal region}                        & $\langle{\rm \sigma}\rangle_{\rm obs}^{95}$[fb]  &  $S_{\rm obs}^{95}$  & $S_{\rm exp}^{95}$ & $CL_{B}$ & $p_0$
   \\
\noalign{\smallskip}\hline\noalign{\smallskip}
M1   &  {$96.2$}  {$(95.4)$} &  {$1951$}   {$(1935)$} &  {${1960}^{+840}_{-320}$}  {$({1950}^{+850}_{-290})$} & $0.49$ &  $0.50$  \\
M2   &  {$28.4$}  {$(28.7)$} &  {$575$}   {$(581)$}   &  {${590}^{+210}_{-120}$}  {$({600}^{+200}_{-120})$} & $0.48$ &  $0.50$  \\
M3   &  {$9.6$}   {$(9.6)$}  &  {$195$}    {$(195)$}  &  {${190}^{+69}_{-53}$}  {$({190}^{+69}_{-54})$} & $0.51$ &  $0.49$  \\
C1 & 1.76 (1.75) &  35.8 (35.5)  & ${37}^{+9}_{-10}$ $({37}^{+10}_{-11})$ & $0.45$ & 0.50  \\
C2 & 0.95 (0.93) &  19.3 (18.9)  & ${22}^{+8}_{-6}$   $({22}^{+9}_{-6})$& $0.35$ & 0.50  \\ 
\noalign{\smallskip}\hline\noalign{\smallskip}
\end{tabular*}
\end{small}
\end{center}
\label{tab:milimits}
\end{table*}

%
%

\subsection{Stop pair production with $\tilde{t}_1 \to  c +  \tilde{\chi}^{0}_{1}$ }

The results are then translated into exclusion limits on the pair production of top squarks  
with $\tilde{t}_1 \to  c +  \tilde{\chi}^{0}_{1}$ (BR=100$\%$) as a function of the stop mass for 
different neutralino masses.   
Expected and observed 95$\%$ CL exclusion limits are set 
using the $CL_s$ approach, for which a simultaneous fit to the signal and control regions is performed 
including statistical and systematic uncertainties. 
Uncertainties on the signal acceptance times efficiency, 
the background predictions, and the luminosity are considered, and 
correlations between systematic uncertainties on signal and background predictions 
are taken into account. The fit accounts for any potential contamination of signal events 
in the control regions which a priori has been estimated to be very small.
In addition, observed limits are computed 
using 
$\pm 1\sigma$ variations on the theoretical predictions for the SUSY  cross sections.
For each SUSY point considered, observed and expected limits are computed separately
for the different monojet-like and $c$-tagged analyses, and the one with the best expected limit is 
adopted as the nominal result. Finally, the 95$\%$ CL observed  
limits corresponding to the $- 1\sigma$ variations on the SUSY theoretical 
cross sections are then  quoted. 

\begin{figure*}[!h]
\centering
\includegraphics[width=0.45\textwidth]{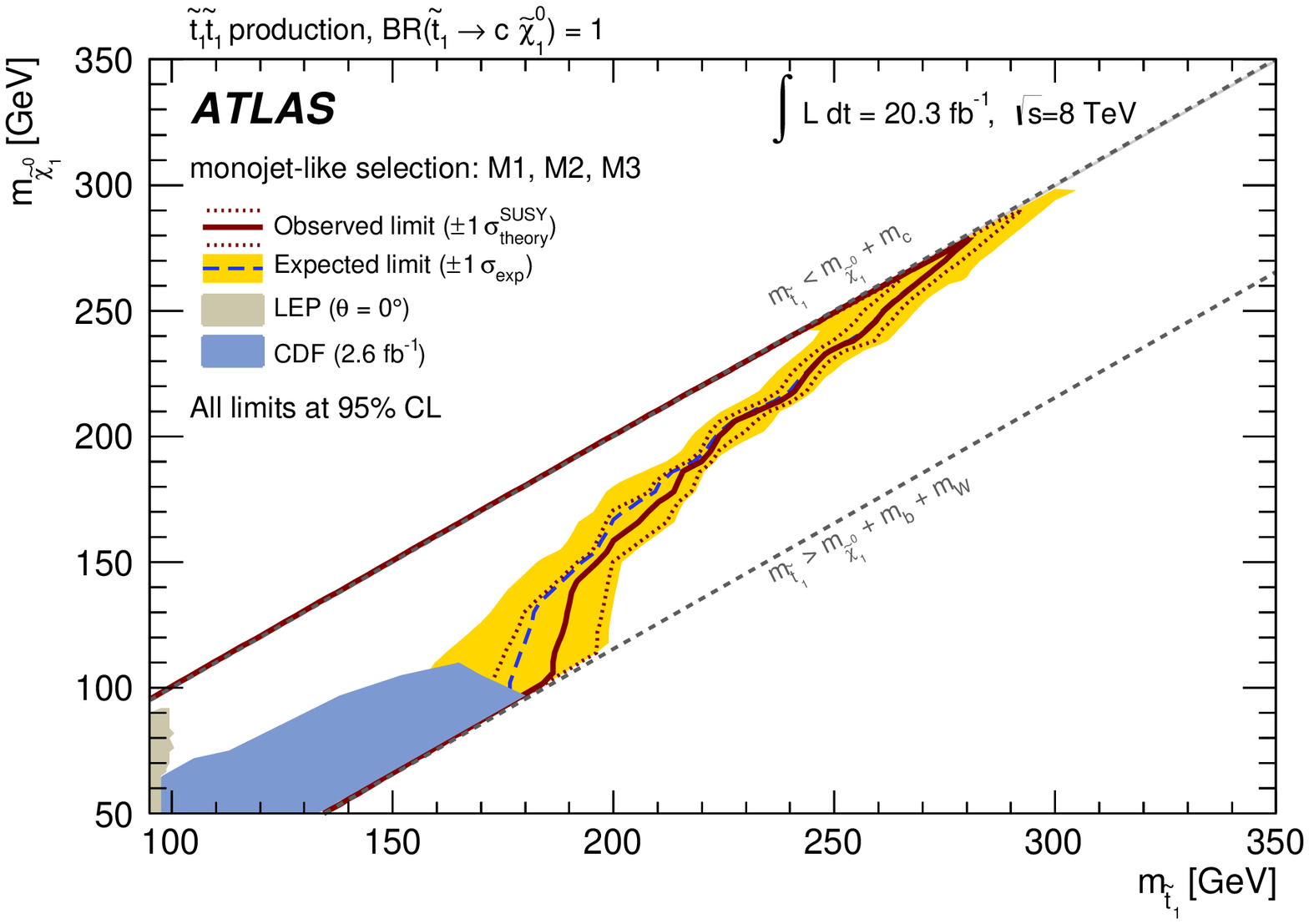}
\includegraphics[width=0.45\textwidth]{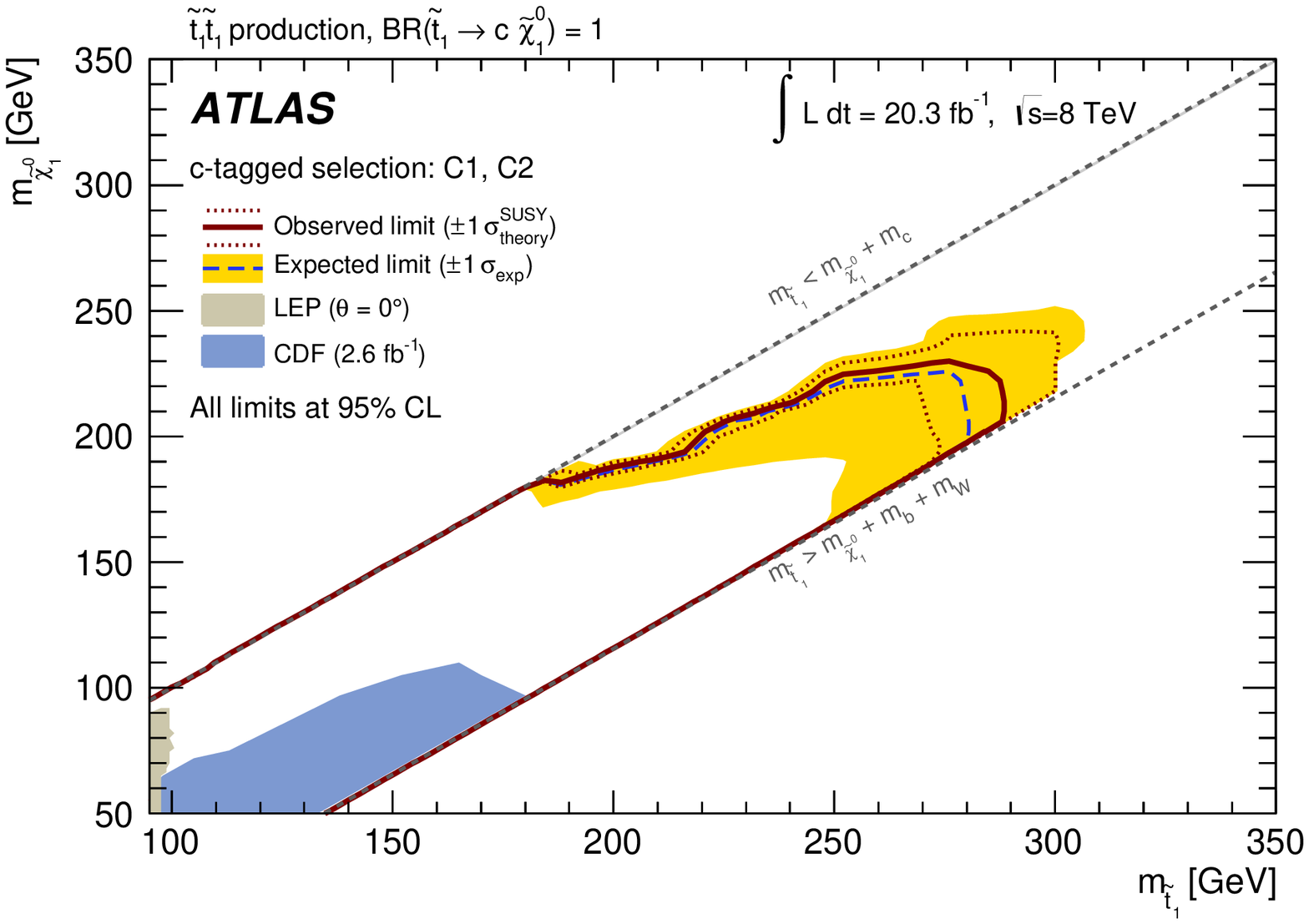}
\caption{
Exclusion plane at 95$\%$ CL as a function of stop and neutralino masses for the decay channel  $\tilde{t}_1 \to  c +  \tilde{\chi}^{0}_{1}$ (BR=100$\%$) as
determined separately for the monojet-like (left) and the $c$-tagged (right) selections.
The observed (red line) and expected (blue line) upper limits from this analysis are compared to previous results from Tevatron
experiments~\cite{Aaltonen:2012tq,Abazov:2008rc}, and from
LEP~\cite{lepsusy} experiments at CERN with squark mixing angle
$\theta = 0^{o}$.
The dotted lines around the observed limit
indicate the range of observed limits corresponding to $\pm 1\sigma$ variations on the NLO SUSY cross-section predictions.  The
shaded area around the
expected limit  indicates the expected $\pm 1\sigma$
ranges of limits in the absence of a signal.  A band for  $\Delta m < 2$~GeV indicates the region in the phase space for which
the stop can become long-lived.
}
\label{fig:ex}
\end{figure*}

\begin{figure}[!htbp]
\centering
\includegraphics[width=0.45\textwidth]{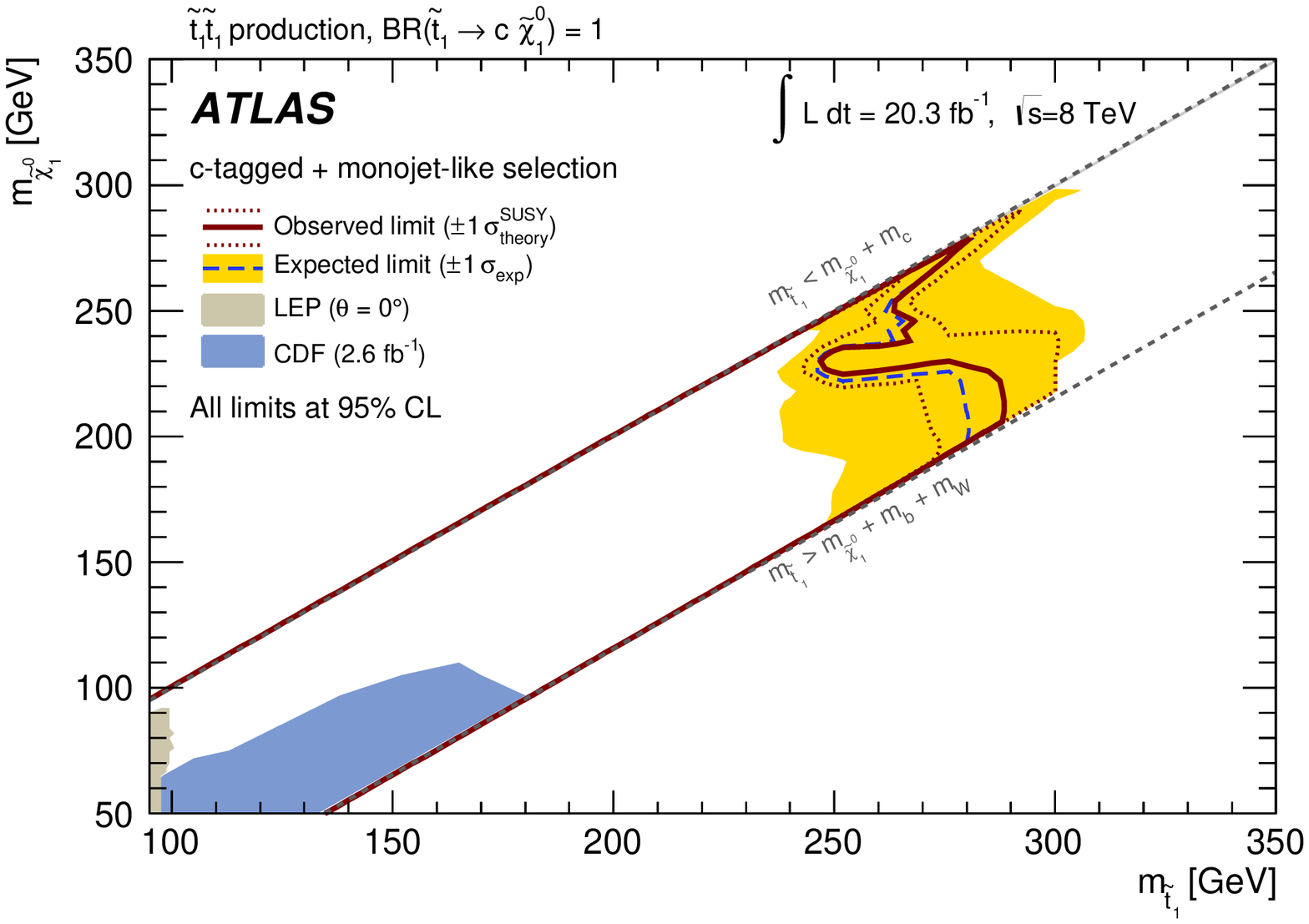}
\caption{ 
Exclusion plane at 95$\%$ CL as a function of stop and neutralino masses for the decay channel  $\tilde{t}_1 \to  c +  \tilde{\chi}^{0}_{1}$ (BR=100$\%$). 
The observed (red line) and expected (blue line) upper limits from this analysis are compared to previous results from Tevatron 
experiments~\cite{Aaltonen:2012tq,Abazov:2008rc}, and from
LEP~\cite{lepsusy} experiments at CERN with squark mixing angle
$\theta = 0^{o}$.  
The dotted lines around the observed limit
indicate the range of observed limits corresponding to $\pm 1\sigma$ variations on the NLO SUSY cross-section predictions.  The 
shaded area around the 
expected limit  indicates the expected $\pm 1\sigma$
ranges of limits in the absence of a signal.  A band for  $\Delta m < 2$~GeV indicates the region in the phase space for which 
the stop can become long-lived.    
}
\label{fig:exclusion}
\end{figure}

%
%
Figure~\ref{fig:ex} shows the results separately for the monojet-like and $c$-tagged analyses, illustrating their complementary regions of sensitivity.
As anticipated, the monojet-like selections drive 
the exclusion limits at very low $\Delta m$ for which the M2 and M3 signal regions 
enhance the sensitivity to large stop and neutralino masses. 
The $c$-tagged results determine the exclusion limits in the rest 
of the plane. Figure~\ref{fig:exclusion} presents the combined results. Masses for the stop up to 240~GeV
are excluded at 95$\%$ CL for arbitrary neutralino masses, within the kinematic boundaries. For neutralino masses 
of about 200~GeV, stop masses below 270~GeV are excluded at 95$\%$ CL.  
In the compressed scenario with the stop and neutralino nearly degenerate in mass, the exclusion extends up to stop masses of 
260~GeV.  The region with $\Delta m < 2$~GeV is not considered in the exclusion since 
in this regime the stop could become long-lived.
These results significantly extend previous 
exclusion limits~\cite{Aaltonen:2012tq,Abazov:2008rc} on the stop and neutralino masses in this channel.
%
%

\subsection{Stop and sbottom pair production with $\tilde{t}_1 \to  b + ff^{'} + \tilde{\chi}^{0}_{1}$  and  $\tilde{b}_1 \to  b +  \tilde{\chi}^{0}_{1}$}

\begin{figure}[!htbp]
 \begin{center}
 \mbox{
   \includegraphics[width=0.45\textwidth]{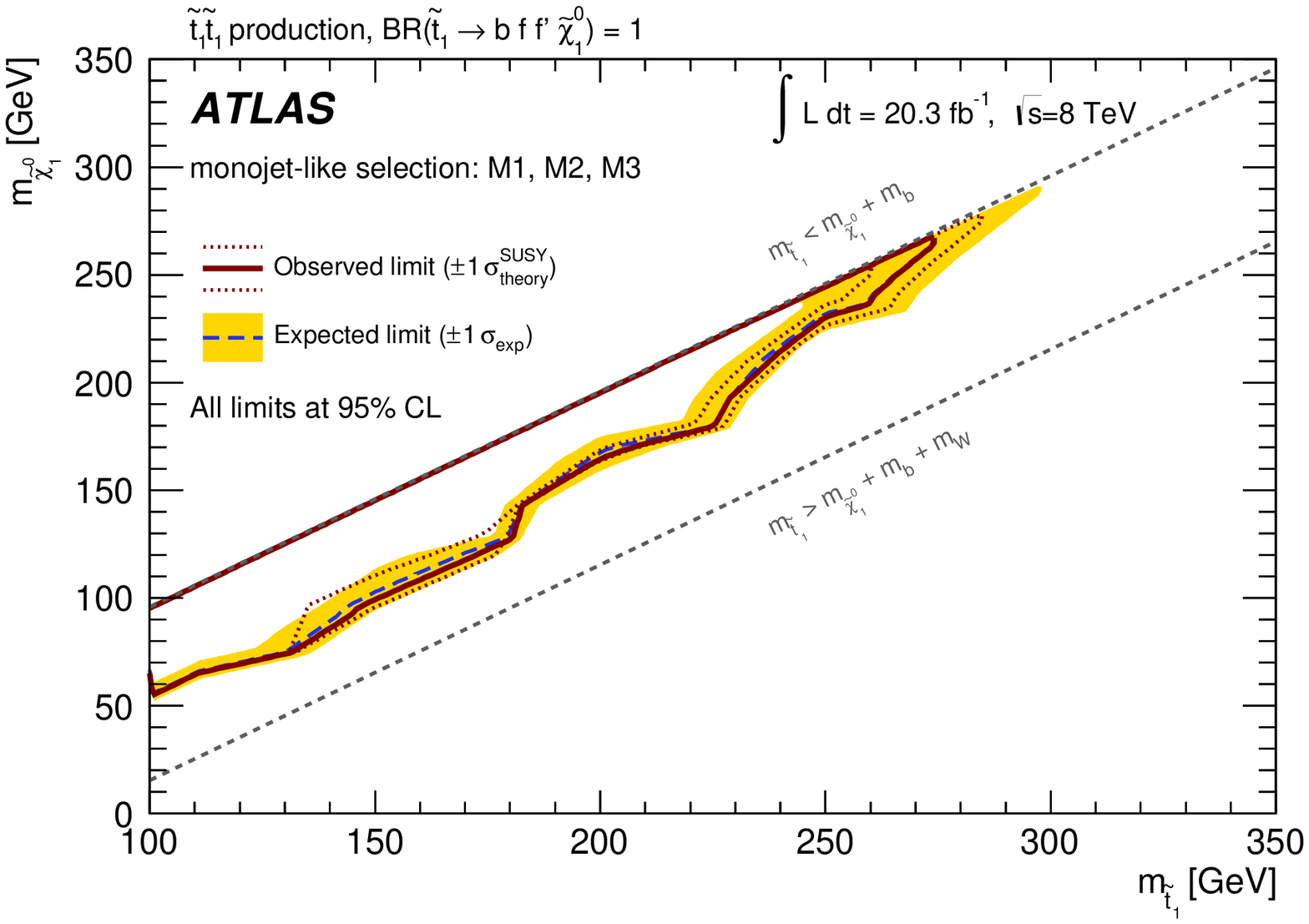}
}
\end{center}
\caption{
Exclusion plane at 95$\%$ CL as a function of stop and neutralino masses for the decay channel  $\tilde{t}_1 \to  b + ff^{'} +  \tilde{\chi}^{0}_{1}$ (BR=100$\%$). 
The dotted lines around the observed limit
indicate the range of observed limits corresponding to $\pm 1\sigma$ variations on the NLO SUSY cross-section predictions.  The 
shaded area around the 
expected limit  indicates the expected $\pm 1\sigma$
ranges of limits in the absence of a signal.  A band for  $m_{\tilde{t}_1} - m_{\ninoone} - m_{b} < 2$~GeV indicates the region in the phase space for which 
the stop can become long-lived.
}
 \label{fig:4body}
 \end{figure}

The monojet-like results are also interpreted in terms of 
exclusion limits on the stop pair production in the four-body decay mode 
$\tilde{t}_1 \to  b + ff^{'} + \tilde{\chi}^{0}_{1}$ (BR=100$\%$) 
and the sbottom pair production with  $\tilde{b}_1 \to  b +  \tilde{\chi}^{0}_{1}$ (BR=100$\%$),    
using the same $CL_s$ approach as explained above. As already mentioned, this is particularly relevant 
in a mass-degenerate scenario in which the
decay products of the squarks are too soft to be identified in the final state, 
and the signal selection relies on the presence of an ISR jet.
Figure~\ref{fig:4body} shows the
expected and observed 95$\%$ CL exclusion limits 
as a function of the stop and neutralino masses for the  $\tilde{t}_1 \to  b + ff^{'} +  \tilde{\chi}^{0}_{1}$ decay channel.
For $\Delta m \sim m_b$, stop masses up to 255~GeV are excluded at 95$\%$ CL.  Top squarks 
with mass of about 150 GeV and 200 GeV are excluded for $m_b < \Delta m < 50$~GeV and  $m_b < \Delta m < 35$~GeV, respectively.  

\begin{figure}[!htbp]
 \begin{center}
 \mbox{
   \includegraphics[width=0.42\textwidth]{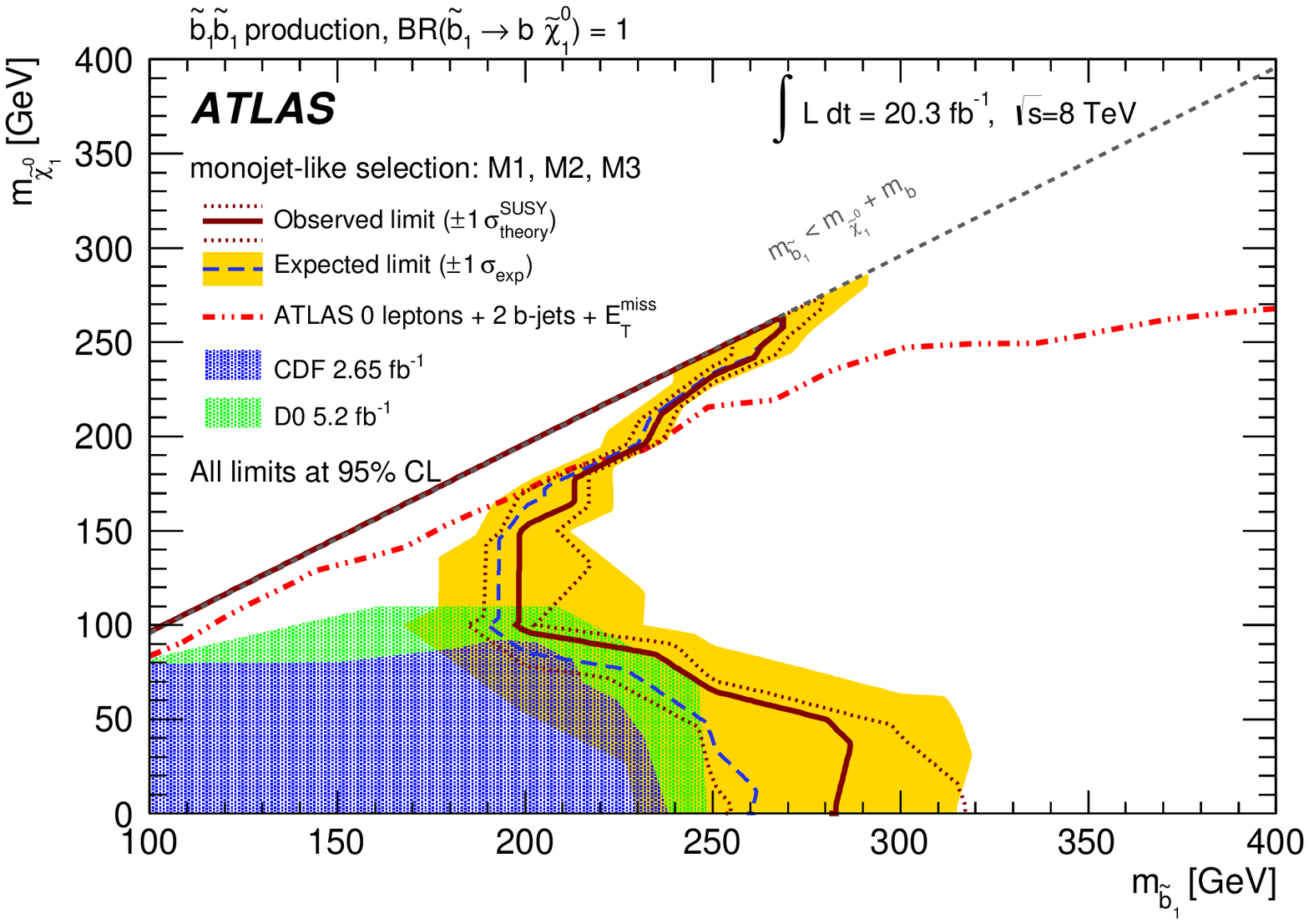}
}
\end{center}
\caption{
Exclusion plane at 95$\%$ CL as a function of sbottom and neutralino masses for the decay channel  $\tilde{b}_1 \to  b +  \tilde{\chi}^{0}_{1}$ (BR=100$\%$). 
The observed (red line) and expected (blue line) upper limits from this analysis are compared to previous results from 
CDF~\cite{Aaltonen:2010dy}, D0~\cite{Abazov:2010wq}, and ATLAS~\cite{Aad:2013ija}. For the latter, the area below 
the dashed-dotted line is excluded. 
The dotted lines around the observed limit
indicate the range of observed limits corresponding to $\pm 1\sigma$ variations on the NLO SUSY cross-section predictions.  The 
shaded area around the 
expected limit  indicates the expected $\pm 1\sigma$
ranges of limits in the absence of a signal.  
A band for  $m_{\tilde{b}_1} - m_{\ninoone} - m_{b} < 2$~GeV indicates the region in the phase space for which 
the sbottom can become long-lived.
}
 \label{fig:sbottom}
 \end{figure}

\noindent
Finally, Fig.~\ref{fig:sbottom} presents the  expected and observed 95$\%$ CL exclusion limits 
as a function of the sbottom and neutralino masses for the  $\tilde{b}_1 \to  b + \tilde{\chi}^{0}_{1}$ decay channel, compared to 
previous results. In the scenario with $m_{\tilde{b}_1} - m_{\ninoone} \sim m_b$, this analysis extends the 95$\%$ CL 
exclusion limits up to a sbottom mass of 255~GeV. 

\section{Conclusions}
\label{sec:sum}

In summary, this paper presents results of a search for
stop pair production in the decay channel $\tilde{t}_1 \to c + \tilde{\chi}^{0}_{1}$
using 20.3~fb${}^{-1}$ of proton-proton collision data at $\sqrt{s}=8$~TeV
recorded with the ATLAS experiment at the LHC. Two different analysis  strategies based on monojet-like and 
$c$-tagged event selections are 
carried out that optimize the sensitivity across the stop--neutralino mass plane. Good agreement is observed between the data and the SM predictions. 
The results are translated into 95$\%$ CL exclusion limits on the stop and neutralino masses.
A stop mass of about 240~GeV is excluded at 95$\%$ confidence
level for $m_{\tilde{t}_1}$--$m_{\tilde{\chi}_{1}^{0}} < 85$~GeV, 
as the maximum mass difference
in which the decay mode $\tilde{t}_1 \to c + \tilde{\chi}_{1}^{0}$ dominates. 
Stop masses up to 270~GeV are excluded for a neutralino mass of 200~GeV. 
In a scenario with the stop  and the lightest neutralino nearly 
degenerate in mass, stop  
masses up to 260~GeV are excluded.  
The results from the monojet-like analysis are also 
re-interpreted in terms of
stop pair production in the four-body decay channel $\tilde{t}_1 \to b + ff^{'} + \tilde{\chi}^{0}_{1}$ and 
sbottom pair production with $\tilde{b}_1 \to b + \tilde{\chi}^{0}_{1}$, leading to 
a similar exclusion for the mass-degenerate scenario. The results in this paper 
significantly extend previous results~\cite{lepsusy,Aaltonen:2012tq,Abazov:2008rc,Aaltonen:2010dy,Abazov:2010wq,Aad:2013ija} at colliders. 


\section*{Acknowledgments}

We thank CERN for the very successful operation of the LHC, as well as the
support staff from our institutions without whom ATLAS could not be
operated efficiently.

We acknowledge the support of ANPCyT, Argentina; YerPhI, Armenia; ARC,
Australia; BMWF and FWF, Austria; ANAS, Azerbaijan; SSTC, Belarus; CNPq and FAPESP,
Brazil; NSERC, NRC and CFI, Canada; CERN; CONICYT, Chile; CAS, MOST and NSFC,
China; COLCIENCIAS, Colombia; MSMT CR, MPO CR and VSC CR, Czech Republic;
DNRF, DNSRC and Lundbeck Foundation, Denmark; EPLANET, ERC and NSRF, European Union;
IN2P3-CNRS, CEA-DSM/IRFU, France; GNSF, Georgia; BMBF, DFG, HGF, MPG and AvH
Foundation, Germany; GSRT and NSRF, Greece; ISF, MINERVA, GIF, I-CORE and Benoziyo Center,
Israel; INFN, Italy; MEXT and JSPS, Japan; CNRST, Morocco; FOM and NWO,
Netherlands; BRF and RCN, Norway; MNiSW and NCN, Poland; GRICES and FCT, Portugal; MNE/IFA, Romania; MES of Russia and ROSATOM, Russian Federation; JINR; MSTD,
Serbia; MSSR, Slovakia; ARRS and MIZ\v{S}, Slovenia; DST/NRF, South Africa;
MINECO, Spain; SRC and Wallenberg Foundation, Sweden; SER, SNSF and Cantons of
Bern and Geneva, Switzerland; NSC, Taiwan; TAEK, Turkey; STFC, the Royal
Society and Leverhulme Trust, United Kingdom; DOE and NSF, United States of
America.

The crucial computing support from all WLCG partners is acknowledged
gratefully, in particular from CERN and the ATLAS Tier-1 facilities at
TRIUMF (Canada), NDGF (Denmark, Norway, Sweden), CC-IN2P3 (France),
KIT/GridKA (Germany), INFN-CNAF (Italy), NL-T1 (Netherlands), PIC (Spain),
ASGC (Taiwan), RAL (UK) and BNL (USA) and in the Tier-2 facilities
worldwide.

\clearpage
\bibliographystyle{apsrev4-1}

%


\onecolumngrid
\clearpage 
\begin{flushleft}
{\Large The ATLAS Collaboration}

\bigskip

G.~Aad$^{\rm 84}$,
B.~Abbott$^{\rm 112}$,
J.~Abdallah$^{\rm 152}$,
S.~Abdel~Khalek$^{\rm 116}$,
O.~Abdinov$^{\rm 11}$,
R.~Aben$^{\rm 106}$,
B.~Abi$^{\rm 113}$,
M.~Abolins$^{\rm 89}$,
O.S.~AbouZeid$^{\rm 159}$,
H.~Abramowicz$^{\rm 154}$
% ...
\bigskip
\\
$^{1}$ Department of Physics, University of Adelaide, Adelaide, Australia\\
% ...
$^{11}$ Institute of Physics, Azerbaijan Academy of Sciences, Baku, Azerbaijan\\
% ...
$^{84}$ CPPM, Aix-Marseille Universit{\'e} and CNRS/IN2P3, Marseille, France\\
% ...
$^{89}$ Department of Physics and Astronomy, Michigan State University, East Lansing MI, United States of America\\
% ...
$^{106}$ Nikhef National Institute for Subatomic Physics and University of Amsterdam, Amsterdam, Netherlands\\
% ...
$^{112}$ Homer L. Dodge Department of Physics and Astronomy, University of Oklahoma, Norman OK, United States of America\\
$^{113}$ Department of Physics, Oklahoma State University, Stillwater OK, United States of America\\
% ...
$^{116}$ LAL, Universit{\'e} Paris-Sud and CNRS/IN2P3, Orsay, France\\
% ...
$^{152}$ Institute of Physics, Academia Sinica, Taipei, Taiwan\\
% ...
$^{154}$ Raymond and Beverly Sackler School of Physics and Astronomy, Tel Aviv University, Tel Aviv, Israel\\
% ...
$^{159}$ Department of Physics, University of Toronto, Toronto ON, Canada\\
% ...
\end{flushleft}

\clearpage
\appendix


%
%

\end{document}